\documentclass[11pt, a4paper]{article}
\pdfoutput=1

\usepackage{amsmath}
\usepackage{amsfonts}
\usepackage{amssymb}
\usepackage{latexsym}
\usepackage{graphicx}
\usepackage{color}
\usepackage{mathrsfs}
\usepackage{slashed}
\usepackage{hyperref}
\usepackage{datetime}
\usepackage{slashed}
\usepackage{multirow}
\usepackage{makecell}
\usepackage{wasysym}

\numberwithin{equation}{section}

\hypersetup{colorlinks, citecolor=bluscuro, linkcolor=bluscuro, urlcolor=bluscuro}
\definecolor{rossos}{rgb}{0.8,0.2,0.3}
\definecolor{bluscuro}{rgb}{0.15, 0.2, 0.9}
\definecolor{verdes}{rgb}{0.1, 0.5, 0.1}
\definecolor{myred}{rgb}{0.85, 0, 0}
\definecolor{myblue}{rgb}{0, 0, 0.7}
\definecolor{mygreen}{rgb}{0, 0.45, 0.1}

\makeatother   
\newcommand{\GeV}{{\rm \,GeV}}

\def\fA{\mbox{{\bf M4$_{5}$}}}
\def\fB{\mbox{{\bf M4$_{14}$}}}
\def\oA{\mbox{{\bf M1$_{5}$}}}
\def\oB{\mbox{{\bf M1$_{14}$}}}

\def\gsim{\lower.7ex\hbox{$\;\stackrel{\textstyle>}{\sim}\;$}}
\def\lsim{\lower.7ex\hbox{$\;\stackrel{\textstyle<}{\sim}\;$}}

 \def\be   {\begin{equation}}   \def\ee   {\end{equation}}
 \def\ba   {\begin{array}}      \def\ea   {\end{array}}
 \def\bea  {\begin{eqnarray}}   \def\eea  {\end{eqnarray}}
 \def\bean {\begin{eqnarray*}}  \def\eean {\end{eqnarray*}}
 \def\nn{\nonumber}

\def\cw{c_{\textrm w}}
\def\sw{s_{\textrm w}}
\newcommand{\BR}{{\rm \,BR}}

\def\lagr{\mathscr{L}}  

\def\Xtt{{X_{\hspace{-0.09em}\mbox{\scriptsize2}\hspace{-0.06em}{\raisebox{0.1em}{\tiny\slash}}\hspace{-0.06em}\mbox{\scriptsize3}}}}
\def\Xft{{X_{\hspace{-0.09em}\mbox{\scriptsize5}\hspace{-0.06em}{\raisebox{0.1em}{\tiny\slash}}\hspace{-0.06em}\mbox{\scriptsize3}}}}
\def\Tt{\widetilde{T}}

\def\barXtt{{\overline{X}_{\hspace{-0.09em}\mbox{\scriptsize2}\hspace{-0.06em}{\raisebox{0.1em}{\tiny\slash}}\hspace{-0.06em}\mbox{\scriptsize3}}}}
\def\barXft{{\overline{X}_{\hspace{-0.09em}\mbox{\scriptsize5}\hspace{-0.06em}{\raisebox{0.1em}{\tiny\slash}}\hspace{-0.06em}\mbox{\scriptsize3}}}}

\def\primeXtt{{{X}'_{\hspace{-0.09em}\mbox{\scriptsize2}\hspace{-0.06em}{\raisebox{0.1em}{\tiny\slash}}\hspace{-0.06em}\mbox{\scriptsize3}}}}
\def\barprimeXtt{{\overline{X}'_{\hspace{-0.09em}\mbox{\scriptsize2}\hspace{-0.06em}{\raisebox{0.1em}{\tiny\slash}}\hspace{-0.06em}\mbox{\scriptsize3}}}}

\begin{document}

\begin{flushright} CERN-PH-TH/2012-323\\
SISSA 31/2012/EP\end{flushright}

\vspace{1.5cm}
\begin{center}

{\LARGE \textbf {
A First Top Partner Hunter's Guide
}}
\\ [1.2cm]
{ 
\textsc{Andrea De Simone}$^{a,\,b}$, 
\textsc{Oleksii Matsedonskyi}$^c$,
\textsc{Riccardo Rattazzi}$^d$,
\textsc{Andrea Wulzer}$^{c\,,e}$}
\\[1cm]

$^a$
\textit{CERN, Theory Division, CH-1211 Geneva 23, Switzerland}\\
$^b$
\textit{SISSA and INFN, Sezione di Trieste, Via Bonomea 265, I-34136 Trieste, Italy}\\
$^c$
\textit{Dipartimento di Fisica e Astronomia and INFN, Sezione di Padova,\\
via Marzolo 8, I-35131 Padova, Italy}\\
$^d$
\textit{Institut de Th\'eorie des Ph\'enom\`enes Physiques,\\
 \'Ecole Polytechnique F\'ed\'erale de Lausanne, CH-1015 Lausanne,
Switzerland}\\
$^e$
\textit{Institute for Theoretical Physics, ETH Zurich, CH-8093 Zurich, Switzerland}

\end{center}

\vspace{1cm}
\begin{center} 
\textbf{Abstract}
\begin{quote}
We provide a systematic effective lagrangian description of the phenomenology of the lightest top-partners
in composite Higgs models. Our construction is  based on symmetry, on selection rules and on plausible dynamical assumptions. The structure of the resulting simplified models depends on the quantum numbers of the lightest top partner and of the operators involved in the generation of the top Yukawa. In all cases the phenomenology is conveniently described by a small number of parameters, and the results of experimental searches are readily interpreted as a test
of naturalness. We recast presently available experimental bounds on  heavy fermions into bounds on top partners: LHC has already stepped well inside the natural region of parameter space.

\end{quote}
\end{center}

\def\thefootnote{\arabic{footnote}}
\setcounter{footnote}{0}
\pagestyle{empty}

\newpage
\pagestyle{plain}
\setcounter{page}{1}

\section{Introduction}
The exploration of the weak scale at the Large Hadron Collider is  set to unveil the dynamics
of electroweak symmetry breaking. A giant step in that direction was achieved this year with the discovery
of a bosonic resonance, whose features are remarkably compatible with those of the Standard Model Higgs boson. 
Whether we like it or not, the main question now facing us concerns the role of naturalness in the dynamics of the newly discovered boson. Theoretically we can think of two broad scenarios that concretely realize naturalness: supersymmetry and compositeness. In the case of supersymmetry, the implications  and the search strategies  have been worked out in much greater detail than in the case of compositeness. That is explained partly by the undisputable  theoretical appeal of supersymmetry
(gauge coupling unification, connection with string theory, etc.) and partly by the comfort of dealing with a perturbative set up. The  difficulty in dealing with strong dynamics has instead, and for a long time, slowed down progress in the exploration of compositeness, and, in particular,   progress on its objective phenomenological difficulties (mostly flavor, but also  precision tests). Interesting ideas were indeed put forward early on \cite{Kaplan:1983fs,partcomp}, but the absence of a weakly coupled approach prevented more concrete scenarios to appear. However, in the last decade, thanks in particular to the holographic perspective on compositeness,  {\it semi-perturbative} scenarios have been depicted and studied \cite{Agashe:2004rs}\footnote{By {\it semi-perturbative} here we mean that in these models, typified by  warped compactifications, there exists a sufficiently interesting subset of questions, even involving physics at energies well above the weak scale, that can be addressed using perturbation theory.}. Even though a very compelling single model did not cross our horizon, we believe we have learned how to broadly depict interesting scenarios, while remaining sufficiently agnostic on the details (see for instance \cite{silh}).
The first aspect of an interesting set up is that the Higgs  is  a pseudo-NG-boson associated with the spontaneous breakdown of an approximate global symmetry. The second aspect is that flavor arises from {\it partial compositeness}: the quarks and leptons acquire a mass by mixing with  composite fermions. Partial compositeness,
although much more convincing than the alternatives, does not, by itself,
 lead to a fully realistic flavor scenario. This is  because of constraints from  $\epsilon_K$ \cite{Csaki:2008zd}, electric dipole moments  and lepton flavor violation (see Ref.~\cite{KerenZur:2012fr} for a recent appraisal). In a realistic scenario partial compositeness should likely be supplemented by additional symmetries. In any case, and regardless of details, a robust feature is that the Higgs potential is  largely determined by the dynamics associated with the top quark and the composite states it mixes to, the so-called top partners. That is in a sense obvious and expected,  as the top quark, because of its large coupling, color multiplicity and numerics already contributes the leading quadratically divergent correction to the Higgs mass within the SM. It is nonetheless useful to have depicted a scenario that concretely realizes that expectation. The naturalness of electroweak symmetry breaking depends then on the mass of the fermionic top-partners. That is in close analogy with the supersymmetric case, where naturalness is largely controlled by the mass of the bosonic  top partners, the stops.
 
 \begin{figure}[t]
\centering
 \includegraphics[scale=0.5]{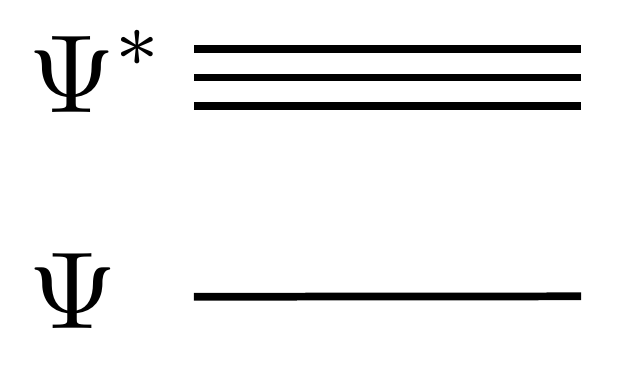}  
\label{fig0}
\caption{\small{Schematic picture of the spectrum.}}
\end{figure}
 
 The case of light stops in supersymmetry is being actively considered  both theoretically and experimentally.
 One main goal is to effectively cover all  regions of parameter space, without being swamped by the less relevant parameters. Simplified models or motivated  assumptions like ``natural susy'' \cite{Barbieri:2009ev}
offer a convenient way to achieve that goal.  In the reduced
parameter space (featuring stop mass parameters and possibly the gluino or lightest neutralino mass),
the constraints from of experimental searches offer a direct and  largely model independent  appraisal of naturalness. The goal of this paper is to provide a similar simplified approach to describe the results of experimental searches for top partners. 
We will focus on the composite Higgs scenario based on the minimal coset $SO(5)/SO(4)$. The basic simplifying assumption is that the  spectrum has the structure depicted in figure~\ref{fig0}, where one $SO(4)$ multiplet of colored Dirac fermions $\Psi$ is parametrically lighter than the other states. As already illustrated in Ref.~\cite{Contino:2011np} for the case of bosonic resonances, in that limit one expects the dynamics of $\Psi$ to be described by a weakly coupled effective lagrangian. Therefore the simplified model, at leading order in  an expansion in loops and derivatives, can  be consistently described by a finite number of parameters. Moreover symmetry and selection rules, via the Callan-Coleman-Wess-Zumino (CCWZ) 
 \cite{ccwz} construction, reduce the number of relevant parameters. It is obviously understood that the limiting situation presented by the simplified model is not expected to be precisely realized in a realistic scenario. However,  a realistic situation where the splitting with the next-to-lightest multiplet is of the order $M_\Psi$ is qualitatively already well described by the simplified model. Only if the splitting were parametrically smaller than $M_\Psi$ would there be dramatic changes. We should also stress that our models are truly minimal, in that they do not even possess sufficient structure (states and couplings) to make the Higgs potential calculable. In principle we could add that structure.
For instance by uplifting our multiplet $\Psi$ to a full split $SO(5)$ multiplet, like in a two site model, we could make the Higgs potential only logarithmically divergent, thus controlling its size in leading log approximation, and making the rough connection between $M_\Psi$ and naturalness more explicit along the lines of \cite{Matsedonskyi:2012ym}
(see also \cite{gillioz, babis, dissertori} for a similar construction). 
We could even go as far as making the one loop Higgs potential finite with a three site model 
\cite{panico,DeCurtis:2011yx}, or by imposing phenomenological Weinberg sum-rules \cite{Marzocca:2012zn, pomarolriva}. 
However in these less minimal models the first signals at the LHC would still be dominated by the lightest $SO(4)$ multiplet, whatever it may be. The point is that while the contribution of the heavier multiplets  does not decouple when focussing on a UV sensitive quantity like the Higgs potential, it does decouple when considering the near threshold production of the lightest states.  For the purpose of presenting the results of the LHC searches in an eloquent way, the simplified model is clearly the way to go. There already exists a literature on simplified top partner models in generic composite Higgs scenarios \cite{Contino:2006nn,continoservant,mrazecwulzer}, where the role of symmetry is not fully exploited. Focussing on the minimal composite Higgs model based on $SO(5)/SO(4)$, our paper aims at developing a systematic approach where all possible top partner models are constructed    purely on the basis of symmetry and selection rules. 

In the end  we shall derive exclusion plots in a reduced parameter space, which in general 
involves the mass and couplings of the top-partner $\Psi$. Now, even though these are not the parameters of a fundamental model,  given their overall size, we can roughly estimate how natural the Higgs sector is expected to be. We can then read  the results of searches as a test of the notion of naturalness. To make that connection, even if qualitative, we must  specifiy the dynamics that gives rise to the top Yukawa.  As  discussed in \cite{Panico:2012uw}, there are several options, each 
leading to a different structure of the Higgs potential and thus to a different level of tuning. The common feature of 
all scenarios is that the top partners need to be light for a reasonably natural theory, the way the tuning scales 
with the top-partners' mass is instead different in each case. In this paper we focus on the possibility that the right handed 
top quark $t_R$ is a $SO(4)$ singlet belonging to the strong sector, therefore the top Yukawa simply arises
from an $SO(5)$ breaking  perturbation of the form
\be
\lambda_L q_L {\cal O}_R+{\rm {h.c.}}\, .
\ee
Here  ${\cal O}_R$ is a  composite operator, which in the low energy theory maps to $Ht_R$, thus giving rise to
a top Yukawa coupling $y_t\sim \lambda_L$. The operator $O_R$ however also interpolates in general for massive states, the top partners. Now, from simple power counting, and also from explicit constructions \cite{panico}, at leading order in the breaking parameter $\lambda_L$ we expect the Higgs potential to have the form 
  \be
\label{powercount}
V(h)=\frac{3y_t^2 m_*^2}{16\pi^2}\left \{a h^2 +\frac{b}{2}\frac{h^4}{f^2}+\frac{c}{3!}\frac{h^6}{f^4}+\dots \right \} \, .
\ee
 where $a,b,c,\dots$ are coefficients expected to be $O(1)$, $f$ is the decay constant of the $\sigma$-model, while   $m_*$ broadly indicates the mass scale of the top partners. Then, since $\Psi$ is, ideally, the lightest top-partner we have  $M_\Psi \lsim m_*$. Given  $m_*$ and $f$, the measured values $v\equiv \langle h\rangle= 246$ GeV and $m_h= 125$ GeV, may require a tuning of $a$ and $b$ below their expected $O(1)$ size. More explicitly one finds
 \be
a= \frac{m_h^2}{m_*^2}\frac{4\pi^2}{3 y_t^2}\simeq\left(\frac{430 \,{\rm GeV}}{m_*}\right )^2
\label{a}
\ee
and,  defining the top-partner coupling as $g_*\equiv m_*/f$ according to  Ref.~\cite{silh},
\be
b= \frac{m_h^2}{m_t^2}\frac{2\pi^2}{3 g_*^2}\simeq \frac{4}{ g_*^2}\, .
\label{b}
\ee
By these equations we deduce that in the most natural scenario the top partners should not only be light (say below a TeV) but also not too strongly coupled. While of course the whole discussion is very qualitative, we still believe eqs.~(\ref{a})-(\ref{b})  give a valid rule of thumb for where the top partners should best be found. It is with eqs.~(\ref{a})-(\ref{b})   in mind that one should interpret the results of the searches for top partners. Notice that while naturalness favors sub-TeV 
fermionic resonances, electroweak precision constraints favor instead bosonic resonances above 2-3 TeV. A technically natural and viable model should therefore be more complex than a generic composite model described by a single scale.
This situation closely resembles that of supersymmetric models, where the light squark families and the gluinos are pushed up by direct searches, while technical naturalness demands the stops to be as light as possible.

This paper is organized as follows. In Section \ref{sect:models} we discuss the structure of the models and their main features
such as the mass spectrum and the couplings of the top partners.
Then, in Section \ref{sec:TPP} we turn to analyse the phenomenology of the top partners, their production mechanisms 
and decay channels, highlighting the most relevant channels to focus LHC searches on. 
The bounds on the model parameters are derived in Section \ref{sect:bounds}, using the LHC
data available at present
\footnote{
While this work was being completed ATLAS \cite{ATLASnew} and CMS \cite{CMSnew}
presented dedicated searches for top partners,
which we did not include in our analysis. From a preliminary investigation we expect mild changes in our results 
from these new data because both the ATLAS and the CMS searches are optimized to detect pair production. 
As we will discuss in the conclusions, a radical improvement of the bounds could perhaps be achieved, with the 
present energy and luminosity, but only with searches dedicated to single production.
}.
Finally, our concluding remarks are collected in Section \ref{conclusions}.

\section{The Models}
\label{sect:models}

Our first goal is to develop a simplified description of the top partners, suited for studying the phenomenology of their production at the LHC. These simplified models should 
capture the robust features of  more complete explicit constructions\footnote{See \cite{Panico:2012uw} for a complete calculable model with totally 
composite $t_R$, analogous holographic 5d models could be formulated following the approach of Ref.~\cite{Agashe:2004rs}.} or, better, of a putative general class of underlying theories. In particular,  robust, and crucial,  features  are the pNGB nature of the Higgs and the selection rules associated with the
small breaking of the corresponding global symmetry.
 We will see below that these features
strongly constraints the structure of the spectrum and of the couplings of the top partners, similarly to what was found in Ref.~\cite{panico} for the 
case of partial $t_R$ compositeness. 

We thus assume that the Higgs is the pNGB of the minimal coset ${\textrm{SO(5)}}/{\textrm{SO(4)}}$ and construct Lagrangians that respect the non-linearly 
realized ${\textrm{SO(5)}}$ invariance. We follow the standard CCWZ construction \cite{ccwz}, whose detailed formulation for our coset is described in Appendix~A. 
The CCWZ methodology has been first employed to model the top partners  in Ref.~\cite{Marzocca:2012zn}. 
The central objects are the Goldstone boson $5\times 5$ matrix $U$ and the $d_\mu$ and $e_\mu$ symbols constructed out of $U$ and  its derivative. The 
top partner field $\Psi$ has definite transformation properties under the unbroken ${\textrm{SO(4)}}$  group. We will consider two cases, $\Psi$ transforming in the 
$r_\Psi={\mathbf{4}}$ or $r_\Psi=\mathbf{1}$ of ${\textrm{SO(4)}}$. 

In our construction the right-handed top quark $t_R$ emerges as a chiral bound state of the strong dynamics. $t_R$
 must thus belong to a complete multiplet 
of the unbroken subgroup ${\textrm{SO(4)}}$, and, given we do not want extra massless states,  it must be a singlet.
That does not yet fully specify its quantum numbers. This is because, in order to reproduce the correct hypercharge, one must  enlarge the global symmetry by including an extra unbroken 
 ${\textrm{U(1)}}_X$ factor and define the hypercharge as $Y\,=\,T_R^3\,+\,X$, where $T_R^3$ is the third ${\textrm{SU(2)}}_R$ generator 
 of ${\textrm{SO(5)}}$.\footnote{See Appendix~A for the explicit form of the generators.}
Therefore the coset is actually  \mbox{${\textrm{SO(5)}}\times{\textrm{U(1)}_X}/ {\textrm{SO(4)}}\times{\textrm{U(1)}_X}$},   $t_R$ has  $X$ charge equal to
$2/3$ while the Higgs is $X$ neutral  (its hypercharge coincides with its $T_R^3$ charge).

A second assumption concerns the coupling of the elementary fields, {\it{i.e.}} the SM gauge fields $W_\mu$ and $B_\mu$ and the elementary left-handed doublet 
$q_L=(t_L,b_L)$, 
to the strong sector \footnote{The light quark families and the leptons will not be considered here because their couplings are most likely very weak.}. The EW 
bosons are coupled by gauging the SM subgroup of ${\textrm{SO(5)}}\times{\textrm{U(1)}_X}$.  The $q_L$ is assumed 
to be coupled \emph{linearly} to the strong sector, following the hypothesis of partial compositeness \cite{partcomp}. In the UV Lagrangian this coupling has therefore the form
\begin{equation}
{\mathcal L}_{\textrm{mix}}^{\textrm{UV}}=y\, \overline{q}_L^\alpha\Delta^*_{\alpha\,I_{\mathcal{O}}}{\mathcal O}^{I_{\mathcal{O}}} +{\textrm h.c.}\equiv y \left(\overline{Q}_L\right)_{I_{\mathcal{O}}}{\mathcal O}^{I_{\mathcal{O}}} +{\textrm h.c.}\,,
\label{lmix}
\end{equation} 
where ${\mathcal{O}}$ is an operator of the strong sector that transforms in some representation ${{r}}_{\mathcal{O}}$ of ${\textrm{SO(5)}}\times{\textrm{U(1)}_X}$.
The choice of ${{r}}_{\mathcal{O}}$ is, to some extent, free. Minimality, and the aim of reproducing explicit models considered in the literature, led us to consider 
two cases: ${{r}}_{\mathcal{O}}={\mathbf{5}}_{\mathbf{2/3}}$ and ${\mathbf{r}}_{\mathcal{O}}={\mathbf{14}}_{\mathbf{2/3}}$ \footnote{ Another possible option considered in the literature is ${{r}}_{\mathcal{O}}={\mathbf{4}}_{\mathbf{1/6}}$. However
this option is not available once $t_R$ is chosen to be a $SO(4)$ singlet: the top would not acquire a mass. It should also be remarked that, regardless of the nature of $t_R$, 
${{r}}_{\mathcal{O}}={\mathbf{4}}_{\mathbf{1/6}}$ is disfavored when considering dangerous tree level corrections to the $Zb\bar b$ vertex \cite{zbb,Mrazek:2011iu}.}.
 Notice that the 
${\textrm{U(1)}}_X$ charge of the operators must be equal to the one of the $t_R$ in order for the top mass to be generated after EWSB. 
In total, depending on whether the 
top partners will be in the ${\mathbf{4}}_{\mathbf{2/3}}$ or in the ${\mathbf{1}}_{\mathbf{2/3}}$ of the unbroken $SO(4)$, we will discuss four models named \fA, 
\fB \ and \oA, \oB \ respectively. The classification of the various models is summarized in Table~\ref{models}.

The explict breakdown of $SO(5)$ due to  $y$ in eq.~(\ref{lmix}) gives rise to a leading contribution to the Higgs potential $V(h)$.
However, in order to be able to tune the Higgs vacuum expectation value $v$ to be much smaller that its natural scale $f$, one may need to tune among themselves contributions to $V(h)$ with a different functional dependence on $h/f$. In the case of ${\mathbf{r}}_{\mathcal{O}}={\mathbf{14}}_{\mathbf{2/3}}$, the top Yukawa seed $y$ itself gives rise to two independent structures, whose coefficients can  be so tuned  that $v/f\ll 1$.
 On the other hand, in the case of ${\mathbf{r}}_{\mathcal{O}}={\mathbf{5}}_{\mathbf{2/3}}$,
the leading contribution to the potential consist of just one structure $\propto \sin^2h/f\cos^2h/f$, with well defined, non-tunable, minima and maxima. In the latter case then, in order to achieve $v\ll f$, one should assume there exists
an additional of $SO(5)$ breaking coupling whose contribution to the potential
 competes with that of the top. If this additional coupling  does not involve the SM fields, which seems resonable, then its contribution to $V$ will arise at tree level. In order not to outcompete  the top contribution, which arises at loop level, then this coupling should be  so  suppressed that its relative impact on strong sector quantities is of order $O(y^2/16\pi^2$). The latter should be compared to the effects of relative size $(y/g_\Psi)^2$ induced at tree level by the mixing in eq.~(\ref{lmix}) and accounted for in this paper.
 We conclude that, even when an extra $SO(5)$ breaking coupling is needed, it is not likely  to affect the phenomenology of  top partners in a quantitatively significant way.

Now back to the top partners. Our choices of their quantum numbers correspond to those obtained in explicit constructions. However our choice could also be motivated 
on general grounds by noticing  the  operators ${\mathcal{O}}$ interpolate  for particles with the corresponding quantum 
numbers. By decomposing ${\mathcal{O}}$ under the unbroken ${\textrm{SO(4)}}$ we obtain, respectively, 
\mbox{${\mathbf{5}}_{\mathbf{2/3}}={\mathbf{4}}_{\mathbf{2/3}}+ {\mathbf{1}}_{\mathbf{2/3}}$} and 
${\mathbf{14}}_{\mathbf{2/3}}={\mathbf{4}}_{\mathbf{2/3}}+ {\mathbf{1}}_{\mathbf{2/3}}+{\mathbf{9}}_{\mathbf{2/3}}$. 
In both cases we expect to  find a ${\mathbf{4}}_{\mathbf{2/3}}$ and/or a $ {\mathbf{1}}_{\mathbf{2/3}}$ in the low-energy spectrum.
 It could be also interesting to study top partners in the ${\mathbf{9}}_{\mathbf{2/3}}$, but this goes beyond the scope of the present paper.

\begin{table}[t]
\begin{center}
\setlength{\tabcolsep}{15pt} \setlength{\extrarowheight}{3pt}
\begin{tabular}{ | c | c | c |  }
\hline 
\ &  $r_{\mathcal{O}}={\mathbf{5}}_{\mathbf{2/3}}$ &     $r_{\mathcal{O}}={\mathbf{14}}_{\mathbf{2/3}}$  \\
  \hline
 $r_\Psi={\mathbf{4}}_{\mathbf{2/3}}$ &  \fA &    \fB  \\
  \hline
  $r_\Psi={\mathbf{1}}_{\mathbf{2/3}}$&  \oA &     \oB \\
  \hline
\end{tabular}\end{center}
\caption{\small{The nomenclature of the four models considered in the present paper, defined by the choices of
 the representations $r_\Psi, r_{\mathcal{O}}$. }}
\label{models}
\end{table}

The coupling of eq.~(\ref{lmix}) breaks the ${\textrm{SO(5)}}\times{\textrm{U(1)}_X}$ symmetry explicitly, but  it must of course  respect the SM group. This fixes unambiguously 
the form of the tensor $\Delta$ and thus of the \emph{embeddings}, $(Q_L)_{I_{\mathcal{O}}}=\Delta_{\alpha\,I_{\mathcal{O}}}q_L^\alpha$, of the elementary $q_L$ in 
${\textrm{SO(5)}}\times{\textrm{U(1)}_X}$ multiplets. For the ${\mathbf{5}}$ and the ${\mathbf{14}}$, respectively  the fundamental and  the two-indices symmetric traceless tensor,  we 
have 
\be
\left(Q_L^{{\mathbf{5}}}\right)_I={1\over \sqrt{2}}\left(\begin{matrix}
i b_L\\
 b_L\\
i t_L\\
- t_L\\
0
\end{matrix}\right)\,,\qquad\qquad
\left(Q_L^{{\mathbf{14}}}\right)_{I,J}={1\over \sqrt{2}}\left(\begin{matrix}
0 & 0 & 0 & 0 & i b_L\\
0 & 0 & 0 & 0 & b_L\\
0 & 0 & 0 & 0 & i t_L\\
0 & 0 & 0 & 0 & -t_L\\
i b_L & b_L & i t_L & -t_L &0\\
\end{matrix}\right)\,.
\label{emb}
\ee
Though explicitly broken, the ${\textrm{SO(5)}}\times{\textrm{U(1)}_X}$ group still gives strong constraints on our theory. Indeed  the elementary-composite 
interactions of eq.~(\ref{lmix})  formally respect the symmetry provided we formally assign suitable transformation properties to the embeddings. Under $g\in{\textrm{SO(5)}}$ 
we have
\be
\left(Q_L^{{\mathbf{5}}}\right)_I\;\rightarrow g_{I}^{\;I'}\left(Q_L^{{\mathbf{5}}}\right)_{I'}\,,\qquad\left(Q_L^{{\mathbf{14}}}\right)_{I\,J}\;
\rightarrow g_{I}^{\;I'} g_{J}^{\;J'}\left(Q_L^{{\mathbf{14}}}\right)_{I'\,J'}\,,
\label{transemb}
\ee
while the ${\textrm{U(1)}_X}$ charge is equal to $2/3$ in both cases. We will have to take into account this symmetry in our constructions.

\subsection{Effective Lagrangians}
\label{efflagr}

Based on the symmetry principles specified above we aim at building phenomenological effective Lagrangians for the $q_L$, the composite $t_R$ 
and the lightest top partner states $\Psi$. The basic idea is that our Lagrangians  emerge from a ``complete'' theory  by integrating out the heavier resonances in the strong sector. We thus need to rely on some qualitative description of the dynamics in order to estimate the importance of the various effective operators. We follow the ``SILH'' approach of Ref.~\cite{silh} and characterize the heavy resonances in terms of a single mass scale $m_*$ and of a single 
coupling $g_*=m_*/f$. As we already suggested in the introduction, parametrizing the strong sector in terms of a single scale is probably insufficient: a $125$~GeV Higgs suggests that the 
mass scale of the fermionic resonances should be slightly lower than that of the vectors. For our purposes the relevant scale $m_*$ should then be identified with the mass scale of the fermionic sector.
We thus adopt the following power-counting rule
\begin{equation}
\displaystyle
{\mathcal L}=\sum\frac{m_*^4}{g_*^2}\left(\frac{y\, q_L}{m_*^{3/2}}\right)^{n_{\textrm{el}}}
\left(\frac{g_* \Psi}{m_*^{3/2}}\right)^{n_{\textrm{co}}}\left(\frac{\partial}{m_*}\right)^{n_\partial}\left(\frac{\Pi}{f}\right)^{n_\pi}\,,
\label{powc}
\end{equation}
where $\Pi=\Pi^{1,\ldots,4}$ denotes the canonically normalized four real Higgs field components and $f$ is the Goldstone decay constant. Notice the presence of the 
coupling $y$ that accompanies (due to eq.~(\ref{lmix})) each insertion of the elementary $q_L$. Analogously the operators involving the SM gauge fields, omitted for 
shortness from eq.~(\ref{powc}), should be weighted by $g_{\textrm{SM}}/m_*$. The $t_R$ is completely composite and therefore it obeys the same power-counting 
rule as the top partner field $\Psi$.

Two terms in our effective Lagrangian will \emph{violate} the power-counting. One is the kinetic term of the elementary fields, which we take to 
be canonical, while eq.~(\ref{powc}) would assign it a smaller coefficient, $(y/g_*)^2$ in the case of fermions and $(g/g_*)^2$ in the case of gauge fields. 
This is because the elementary field kinetic term 
does not emerge from the strong sector, it was already present in the UV Lagrangian with ${\mathcal{O}}(1)$ coefficient. 
Indeed it is precisely because their kinetic coefficient is bigger than what established in eq.~(\ref{powc}), that the elementary fields have a coupling weaker than $g_*$.
The other term violating power-counting is 
the \emph{mass} of the top partners, which we denote by $M_\Psi$. We assume $M_\Psi<m_*$ in order to justify the construction of an effective theory in which only 
the top partners are retained while the other resonances are integrated out. The ratio $M_\Psi/m_*$ is our expansion parameter. We will therefore obtain accurate results 
only in the presence of a large separation, $M_\Psi\ll m_*$, among the lightest state and the other resonances \footnote{
An organizing principle, termed {\it {partial UV completion}} (PUV), to consistently construct an effective lagrangian for a parametrically light resonance was proposed in Ref.~\cite{Contino:2011np}. There the focus was on the  more involved case of vector and scalar resonances. According to PUV,  the couplings involving the lighter resonance should roughly saturate the strength $g_*$ when extrapolated at the scale $m_*$. We refer to Ref.~\cite{Contino:2011np} for a more detailed discussion. The effective lagrangians we construct in this paper automatically  satisfy PUV in the range of parameters
suggested by the power counting rule in eq.~(\ref{powc}).}. However already for a moderate separation, 
$M_\Psi\lesssim m_*$, or even extrapolating towards $M_\Psi\simeq m_*$, our models should provide a valid qualitative description of the relevant physics. Nevertheless 
for a more careful study of the case of small separation our setup should be generalized by incorporating more resonances in the effective theory.

\subsubsection{Top partners in the fourplet}

First we consider models \fA \ and \fB, in which the top partners are in the  ${\mathbf{4}}_{\mathbf{2/3}}$. In this case the top partner field is 
\be
\Psi={1\over \sqrt{2}}\left(\begin{matrix}
iB-i\Xft\\
B+\Xft\\
iT+i\Xtt\\
-T+\Xtt
\end{matrix}\right)\,,
\label{4plet}
\ee
and it transforms, following CCWZ, as 
\be
\Psi_i\to  h(\Pi ; g)_i^{\;j}\Psi_j\,,
\ee
under a generic element $g$ of ${\textrm{SO(5)}}$. The $4\times4$ matrix $h$ is defined by eq.s~(\ref{gtrans}) and (\ref{hd}) and provides a non-linear 
representation of the full ${\textrm{SO(5)}}$. The four $\Psi$ components decompose into two SM doublets $(T,B)$ and $(\Xft,\Xtt)$ of hypercharge $1/6$ and $7/6$ 
respectively. The first doublet has therefore the same quantum numbers as the $(t_L,b_L)$ doublet while the second one contains a state of exotic charge $5/3$ plus another 
top-like quark $\Xtt$.

When the $q_L$ is embedded in the  ${\mathbf{5}}_{\mathbf{2/3}}$, {\it{i.e.}} in model \fA, the leading order Lagrangian is 
\bea
\lagr^{\textrm{\fA}}&=&i\,\bar q_L \slashed{D}\, q_L+i\,\bar t_R \slashed{D}\, t_R
+i\,\bar\Psi(\slashed{D}+i\slashed{e})\Psi-M_\Psi\bar\Psi\Psi\nn\\
&&
+\left[ i\,c_1 \left(\bar{\Psi}_R\right)_i \,\gamma^\mu d_\mu^i \, t_R
+ y f \, (\overline{Q}_L^{{\mathbf{5}}})^{I} U_{I\, i} \,\Psi_R^i
+ y \,c_2 f \, (\overline{Q}_L^{{\mathbf{5}}})^{I} U_{I\, 5} \,t_R
+\textrm{h.c.}\right]\,,
\label{eq:lagrangian2}
\eea
where $c_{1,2}$ are  coefficients expected to be of order $1$. 
The above Lagrangian with totally composite $t_R$ was first written in Ref.~\cite{Marzocca:2012zn}.
Notice the presence of the $\slashed{e}=e_\mu\gamma^\mu$ term which accompanies the derivative of the top partner 
field: it reconstructs the CCWZ covariant derivative defined in eq.~(\ref{covder}) and is essential  to respect $SO(5)$. In the second line of the equation 
above we find, first of all, a \emph{direct} interaction, not mediated by the coupling $y$, among the composite $t_R$ and the top partners. This term is entirely generated by the 
strong sector and would have been suppressed in the case of partial $t_R$ compositeness. It delivers, looking at the explicit form of $d_\mu$ in eq.~(\ref{dande}), couplings involving
the top, the partners and  the SM gauge fields. These  will play an important role in the single production and in  the decay of the top partners. The last two terms give rise,
in particular, to  the top quark mass but also to trilinear couplings contributing to the single production of  top partners.
Notice that 
the indices of the embedding $Q_L^{{\mathbf{5}}}$ \emph{can not} be contracted directly with those of $\Psi$ because
they live in different spaces. The embeddings transform linearly under 
$\textrm{SO}(5)$ as reported in eq.~(\ref{transemb}) while $\Psi$ transforms under  the non-linear representation $h$. 
For this reason one insertion of the Goldstone matrix, transforming according to eq.~(\ref{gtrans}), is needed.

For brevity we  omitted from eq.~(\ref{eq:lagrangian2}) the kinetic term of the gauge fields and of the Goldstone Higgs, the latter is given for reference in 
eq.~(\ref{hkt}). Moreover we have not yet specified the covariant derivatives $D_\mu$ associated with the SM gauge group, these are obviously given by
\bea
D_\mu q_L&=&\left(\partial_\mu-ig W_\mu^i {\sigma^i\over 2}-i{1\over 6}g' B_\mu-i\,g_SG_\mu\right)q_L\,, \\
D_\mu t_R&=&\left(\partial_\mu-i{2\over 3}g' B_\mu-i\,g_SG_\mu\right)t_R \,, \\
D_\mu \Psi&=&\left(\partial_\mu-i {2\over 3} g' B_\mu -i\,g_SG_\mu\right)\Psi\,.
\label{cder}
\eea
where $g,g'$ and $g_S$ are the ${\textrm{SU}}(2)_L\times {\textrm{U}}(1)_Y$ and ${\textrm{SU}}(3)_c$ gauge couplings. We remind the reader that the top 
partners form  a color triplet, hence the gluon in the above equation.

The Lagrangian is very similar for model \fB, where the $q_L$ is embedded in the symmetric traceless $Q_L^{{\mathbf{14}}}$. We have
\bea
\lagr^{\textrm{\fB}}&=&i\,\bar q_L \slashed{D}\, q_L+i\,\bar t_R \slashed{D}\, t_R
+i\,\bar\Psi (\slashed{D}+i\slashed{e})\Psi-M_\Psi\bar\Psi\Psi\nn\\
&+&\!\!\!\!\left[ i\, c_1 \left(\bar{\Psi}_R\right)_i \,\gamma^\mu d_\mu^i \, t_R
+ y f \, (\overline{Q}_L^{{\mathbf{14}}})^{I\,J} U_{I\, i} U_{J\, 5} \,\Psi_R^i
+ {y c_2 \over 2} f \, (\overline{Q}_L^{{\mathbf{14}}})^{I\,J} U_{I\, 5}U_{J\, 5} \,t_R
+\textrm{h.c.}\right],\quad\quad
\label{eq:lagrangian214}
\eea
notice that the two indices of $Q_L^{{\mathbf{14}}}$ are symmetric and therefore the term that mixes it with $\Psi$ is unique. The factor ${1 \over 2}$ introduced in the last term is merely conventional.

In both models \fA \  and \fB\  the 
leading order Lagrangian contains four parameters, $\{ M_\psi$, $y$, $c_1$, $c_2\}$, on top of the Goldstone decay constant $f$. One parameter will however have to 
be fixed to reproduce the correct top mass, while the remaining three parameters could be traded for two physical masses, for instance $m_{\Xft}$ and $m_B$, 
and the coupling $c_1$. It will 
often be convenient to associate the mass $M_\Psi$ with a coupling $g_\psi$ 
$$
g_\Psi\equiv\frac{M_\Psi}{f}\,.
$$
We will see below that $c_1\times g_\Psi$ controls the strength of the interactions between the top partners  and the  Goldstone bosons at energy $\sim M_\Psi$. In particular it controls the on-shell couplings relevant for single production and for two body decays. Notice that,  as a function of energy, the effective strength of this trilinear interaction is instead $\sim c_1 E/f$. For $c_1=O(1)$, as suggested by power counting, the effective coupling is of order $g_*\equiv m_*/f$ at the energy scale of the heavier resonances, in accord with the principle of partial UV completion proposed in Ref.~\cite{Contino:2011np}. Power counting and partial UV completion then equivalently imply $c_1=O(1)$ and therefore  $c_1 g_\Psi < g_*$.
This result  obviously 
follows from the fact that the Higgs is a derivatively coupled pNGB. It would be lost if the Higgs was instead treated as a generic resonance. In the latter case 
the expected coupling would be independent of the mass and it would be larger, of order $g_*$. Moreover notice that, although on shell it leads to an effective Yukawa vertex, the interaction associated with $c_1$ does not affect the spectrum when $H$ acquires a vacuuum expectation value. That again would not be true
if we did not account for the pNGB nature of $H$. The pNGB nature of $H$ is not accounted for in the first thorough  work on simplified top partner models   \cite{continoservant} and in the  following studies (see in particular  \cite{mrazecwulzer,AguilarSaavedra:2009es}).

Notice that, a priori, one of the four  parameters describing the simplified model
could be complex. This is because we have at our disposal only $3$ chiral rotations to eliminate the phases from the Lagrangians
 (\ref{eq:lagrangian2}) and (\ref{eq:lagrangian214}). 
Nevertheless we are entitled to keep all 
the parameters real if we demand the strong sector respects  a CP symmetry defined in Appendix~A.
It is easy to check that CP requires the non-derivative couplings to be  real while the coefficient of the term involving to
$d_\mu$ must be  purely imaginary.  CP conservations is an additional hypothesis 
of our construction, however  the broad phenomenology does not significantly depend on it.

\subsubsection{Top partners in the singlet}

The Lagrangian is even simpler if the top partners are in the ${\mathbf{1}}_{\mathbf{2/3}}$. In this case we only have one exotic top-like state which we denote as 
${\widetilde{T}}$. For the two models, \oA\ and \oB \ that we aim to consider the Lagrangian reads, respectively
\bea
\lagr^{\textrm{\oA}}&&=\bar q_L \, i\slashed{D}\, q_L+\bar t_R \, i\slashed{D}\, t_R
+i\bar\Psi i\slashed{D}\Psi-M_\Psi\bar\Psi\Psi\nn\\
&&
+\left[ y f \, (\overline{Q}_L^{{\mathbf{5}}})^{I} U_{I\, 5}  \Psi_R
+ y \,c_2 f \, (\overline{Q}_L^{{\mathbf{5}}})^{I} U_{I\, 5} \,t_R
+\textrm{h.c.}\right]\,,\nn\\
\lagr^{\textrm{\oB}}&&=\bar q_L \, i\slashed{D}\, q_L+\bar t_R \, i\slashed{D}\, t_R
+i\bar\Psi i\slashed{D} \Psi-M_\Psi\bar\Psi\Psi\nn\\
&&
+\left[ {y  \over 2} f \, (\overline{Q}_L^{{\mathbf{14}}})^{I\,J} U_{I\, 5} U_{J\, 5}  \Psi_R
+ {y \,c_2  \over 2} f \, (\overline{Q}_L^{{\mathbf{14}}})^{I\,J} U_{I\, 5} U_{J\, 5} \,t_R
+\textrm{h.c.}\right]\,.
\label{eq:lagrangian211}
\eea
Notice that we could have also written a direct mixing among $t_R$ and $\Psi$ because the two fields now have identical quantum numbers. However this 
mixing can obviously be removed by a field redefinition. Models \oA\ and \oB,\ apart from $f$, contain three parameters, $\{M_{\psi},\,y,\,c_2\}$, one of which must again be fixed 
to reproduce the top mass. We are left with two free parameters that correspond to the coupling $c_2$ and to the mass $m_{\widetilde{T}}$ of the partners. 
Notice that in this case all the parameters can be made real by chiral rotations without need of imposing the CP symmetry. The latter symmetry is automatically 
respected in models \oA \ and \oB.

In order to complete the definition of our models let us discuss the theoretically expected size of their parameters.
From the discussion  in the introduction and from experience with concrete models, one can reasonably argue that the favorite range for $M_\Psi$ is between $500$~GeV and $1.5$~TeV, while $g_\Psi$ is favored  in the range $1\lsim g_\Psi \lsim 3$. It is also worth recalling the favorite range of the decay constant $f\equiv M_\Psi/g_\Psi$, which 
is conveniently traded for the parameter $\xi$ defined in Ref.~\cite{Agashe:2004rs}
\be
\xi=\frac{v^2}{f^2}\,,
\ee
where $v=2m_W/g=246$~GeV is the EWSB scale. Since $\xi$ controls the deviation from the SM at low energies it cannot be too large. Electroweak precision tests suggest 
$\xi\simeq 0.2$ or $\xi\simeq 0.1$, which corresponds to $f\simeq500$~GeV or $f\simeq800$~GeV. Smaller values of $\xi$ would  of course require more tuning. Finally, the strength of the elementary-composite coupling $y$ is 
fixed by the need of reproducing the correct mass of the top quark. We will see in the following section that this implies $y\sim y_t=1$.

\subsection{A first look at the models}

Now that the models are defined let us start discussing their implications. The simplest aspects will be examined in the present section while a more detailed 
analysis of their phenomenology will be postponed to the following one.

\subsubsection{The Spectrum}

We start from model \fA \ and we first focus on the fermionic spectrum. The mass-matrix after EWSB is easily computed form eqs.~(\ref{eq:lagrangian2}) and 
(\ref{emb}) by using the explicit form of $U$ on the Higgs VEV obtained from eq.~(\ref{uvev}). By restricting to the sector of $2/3$-charged states we find  
\be
\displaystyle
\left(\begin{matrix}
\bar t_L\\ \bar T_L\\  \barXtt_L
\end{matrix}\right)^T
\left(\begin{matrix}
- \frac{c_2 y\,f}{ \sqrt{2}} \sin{\epsilon} &y\,f\cos^2{\frac{\epsilon}{2}}&y\,f\sin^2{\frac{\epsilon}{2}} \\
0&-M_\psi &0\\
0&0&-M_\psi\\
\end{matrix}\right) 
\left(\begin{matrix}
t_R\\
T_R\\
{\Xtt}_R
\end{matrix}\right)
\,,
\label{eq:mass231}
\ee
where $\epsilon = \langle h\rangle/f$ is defined as the ratio among the VEV of the Higgs field and the Goldstone decay constant. 
The relation among $ \langle h\rangle$ and the EWSB scale is reported in eq.~(\ref{VEV}), from which we derive
 \be
\xi=\frac{v^2}{f^2}\,=\,\sin^2{\epsilon}\,.
\ee
We immediately notice a remarkable feature of the mass-matrix (\ref{eq:mass231}): only the \emph{first line}, {\it{i.e.}} the terms which involve the $t_L$, is sensitive to EWSB while 
the rest of the matrix remains unperturbed. This is due to the fact that the Higgs is a pNGB and therefore its non-derivative interactions can only originate from the 
breaking of the Goldstone symmetry ${\textrm{SO}}(5)$. The ${\textrm{SO}}(5)$ invariant terms just produce derivative couplings of the Higgs and therefore 
they cannot contribute to the mass-matrix. Since the Goldstone symmetry is broken exclusively by the terms involving the elementary $q_L$ it is obvious that the mass-matrix 
must have the form of eq.~(\ref{eq:mass231}). Notice that this structure would have been lost if we had not taken into account the pNGB 
nature of the Higgs. Indeed if we had treated the Higgs as a generic composite ${\textrm{SO}}(4)$ fourplet, Yukawa-like couplings of order $g_*$ and involving $t_R$ and $\Psi$ would have been allowed. After EWSB those terms would have given rise to $(2,1)$ and $(3,1)$ mass matrix entries  of order  $g_* v$.

The peculiar structure of the mass-matrix has an interesting consequence. It implies that only one linear combination of $T$ and $\Xtt$, with coefficients 
proportional to the  $(1,2)$ and $(1,3)$ entries, mixes with the $q_L$, while the orthogonal combination  does not mix either with the $q_L$ or with any other state. Explicitly, the 
two combinations are
\bea
T'=&&\frac1{\sqrt{\cos^4{\frac{\epsilon}{2}}+\sin^4{\frac{\epsilon}{2}}}}\left[\cos^2{\frac{\epsilon}{2}}T+\sin^2{\frac{\epsilon}{2}}\Xtt\right]\,,\nn\\
\Xtt'=&&\frac1{\sqrt{\cos^4{\frac{\epsilon}{2}}+\sin^4{\frac{\epsilon}{2}}}}\left[\cos^2{\frac{\epsilon}{2}}\Xtt-\sin^2{\frac{\epsilon}{2}}T\right]\,.
\label{XT}
\eea
After this field redefinition the mass-matrix becomes block-diagonal
\be
\displaystyle
\left(\begin{matrix}
\overline{t}_L\\ \overline{T}_L'\\  {\barprimeXtt}_L
\end{matrix}\right)^T
\left(\begin{matrix}
- \frac{c_2 y\, f}{ \sqrt{2}} \sin{\epsilon} &y\,f{\sqrt{\cos^4{\frac{\epsilon}{2}}+\sin^4{\frac{\epsilon}{2}}}} &0 \\
0&-M_\psi &0\\
0&0&-M_\psi\\
\end{matrix}\right) 
\left(\begin{matrix}
t_R\\
T_R'\\
{\primeXtt}_R
\end{matrix}\right)
\,,
\label{eq:mass231prime}
\ee
so that the state  $\primeXtt$ is already a mass eigenstate with mass $m_{\Xtt}=M_\Psi$. But the spectrum also contains a second particle with exactly the same mass. 
Indeed the $\Xft$ cannot mix because it is the only state with exotic charge and therefore it maintains the mass  $m_{\Xft}=M_\Psi$ it had before EWSB. The 
$\Xtt$ and the $\Xft$ are thus \emph{exactly degenerate}. This remarkable property is due to the pNGB nature of the Higgs and it would be generically violated, 
as previously discussed, if this assumption was relaxed. 
This result also depends on $t_R$ being a composite singlet. If $t_R$ was instead a partially composite state mixing to a non-trivial representation of $SO(5)$ (for instance a {\bf 5}) there would be additional entries in the mass matrix. 
\footnote{The top partner's spectrum with partially composite $t_R$ has been worked out in 
Ref.~\cite{panico,Matsedonskyi:2012ym}.}
 In a sense our result depends on $y$ being the only relevant parameter that breaks $SO(5)$ explicitly.

\begin{figure}[t]
\centering
 \includegraphics[scale=0.8]{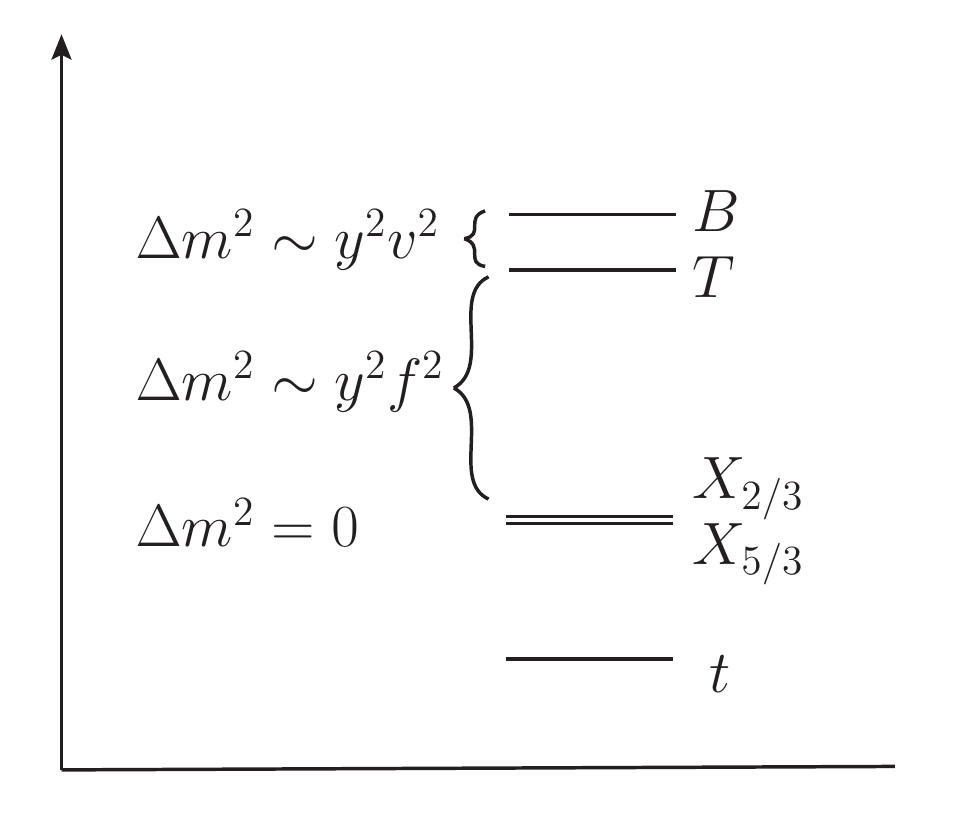}  
  \caption{\small{The typical spectrum of the top partners.}}
\label{spectrum}
\end{figure}

Once the mass-matrix has been put in the block-diagonal form of eq.~(\ref{eq:mass231prime}) it is straightforward to diagonalize it and to obtain exact formulae for the 
rotation matrices and for the masses of the top and of the $T$ partner. 
However the resulting expressions are rather involved and we just report here approximate expressions for the masses. We have
\bea
m_t\simeq &&\frac{c_2y\,f}{\sqrt{2}}\frac{g_\Psi}{\sqrt{g_\Psi^2+y^2}}\sin{\epsilon}
\left[1+{\mathcal{O}}\left(\frac{y^2}{g_\Psi^2} \xi\right)\right]\,,\nn\\
m_T\simeq &&\sqrt{ M_\Psi^2+y^2f^2}
\left[1-\frac{y^2 \left(g_\Psi^2+(1-c_2^2)y^2\right)}{4\left(g_\Psi^2+y^2\right)^2} \sin^2{\epsilon}+\ldots\right]\,.
\label{masss}
\eea
From the above equation we obtain the correct order of magnitude for the top mass if, as anticipated, $y\sim y_t$ and  $g_\Psi\gtrsim1$. In this region 
of the parameter space the corrections to the approximate  formulae are rather small, being suppressed by both a factor $y^2/g_\Psi^2$ 
(which is preferentially smaller than one) and by $\xi\ll1$. However we will consider departures from this theoretically expected region and therefore 
we will need to use the exact formulae in the following sections.

Similarly we can study the sector of $-1/3$ charge states. It contains 
a massless $b_L$, because we are not including the $b_R$ in our model, plus the heavy $B$ particle with a mass
\be
m_B=\sqrt{M_\Psi^2+y^2f^2}\,.
\label{mb1}
\ee
This formula is exact and shows that the bottom sector does not receive, in this model, any contribution from EWSB. By comparing the equation 
above with the previous one we find that the splitting among $T$ and $B$ is typically small
\be
m_B^2-m_T^2\simeq y^2f^2\frac{ g_\Psi^2+(1-c_2^2)y^2}{2\left(g_\Psi^2+y^2\right)}
\sin^2{\epsilon}\,,
\label{mdtb}
\ee
and positive in the preferred region $g_\Psi>y$, although there are points
in the parameter space where the ordering $m_T>m_B$ can occur.
The splitting among the two doublets is instead always positive, 
$m_B^2-m_{\Xft}^2=y^2f^2$. The typical spectrum of the top partners that we have in our model is depicted in figure~\ref{spectrum}.

The situation is not much different in model \fB. The mass-matrix for charge $2/3$ states has again the form of eq.~(\ref{eq:mass231}) 
\be
\displaystyle
\left(\begin{matrix}
\bar t_L\\ \bar T_L\\  \barXtt_L
\end{matrix}\right)^T
\left(\begin{matrix}
- \frac{c_2 y\,f}{ 2 \sqrt{2}} \sin{2\epsilon} &\frac{y\,f}2\left(\cos{\epsilon}+\cos{2\epsilon}\right)
&\frac{y\,f}2\left(\cos{\epsilon}-\cos{2\epsilon}\right) \\
0&-M_\psi &0\\
0&0&-M_\psi\\
\end{matrix}\right) 
\left(\begin{matrix}
t_R\\
T_R\\
{\Xtt}_R
\end{matrix}\right)
\,,
\label{eq:mass232}
\ee
and again it can be put in a block-diagonal form by a rotation among the $T$ and the $\Xtt$ similar to the one in eq.~(\ref{XT}). 
Therefore also in model \fB \ the physical $\Xtt$ has mass $M_\Psi$ and it is degenerate with the 
$\Xft$.
The approximate top and $T$ mass are given in this case by
\bea
m_t\simeq &&\frac{c_2y\,f}{\sqrt{2}}\frac{g_\Psi}{\sqrt{g_\Psi^2+y^2}}{\sin{2\epsilon} \over 2}
\left[1+{\mathcal{O}}\left(\frac{y^2}{g_\Psi^2} \xi\right)\right]\,,\nn\\
m_T\simeq &&\sqrt{ M_\Psi^2+y^2f^2}
\left[1-\frac{y^2 \left(5g_\Psi^2+(5-c_2^2)y^2\right)}{4\left(g_\Psi^2+y^2\right)^2} \sin^2{\epsilon}\right]\,.
\label{masss1}
\eea
Similarly we can compute the mass of the $B$ partner and we find
\be
m_B=\sqrt{M_\Psi^2+y^2f^2\cos^2{\epsilon}}\simeq\sqrt{M_\Psi^2+y^2f^2}-\frac{y^2f^2}{2\sqrt{M_\Psi^2+y^2f^2}}\sin^2\epsilon\,.
\ee
In this case, differently from model \fA\ (see eq.~(\ref{mb1})), the mass of the $B$ is sensitive to EWSB. Apart from this little difference 
the spectrum is very similar to the one of model \fB\ described in figure~\ref{spectrum}.

The models with the singlet are much simpler because there is only one exotic state. The mass matrices read:

\be
\displaystyle
\left(\begin{matrix}
\overline{t}_L\\ \overline{\widetilde T}_L
\end{matrix}\right)^T
\left(\begin{matrix}
- \frac{c_2 y\, f}{ \sqrt{2}} \sin{\epsilon} & -\frac{ y\, f}{ \sqrt{2}} \sin{\epsilon} \\
0&-M_\psi\\
\end{matrix}\right) 
\left(\begin{matrix}
t_R\\
\widetilde T_R
\end{matrix}\right)
\,,
\label{eq:mass1A}
\ee

\be
\displaystyle
\left(\begin{matrix}
\overline{t}_L\\ \overline{\widetilde T}_L
\end{matrix}\right)^T
\left(\begin{matrix}
- \frac{c_2 y\, f}{ 2\sqrt{2}} \sin{2 \epsilon} & -\frac{ y\, f}{ 2\sqrt{2}} \sin{2 \epsilon} \\
0&-M_\psi\\
\end{matrix}\right) 
\left(\begin{matrix}
t_R\\
\widetilde T_R
\end{matrix}\right)
\,,
\label{eq:mass1B}
\ee
for models \oA\ and \oB\ respectively. The mass eigenvalues for model \oA\ are
\bea
m_t\simeq &&\frac{c_2y\,f}{\sqrt{2}} \sin{\epsilon}
\left[1+{\mathcal{O}}\left(\frac{y^2}{g_\Psi^2} \xi\right)\right]\,,\nn\\
m_{\widetilde{T}}\simeq &&M_\Psi
\left[1+\frac{y^2}{4 g_\Psi^2} \sin^2{\epsilon}\right]\,.
\label{masss21}
\eea
For model \oB\ instead we have
\bea
m_t\simeq &&\frac{c_2y\,f}{2\sqrt{2}} \sin{2\epsilon}
\left[1+{\mathcal{O}}\left(\frac{y^2}{g_\Psi^2} \xi\right)\right]\,,\nn\\
m_{\widetilde{T}}\simeq &&M_\Psi
\left[1+\frac{y^2}{4 g_\Psi^2} \sin^2{\epsilon}\right]\,.
\label{masss22}
\eea
As one can see from the last expressions the mass of the $\widetilde T$ receives positive contributions proportional to $y^2$ and hence for a fixed mass of the $\widetilde T$, $y$ must be limited from above. Unlike the models with fourplet partners, in the singlet case $y$ completely controls the couplings of  the $\widetilde T$ with the top and bottom quarks (see Sec.~\ref{gc}). Therefore one can expect that for a given $m_{\widetilde T}$ there exists a maximal allowed coupling of the SM particles with the top partner and hence for small masses the single production of $\widetilde T$ is suppressed. In addition small values of $m_{\widetilde T}$ become unnatural since they require very small $y$ together with a very large $c_2$ needed to recover correct top mass. By minimizing the largest eigenvalue of the mass matrix with respect to $M_{\Psi}$ for fixed $y$ and $f$ one can find a minimal allowed mass of the $\widetilde T$ which is given by
\bea
m_{\widetilde T}^{\text{min, \oA}}&=&m_{t}+{ 1 \over \sqrt 2} y f \sin \epsilon\,,\nn\\
m_{\widetilde T}^{\text{min, \oB}}&=&m_{t}+{ 1 \over 2 \sqrt 2} y f \sin 2\epsilon\,,
\label{minmassttilde}
\eea
for the models \oA\ and \oB\ respectively. The bound given in eq.~(\ref{minmassttilde}) will affect the exclusion plots in the following.

\subsubsection{Trilinear Couplings}
\label{trilinear}

Other interesting qualitative aspects of our models are discovered by inspecting the explicit form of the Lagrangians in 
unitary gauge. These are reported in Appendix~\ref{ferc}, and are written in the ``original'' field basis used to define 
the Lagrangians in eq.s~(\ref{4plet}, \ref{eq:lagrangian2}, \ref{eq:lagrangian214}, \ref{eq:lagrangian211}), 
{\it{i.e.}} before the rotation to the mass eigenstates. Appendix~\ref{ferc} contains, for reference, the complete 
Lagrangian including all the non-linear and the derivative Higgs interactions. However the coupling that  are 
relevant to  the present discussion are the trilinears involving  the gauge fields and the Higgs in the models \fA \ and \fB, 
reported in eq.~(\ref{dcoup}), (\ref{ecoup}), (\ref{mc5}) and (\ref{mc14}).

The first remarkable feature of eq.~(\ref{ecoup}) is that the $Z$ boson couplings with the $B$ is completely standard:
 it is not modified by EWSB effects and coincides with the familiar SM expression $g_Z=g/c_w(T_L^3-Q)$. 
In particular  it coincides with the $Z\bar b_L b_L$ coupling, involving  the elementary $b_L$,  because $b_L$ and $B$ have the same $SU(2)\times U(1)$ quantum numbers. The $Z$-boson coupling to charge $-1/3$ quarks is therefore proportional to the  identity matrix. 
Consequently the $Z$ interactions remain diagonal and canonical even after rotating  to the mass eigenbasis. In particular, in the charge $-1/3$ sector, there will not be a  neutral current vertex  of the form $B\rightarrow Z b$.

This property is due to an accidental parity, $P_{LR}$, defined in Ref.~\cite{Contino:2011np} as the exchange of the Left and 
the Right ${\textrm{SO}}(4)$ generators. This symmetry is an element of ${\textrm{O}}(4)$ and it acts on the top 
partner fourplet of eq.~(\ref{4plet}) and on the Higgs field $\vec{\Pi}$ through the $4\times4$ matrix
\be
P_{LR}^{(4)}=\left(\begin{matrix}
-1&0&0&0\\ 
0&-1&0&0\\ 
0&0&-1&0\\ 
0&0&0&1\\ 
\end{matrix}\right)\,.
\label{plr}
\ee
The action of $P_{LR}$ is readily uplifted to ${\textrm{O}}(5)$ with the $5\times5$ matrix $P_{LR}^{(5)}={\textrm{diag}}
(-1,-1,-1,1,1)$. We see that $P_{LR}$ is not broken by the Higgs VEV, which only appears in the last component of the 
$\vec\Pi$ vector. In Ref.~\cite{zbb}, it was shown  that $P_{LR}$ invariance protects the $Z$ couplings from tree-level corrections at zero momentum transfer. That case applied to $b$ quarks, but the  statement generalizes straighforwardly:   if all the particles with a given charge have the same $P_{LR}$, then, at tree level in the weak interactions,  the neutral current vertices in that charge sector are canonical  and, in particular, diagonal.

The Lagrangians (\ref{eq:lagrangian2}) and (\ref{eq:lagrangian214}) are approximately $P_{LR}$ invariant, with the breaking coming only from the weak gauge couplings and from the weak mixing $y$ between elementary and composites. However at tree level, for which case the elementary fields can be treated as external spectators, even this weak breaking is ineffective in the charge $-1/3$ and $5/3$ sectors. Notice indeed that according to eqs.~(\ref{4plet},\ref{plr}), under $P_{LR}$, $B$ and $X_{5/3}$ are  odd, while $T$ and $-X_{2/3}$ are interchanged. Then,
inspection  of the embedding in eq.~(\ref{emb}) shows that while we cannot assign a consistent $P_{LR}$ to $t_L$, we can instead assign negative $P_{LR}$ to $b_L$. At tree level, $t_L$ will not affect processes involving only  quarks
with charge $-1/3$ and $5/3$, and therefore the associated explicit  breaking of $P_{LR}$ will be ineffective. Analogously the breaking in the gauge sector is seen not to matter at tree level. For a detailed discussion we refer the reader to section 2.4 of Ref.~\cite{Mrazek:2011iu}. This explains the result previously mentioned for the $(b, \,B)$ sector and also predicts that the coupling  of the $\Xft$ must be canonical as well.  This is indeed what we see in eq.~(\ref{ecoup}).

The same argument applies to $\widetilde T_R$ and $t_R$ in the singlet models  \oA\ and \oB\,. The  Z-vertex of those states  is not modified, in particular there is no $\bar{t}_R Z\widetilde T_R$ vertex and the production/decay with Z is always controlled by left-handed coupling. On the other hand, for $\widetilde T_L$ and $t_L$ the argument does not apply, regardless of $P_{LR}$, given $\widetilde T_L$ and $t_L$ do not have the same $SU(2)\times U(1)$ quantum numbers.

Another interesting property concerns the $W$ couplings of the $B$ with the charge $2/3$ states. We see in 
eq.~(\ref{ecoup}) that the linear combination of the $T$ and the $\Xtt$ that couples with the $B$ is exactly 
orthogonal to the physical (mass-eigenstate) $\Xtt'$ field defined in eq.~(\ref{XT}) for model \fA. Only the 
$T'$ couples to $W\,B$, leading after the second rotation to transitions among the physical $t$, $T$ and 
$b$, $B$. Such couplings are instead absent for the physical $\Xtt$ which therefore,  cannot 
decay to $Wb$. This feature is not, for what we can say, the result of a symmetry, but rather an accidental feature 
of model \fA. In model \fB\ instead the coupling is allowed because the physical $\Xtt$ (see the mass-matrix 
in eq.~(\ref{eq:mass232})) is not anymore orthogonal to the combination that couples to $W\,B$. Nevertheless the 
$\Xtt$-$B$ coupling is suppressed by $\langle h\rangle^2/f^2$ and therefore the decay $\Xtt\rightarrow Wb$, 
though allowed in principle, is phenomenologically irrelevant as we will discuss in the following section. 

A final comment concerns the couplings of the physical Higgs field $\rho$. The couplings  from the strong sector  are only  due to the $d_\mu$ term in eq.~(\ref{dcoup})  and are purely
derivative. Therefore, because of charge conservation, they cannot involve 
the $B$ partner. Higgs couplings in the $-1/3$ charge sector could only emerge from the elementary-composite mixings. However they are
accidentally absent in model \fA, as shown by eq.~(\ref{mc5}). Therefore the decay of the $B$ to the Higgs is absent in this model. In 
model \fB, on the contrary, this decay is allowed through the vertex in eq.~(\ref{mc14}).

\section{Top Partners Phenomenology}
\label{sec:TPP}

Let us now turn to discuss the main production mechanisms and decay channels of the top partners in the models under consideration. 
We will first of all, in sect.~\ref{pd}, describe how the cross-sections of the production processes and the partial decay widths can be conveniently 
parametrized analytically in terms of few universal functions, extracted from the Monte Carlo integration. This method, supplemented with tree-level 
event simulations to compute the acceptances associated with the specific cuts of each experimental search, will allow us to explore efficiently the 
multi-dimensional parameter space of our model avoiding a time-consuming scan. Not all the production and decay processes that could be computed 
with this method are equally sizable, however. In sect.~\ref{gc} we will present an estimate of the various processes based on the use of the Goldstone 
boson Equivalence Theorem \cite{equivalence}, this will allow us to classify (in sect.~\ref{mrc}) the channels which are more promising for the search of 
the top partners at the LHC.

\subsection{Production and Decay}
\label{pd}

Given that the partners are colored they can be produced in pairs through the QCD interactions. The pair production cross-section is universal for all the 
partners and it can be parametrized by a function 
\be
\sigma_{\textrm{pair}}(m_X)\,,
\ee
which depends uniquely on the partner's mass $m_X$, for which we have analytical formulae. 
We have constructed $\sigma_{\textrm{pair}}$ by interpolation using the HATHOR code 
\cite{hathor} which incorporates perturbative QCD corrections up to NNLO. The values of the cross-section used in the fit are reported in 
Table~\ref{tab:xsecpair} for the LHC at $7$ and $8$ TeV center of mass energy. In this and all the other simulations we adopted the set of 
parton distribution functions MSTW2008 \cite{mstw}. 

\begin{table}
\begin{center}
\begin{tabular}{ | c | c | c | }
\cline{2-3}
 \multicolumn{1}{c}{}
  &    \multicolumn{2}{|c|}{$\sigma \, [\textrm{fb}]$ @ NNLO} \\
 \multicolumn{1}{c}{}    & \multicolumn{2}{|c|}{pair production} \\
  \hline
   $M\, [\textrm{GeV}]$ & $\sqrt{s}=7$ TeV & $\sqrt{s}=8$ TeV\\
    \hline
 400 & (0.920) 1.41 $\times 10^3$ & (1.50)  2.30 $\times 10^3$  \\
  500 & (218) 330 & (378) 570 \\
 600 & (61.0) 92.3 & (113) 170 \\
 700 & (19.1) 29.0 & (37.9) 56.9 \\
 800 & (6.47) 9.88 & (13.8) 20.8 \\
 900 & (2.30)  3.55 & (5.33) 8.07 \\
 1000 & (0.849) 1.33& (2.14) 3.27\\
 1100 & (0.319) 0.507 & (0.888) 1.37\\
 1200 & (0.122) 0.196 & (0.375) 0.585 \\
 1300 &  (4.62) 7.60 $\times 10^{-2}$ & (0.160) 0.253\\
  \hline
\end{tabular}
\end{center}
\caption{\small{Cross sections for the NNLO pair production of heavy fermions
at $\sqrt{s}=7, 8$ TeV (the LO values are in brackets), with HATHOR \cite{hathor}.}}
\label{tab:xsecpair}
\end{table}

\begin{figure}[t]
	\begin{center}
\scalebox{0.5}{\includegraphics{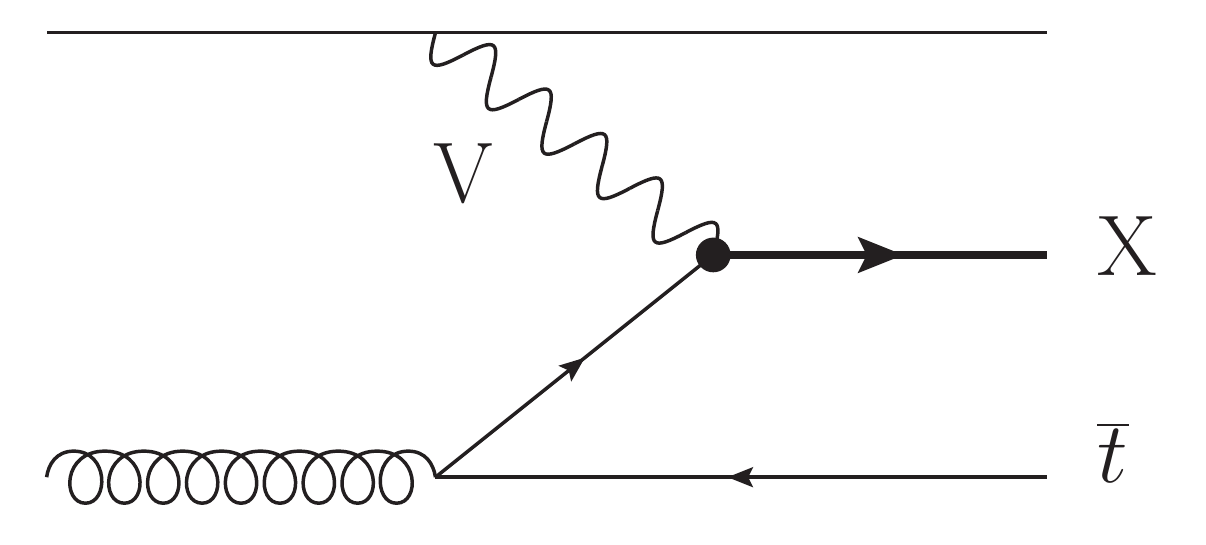}}\hspace{50pt}
\scalebox{0.5}{\includegraphics{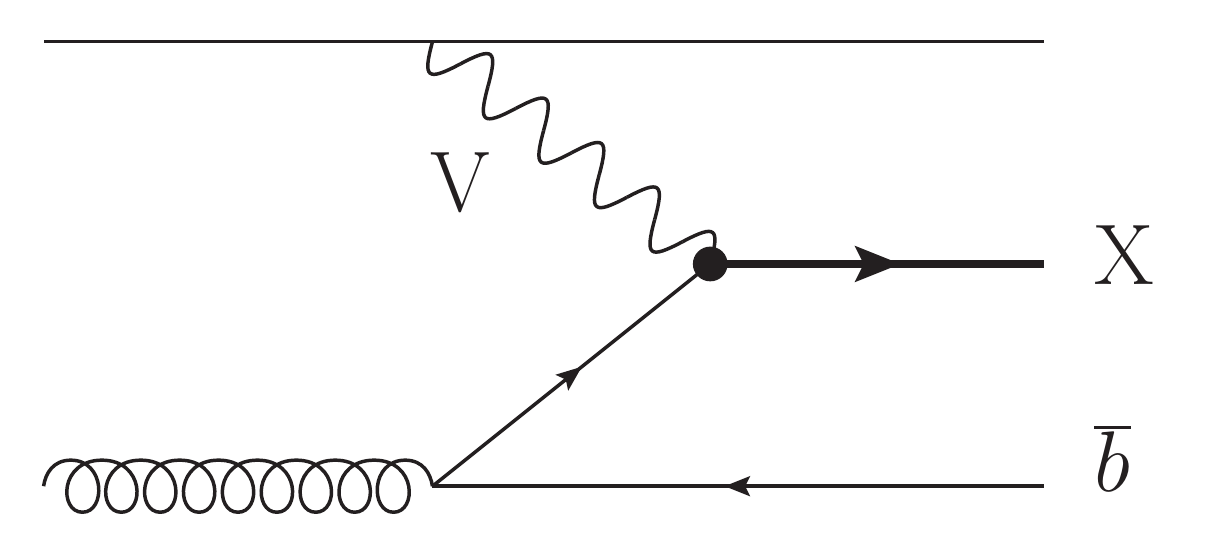}}
	\end{center}
	\caption{\small{The single-production diagrams. }}
	\label{spd}
\end{figure}

The other relevant process is the single production of the top partners in association with either a top or a bottom quark. This originates, as depicted in 
Figure~\ref{spd}, from a virtual EW boson $V=\{W^\pm,\,Z\}$ emitted from a quark line which interacts with a gluon producing the top partner and one 
third-family anti-quark. The possible relevance of single production was first pointed out in Ref.~ \cite{Willenbrock} . The relevant couplings  have the form
\be
\displaystyle
g_{Xt_R}\overline{X}_R\slashed{V}t_R+g_{Xt_L}\overline{X}_L\slashed{V}t_L+
g_{Xb_L}\overline{X}_L\slashed{V}b_L\,,
\label{spc}
\ee
where $X=\{T,B,\Xtt,\Xft,\Tt\}$ denotes generically any of the top partners. At each vertex the EW boson $V$ is understood to be the one of appropriate 
electric charge. Notice that there is no vertex with the $b_R$ because the latter state is completely decoupled in our model, we expect this coupling to be 
negligible even in more complete constructions.

It is important to outline that the couplings $g_{Xt_R}$, $g_{Xt_L}$ and $g_{Xb_L}$ can be computed analytically in our models. They arise from 
the interactions reported in Appendix~B after performing the rotation to the physical basis of mass eigenstates. Since the rotation matrices can be expressed in 
a closed form the explicit formulae for the couplings are straightforwardly derived. 
The result is rather involved and for this reason it will not be reported here, however 
it is easily implemented in a {\sl{Mathematica}} package. 

The single production cross-sections are quadratic polynomials in the couplings, with coefficients that encapsulate the effect of the QCD interactions, 
the integration over the phase-space and the convolution with the parton distribution functions. These coefficients depend uniquely on the 
mass of the partner and can be computed by Monte Carlo integration. Once the latter are known we obtain semi-analytical formulae for the 
cross-sections. The production in association 
with the $\overline{b}$ is simply proportional to $g_{Xb_L}^2$ while the one with $\overline{t}$ would be, a priori, the sum of three terms proportional 
to $g_{Xt_L}^2$, $g_{Xt_R}^2$ and $g_{Xt_L}\cdot g_{Xt_R}$ which account, respectively, for the effect of the 
left-handed coupling, of the right-handed one and of the interference among the two. However in the limit of massless top quark, $m_t\ll m_X$, the processes 
mediated by the left-handed and by the right-handed couplings become physically distinguishable because the anti-top produced in association with $X$ 
will have opposite chirality in the two cases. Therefore in the limit $m_t\rightarrow0$ the interference term can be neglected. Moreover, the coefficients of the 
${g_{Xt_L}}^2$ and ${g_{Xt_R}}^2$ terms will be equal because the QCD interactions are invariant under parity. Thus the cross-sections 
will be very simply parametrized as
\bea
&&\sigma_{{\textrm{sing}}}(X\overline{t})=\left[\left(g_{Xt_L}\right)^2+\left(g_{Xt_R}\right)^2\right]\sigma_{Vt}(m_X)\,,\nonumber\\
&&\sigma_{\textrm{sing}}(X\overline{b})=\left(g_{Xb_L}\right)^2\sigma_{Vb}(m_X)\,,
\label{prod1}
\eea
in terms of few functions $\sigma_{Vt}(m_X)$ and $\sigma_{Vb}(m_X)$. The charge-conjugate processes, in which either $\overline{X}\,t$ or 
$\overline{X}\,b$ are produced, can be parametrized in terms of a similar set of coefficient functions. The only difference  is 
the charge of the virtual $V$ emitted from the light quark line. We thus have
\bea
&&\sigma_{\textrm{sing}}(\overline{X}t)=\left[\left(g_{Xt_L}\right)^2+\left(g_{Xt_R}\right)^2\right]\sigma_{V^\dagger t}(m_X)\,,\nonumber\\
&&\sigma_{\textrm{sing}}(\overline{X}b)=\left(g_{Xb_L}\right)^2\sigma_{V^\dagger b}(m_X)\,,
\label{prod2}
\eea
where $V^\dagger$ denotes the charge conjugate of the vector boson $V$. A similar way of computing cross sections of the $W-b$ fusion type of single-production was carried out in 
Ref.~\cite{Godfrey} where they adapted the fitting functions of Ref.~\cite{Berger} to 
non-SM couplings.

\begin{table}[t]
\begin{center}
\begin{tabular}{ | c | c | c| c | c | }
\cline{2-5}
  \multicolumn{1}{c}{} &  \multicolumn{2}{|c}{$\sigma \, [\textrm{pb}]$  @ NLO}  &
  \multicolumn{2}{|c|}{$\sigma \, [\textrm{pb}]$  @ NLO}  \\
  \multicolumn{1}{c}{} & \multicolumn{2}{|c}{single production of $\overline{t}B+t\overline{B}$} &
 \multicolumn{2}{|c|}{single production of $\overline{b}\widetilde T+b\overline{\widetilde T}$}
  \\
  \hline
    $M\, [\textrm{GeV}]$   & $\sqrt{s}=7$ TeV & $\sqrt{s}=8$ TeV
        & $\sqrt{s}=7$ TeV & $\sqrt{s}=8$ TeV\\
    \hline
 400 & (2.70) 3.10       & (4.32) 4.92      & (32.49) 43.47 & (47.83) 61.43 \\
 500 & (1.49) 1.80       & (2.50) 2.97      & (15.85) 20.44 & (24.10) 33.10 \\
 600 & (0.858) 1.06     & (1.49) 1.84      & (8.53) 12.89   & (13.55) 18.80 \\
 700 & (0.511) 0.637   & (0.928) 1.15   & (4.60) 6.70      & (7.92) 11.34 \\
 800 & (0.313) 0.399   & (0.590) 0.745 & (2.82) 4.01      & (4.58) 7.22 \\
 900 & (0.194) 0.250   & (0.377) 0.497 & (1.60) 2.50      & (2.89) 4.48 \\
 1000 & (0.121) 0.160 & (0.246) 0.325 & (0.956) 1.636 & (1.81) 2.83 \\
 1100 & (0.075) 0.103 & (0.164) 0.215 & (0.604) 0.980 & (1.181) 1.72 \\
 1200 & (0.048) 0.066 & (0.107) 0.146 & (0.377) 0.586 & (0.726) 1.23. \\
 1300 & (0.031) 0.043 & (0.072) 0.098 & (0.234) 0.386 & (0.463) 0.731 \\
  \hline
\end{tabular}
\end{center}
\caption{\small{Cross sections for the NLO single production of $B$ and $\tilde T$ for a unit coupling,
at $\sqrt{s}=7, 8$ TeV (the LO values are in brackets), with MCFM \cite{mcfm}.}}
\label{tab:xsecsingle}
\end{table}

One might question the validity of the zero top mass approximation which allowed us to neglect the interference and parametrize the cross-section as in 
eq.s~(\ref{prod1}) and (\ref{prod2}). We might indeed generically expect relatively large corrections, of the order of $m_t/m_X$. However the corrections 
are much smaller in our case, we have checked that they are around $1\%$ in most of the parameter space of our models. The reason is that the interference is further reduced in our case because the left- and right-handed couplings are never comparable, one of the 
two always dominates over the other. This enhances the leading term,  $g_{Xt_L}^2$ or $g_{Xt_R}^2$, in comparison with the interference 
 $g_{Xt_L}\cdot g_{Xt_R}$. Moreover this implies that eq.s~(\ref{prod1}) and (\ref{prod2}) could be further simplified, in the sum it would be enough to retain 
 the term which is dominant in each case. We will show in the following section that the dominant coupling is $g_{Xt_R}$ in the case of the fourplet (models 
 \fA\ and \fB) and $g_{Xt_L}$ in the case of the singlet (models \oA\ and \oB).

It total, all the single-production processes are parameterized in terms of $5$ universal coefficient 
functions $\sigma_{W^\pm t}$, $\sigma_{Z t}$ and $\sigma_{W^\pm b}$. 
Notice that a possible $\sigma_{Zb}$ vanishes because flavor-changing neutral couplings are forbidden in the charge $-1/3$ sector as explained in the previous section. 
As such, the single production of the $B$ in association with a bottom quark does not take place. We have computed the coefficient functions $\sigma_{W^\pm t}$ and $\sigma_{W^\pm b}$, including 
the QCD corrections up to NLO, using the MCFM code~\cite{mcfm}. To illustrate the results, we report in Table \ref{tab:xsecsingle} the single production cross-section with  coupling set to unity, for different values of the 
heavy fermion mass, and for  the $7$ and $8$~TeV LHC. The values in the table correspond to the sum 
of the cross sections for producing the heavy fermion and its antiparticle, on the left side we show the results for $t\,B$ production, on the right one we consider 
the case of $b\,\widetilde{T}$. In our parametrization of eq.s~(\ref{prod1}) and (\ref{prod2}) the cross-sections in the table correspond respectively to 
$\sigma_{W^+ t}+\sigma_{W^- t}$ and to $\sigma_{W^+ b}+\sigma_{W^- b}$. We see that the production with the $b$ is one order of magnitude larger than 
the one with the $t$, this is not surprising because the $t$ production has a higher kinematical threshold and therefore it is suppressed by the steep fall 
of the partonic luminosities. The values in the table do not yet correspond to the physical 
single-production cross-sections, they must still be multiplied by the appropriate couplings. 

The last coefficient function $\sigma_{Z t}$ cannot be computed in MCFM and therefore to extract it we used a LO cross section computed with {\sc{MadGraph 5}}~\cite{madgraph} using the model files produced with  {\sc{FeynRules}} package~\cite{feynrules}. To account for QCD corrections in this case we used the k-factors computed with MCFM for the $t\, B$ production process. 

\begin{figure}[t]
\centering
 \includegraphics[width=0.5\textwidth]{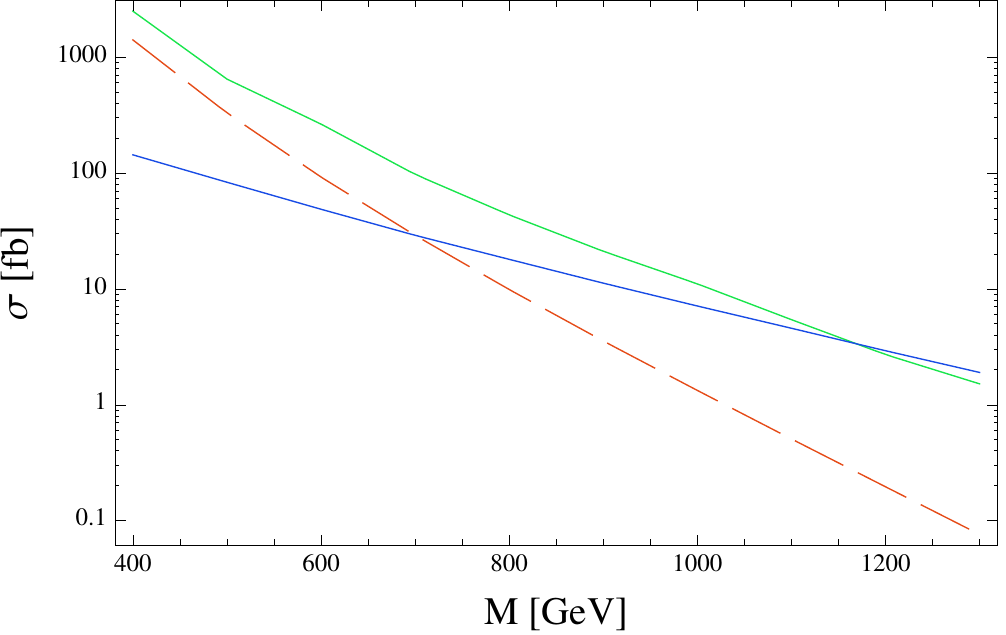}      
  \caption{\small{In red dashed: the cross sections of pair production. In green and blue the single production of the 
  $\widetilde{T}$(in association with a $b$) and of the 
  $X_{5/3}$(in association with a $t$), respectively in model
   \oA\ and \fA. 
  The point chosen in the parameter space is $\xi=0.2$, $c_1=1$ and $y=1$. The value of $c_2$ is fixed, at each value of $M_\Psi$, in order to reproduce the 
  top quark mass.}}
\label{fig:prod}
\end{figure}

In order to quantify the importance of single production we 
plot in figure~\ref{fig:prod} the cross-sections for the various production mechanisms in our models as a function of the mass of the partners and for a 
typical choice of  parameters. We see that the single production rate can be very sizeable and that it dominates over the QCD pair production already
at moderately high mass. This is again due to the more favorable lower kinematical threshold, as carefully discussed in Ref.~\cite{mrazecwulzer}.

Let us finally discuss the decays of the top partners. The main channels are two-body decays to vector bosons and third-family quarks, mediated by the 
couplings in eq.~(\ref{spc}). For the partners of charge $2/3$ and $-1/3$ also the decay to the Higgs boson is allowed, and competitive with the others in 
some cases. This originates from the interactions of the partners with the Higgs reported in Appendix~B, after the rotation to the physical basis of mass 
eigenstates. The relevant couplings can be computed analytically similarly to the $g_{t_{L,R}X}$ and $g_{b_{L}X}$. Thus we easily obtain analytical tree-level 
expressions for the partial widths and eventually for the branching fractions. In principle cascade decays $X\rightarrow X' V$ or $X'H$ are also allowed, however 
these are never sizable in our model as we will discuss in sect.~\ref{mrc}.

\subsection{Couplings to Goldstone Bosons}
\label{gc}

Let us now turn to classify the relative importance of the various production mechanisms and decay channels described in the previous section. Since the partners 
are much heavier than the EW bosons, $m_X\gg m_W$, their dynamics is conveniently studied by using the Equivalence Theorem, which applies at energies $E\gg m_W$.
 To this end, we will momentarily abandon the unitary gauge and describe our model in the $R_\xi$-gauge where the 
Goldstone degrees of freedom associated with the unphysical Higgs components are reintroduced. The Higgs field is now parameterized as 
\footnote{Notice that the Goldstone fields $\phi^{\pm,0}$ in eq.~(\ref{Hdoublet})   are not canonically normalized. Indeed  
the non-linearities in the Higgs kinetic term of eq.~(\ref{hkt}) lead to a kinetic coefficient equal to $\sin{\epsilon}/\epsilon$, with 
$\epsilon=\langle h\rangle/f$. However this is irrelevant for the purpose of the present discussion.}
\be
\displaystyle
H=\left(\begin{array}{c} h_u\\ h_d\end{array}\right)=\left(\begin{array}{c} \phi^+\\ \frac1{\sqrt{2}}\left(\langle{h}\rangle+\rho+i\phi^0\right)\end{array}\right)\,.
\label{Hdoublet}
\ee
The Equivalence Theorem states that, at high energies, the longitudinal components of the $W^\pm$ and of the $Z$ bosons are described, respectively, by the 
charged and the neutral Goldstone fields $\phi^\pm$ and $\phi^0$. The transverse polarizations are instead well described by vector fields  $W^\pm_\mu$ and $Z_\mu$,  in the absence of  symmetry breaking. However the transverse components give a negligible contribution to our processes, and this is for two reasons. First, their 
interactions emerge from the SM covariant derivatives and therefore these are proportional to the EW couplings $g$ or $g'$. We will see below that the couplings 
of the longitudinal, {\it{i.e.}} of the Goldstones, are typically larger than that. Second, the transverse components can not mediate, before EWSB, any transition 
between particles in different multiplets of the  gauge group. Indeed the couplings of the $W_\mu^\pm$ and $Z_\mu$ fields are completely fixed 
by gauge invariance and therefore they are diagonal in flavor space. Only after EWSB do states from different multiplets mix and  flavor-changing couplings like in eq.~(\ref{spc}) arise. Therefore these effects must be suppressed by a power of $\epsilon=\langle{h}\rangle/f$. 
This means that the transverse gauge bosons basically do not participate to the production and decay of the top partners: the decay will  mostly be to longitudinally polarized vectors,
while the virtual $V$ exchanged in  single production diagram will be dominantly longitudinally polarized.

For our purposes, we can thus  simply ignore the vector fields and concentrate  on  the Goldstones. In the models with the fourplet, 
\fA\ (\ref{eq:lagrangian2}) and \fB\ (\ref{eq:lagrangian214}), the first source of Goldstone couplings is the term 
$ i\, c_1 \left(\bar{\Psi}_R\right)_i \,\slashed{d}^i \, t_R$. One would naively expect this interaction to be the dominant one because it originates entirely from the 
strong sector without paying any insertion of the elementary-composite coupling $y$. Before EWSB the couplings are
\be
\displaystyle
i\,\frac{\sqrt{2}c_1}{f}\left[
-\overline{T}\gamma^\mu t_R\partial_\mu\left(\frac{\rho-i\phi^0}{\sqrt{2}}\right)
+\overline{B}\gamma^\mu t_R\partial_\mu\phi^-
+{\barXtt}\gamma^\mu t_R\partial_\mu \left(\frac{\rho+i\phi^0}{\sqrt{2}}\right)
+{\barXft}\gamma^\mu t_R\partial_\mu\phi^+
\right]\,+\,{\textrm{h.c.}}\,.
\label{gc0}
\ee
It is not difficult to check that the interactions above respect not only the SM but also the full ${\textrm{SO}}(4)$ symmetry of the strong sector. 
Eq.~(\ref{gc0}) contains derivative operators, therefore it is not yet suited to read out the actual strength of the interactions. However it can be 
simplified, provided we work at the tree-level order, by making use of the equations of motion of the fermion fields. \footnote{When considering a perturbation described by a small parameter $\eta$ to a Lagrangian, the use of the equations of motion of the unperturbed theory is equivalent to permorming field redefinitions of the form $\Phi\to \Phi+\eta F[\Phi,\partial]$. For example, to deal with the first term of 
eq.~(\ref{gc0}),  the relevant redefinition is 
$$
\displaystyle
\begin{matrix}
T_R\;\rightarrow\;T_R+\frac{\sqrt{2}c_1}{f}h_d^\dagger t_R\\
t_R\;\rightarrow\;t_R-\frac{\sqrt{2}c_1}{f}h_d T_R
\end{matrix}\,.
$$
This eliminates the derivative interaction and makes the first term of eq.~(\ref{gc1}) appear. It also leads to new interactions with more fields that 
however are irrelevant for our processes at the tree-level.
}
After integrating by parts and neglecting the top mass, we find
\be
\displaystyle
\frac{\sqrt{2}c_1}{f}\left[
-m_T\left(\frac{\rho-i\phi^0}{\sqrt{2}}\right)\overline{T} t_R
+m_B \phi^- \overline{B}t_R
+ m_\Xtt \left(\frac{\rho+i\phi^0}{\sqrt{2}}\right){\barXtt} t_R
+m_\Xft \phi^+{\barXft}t_R
\right]\,+\,{\textrm{h.c.}}\,,
\label{gc1}
\ee
showing that  the strength of the interaction is controlled by the masses of the heavy fermions. Neglecting the elementary-composite coupling 
$y$, the masses all equal  $M_\Psi$, and the coupling, modulo an $O(1)$ coefficient,  is given by $ g_\Psi=M_\Psi/f$, as anticipated in the previous section. Once again we remark that 
this feature follows from the Goldstone boson nature of the Higgs. Indeed if the Higgs were  a generic resonance, not a Goldstone,  then it could  more plausibly have a Yukawa  $g_*\overline{\Psi}^i\Pi_it_R$ vertex with strength dictated by the
strong sector coupling $g_*$.
 
Those of eq.~(\ref{gc1}) are the complete Goldstone interactions in the limit of a negligible elementary-composite coupling $y$. However we can not rely 
on this approximation because we will often be interested in relatively light top partners, with $g_\Psi\leq y\simeq y_t$. It is straightforward to incorporate the 
effect of $y$,  due to the mixing terms in eq.s~(\ref{eq:lagrangian2}) and (\ref{eq:lagrangian214}) for model \fA\ and \fB, respectively. After diagonalizing 
the mass-matrix, again neglecting EWSB, the Goldstone interactions for both models become 
\begin{equation}
\begin{tabular}{ | c | c | }
\cline{2-2}
\multicolumn{1}{c|}{}   & \fA, \fB \\
\hline
$ \phi^+{\barXft}_L \,t_R$ & $ \sqrt{2} c_1 g_{\psi}$  \\
$(\rho+i \phi^{0}){\barXtt}_L \,t_R $ & $c_1 g_{\psi}$  \\
$(\rho-i \phi^{0})\overline{T}_L \,t_R $  & $-c_1 \sqrt{y^2 + g_{\psi}^2} + {c_2 y^2 \over \sqrt 2  \sqrt{y^2 + g_{\psi}^2}}$  \\
$ \phi^- \overline{B}_L \,t_R$  & $c_1 \sqrt{2} \sqrt{y^2 + g_{\psi}^2} - {c_2 y^2 \over \sqrt{y^2 + g_{\psi}^2}}$  \\
 \hline
\end{tabular}
\label{coup4}
\end{equation}
which reduces to eq.~(\ref{gc1}) for $y\ll g_\Psi$. Notice that eq.~(\ref{coup4}) only contains couplings with the right-handed top quark. This is not surprising 
because the top partners live in SM doublets and therefore their only allowed Yukawa-like interactions are with the $t_R$ singlet. The couplings with the 
$q_L$ doublet emerge only after EWSB and are suppressed by one power of $\epsilon$. Therefore they typically do not play a mayor role in the phenomenology.
Obviously the SM symmetry is respected in eq.~(\ref{coup4}), this explains the $\sqrt{2}$ suppression of the $\Xtt$ and of the $T$ couplings compared with the ones 
of the $\Xft$ and of the $B$.

The situation is different in the models with the singlet, \oA\ and \oB\ (\ref{eq:lagrangian211}). In that case there is no direct contribution from the strong sector to 
the Goldstone coupling and all the interactions are mediated by $y$. The couplings are
\begin{equation}
\begin{tabular}{ | c | c| }
\cline{2-2}
\multicolumn{1}{c|}{}  & \oA, \oB \\
\hline
$(\rho+i \phi^{0})\overline{\widetilde{T}}_R \,t_L $ & $ { y \over \sqrt{2}} $ \\
$ \phi^+ \overline{\widetilde{T}}_R \,b_L$ & $y$ \\
 \hline
\end{tabular}
\label{coup1}
\end{equation}
The top partner $\Tt$ now is in a SM singlet, therefore the interactions allowed before EWSB are the ones with the left-handed doublet. The $\sqrt{2}$ 
suppression of the coupling with the top is due, once again, to the SM symmetry. 
One important implication of eq.~(\ref{coup1}) is that the $\Tt$, contrary to the partners in the fourplet, can be copiously produced singly in association 
with a bottom quark. We will discuss this and other features of our models in the following section.

\subsection{The Most Relevant Channels}
\label{mrc}

We discuss here the most relevant production and decay processes of each top partner, identifying the best channels where these particles 
should be looked for at the LHC. Obviously one would need an analysis of the backgrounds to design concrete experimental searches 
for these promising channels and to establish their practical observability. We leave this to future work and limit ourselves to study, in section~4, 
the constraints on the top partners that can be inferred from presently available LHC searches of similar particles

Let us first consider  the models \fA, \fB\ and analyze separately each of the new fermions. 

\begin{itemize}

\item $\Xft$

$\Xft$, together with  $\Xtt$, is the lightest top partner,  it is therefore the easiest to produce. Production can occur in pair, via QCD interactions, or in association 
with a top quark through its coupling with a top and a $W^+$. The coupling, see eq.~(\ref{coup4}), is controlled by  $g_\psi=m_\Xft/f$, which grows with mass at fixed $f$.  We thus expect  
single production to  play an important role at high mass, where it is enhanced with respect to pair production by both  kinematics and  a larger coupling (at fixed $f$). This is confirmed, for a particular but typical choice of  parameters, by the plot in Figure~\ref{fig:prod}.

Since it is the lightest partner,  $\Xft$ decays to $W^+t$ with unit branching ratio. The relevant channel for its observation is
$\Xft\rightarrow tW$ in association with a second top quark of opposite charge. The latter  is present  in  both single and  pair production processes. 
This results in clean signals consisting of either same-sign dileptons or trileptons plus jets. In the following section we will recast the LHC searches for these signals  and obtain a limit on $\Xft$ production. In addition to two top quarks and a $W$,  pair production  also leads to a second hard $W$ while  single production (see Figure~(\ref{spd})) features
 a light-quark jet associated with virtual $W$ emission.

Notice that the light-quark  jet in single production is typically forward with a $p_T\lesssim m_W$ because the emission of the virtual $W$ is enhanced 
in this kinematical region \cite{mrazecwulzer} . In practice this jet has the same features of the``tag jets" in VBF Higgs production and in $WW$--scattering. The 
events are thus characterized by a forward isolated jet in one of the hemispheres. The relevant kinematical distributions are shown in 
Figure~(\ref{fig:forwardjet}) for the production of a $600$~GeV partner. Like in VBF or $WW$-scattering, one might hope to employ the forward jet
 as a tag to discriminate single production form the background.  Ref.~\cite{mrazecwulzer} argued that the main source 
of forward jets in the background, QCD initial state radiation, tends to produce more central and less energetic jets, however further investigations are 
needed. Present LHC searches  are  designed for pair- rather than  for  single-production.  Because of the $\eta^{jet}$ and $p_T^{jet}$ cuts that they adopt, they are thus weakly sensitivity to 
forward jets. We believe that it would be worth to explore the possible relevance of forward jets in designing 
the searches for  top partners. 

\begin{figure}[t]
\centering
  \includegraphics[width=0.49\textwidth]{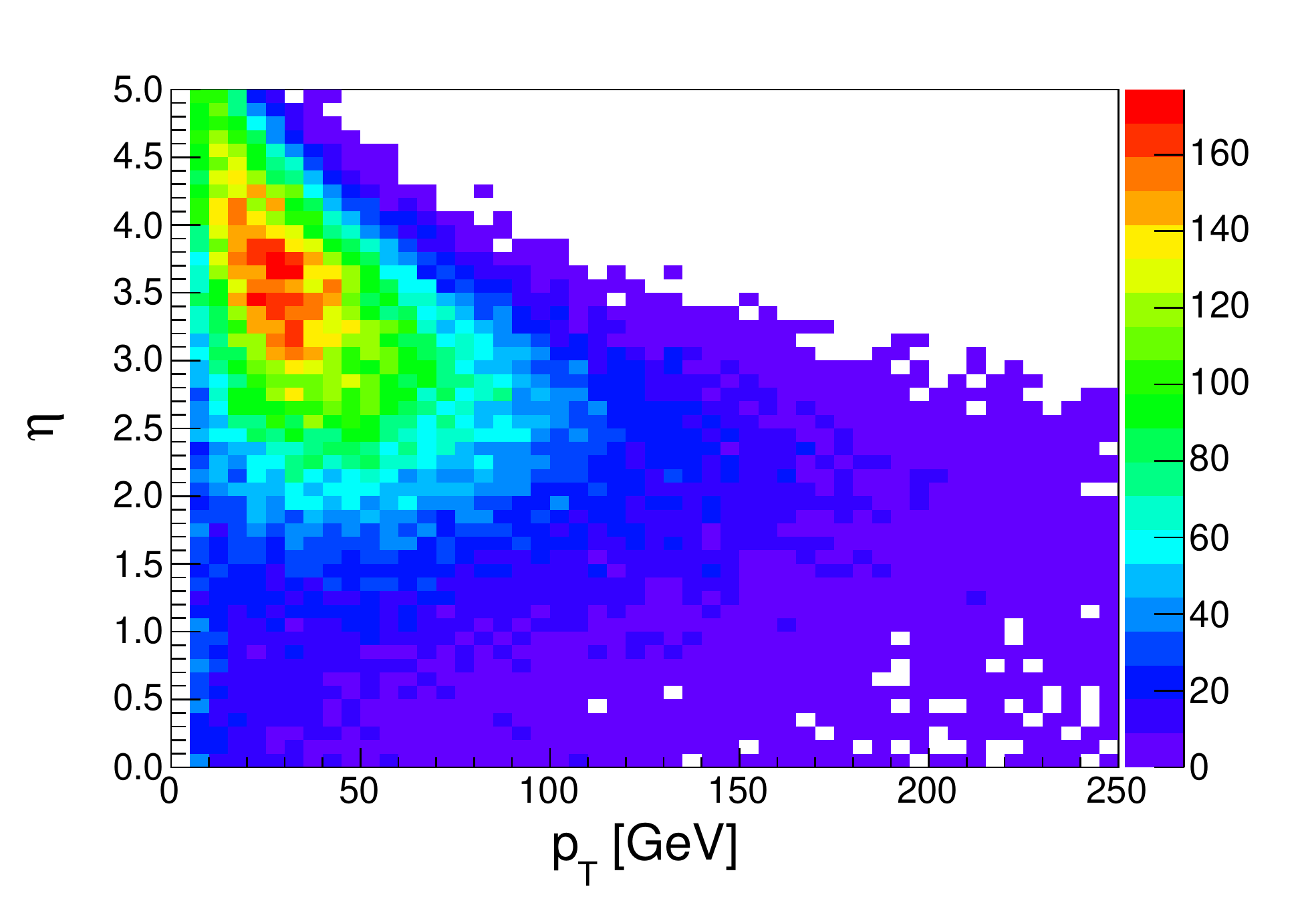}   \hfill
  \includegraphics[width=0.49\textwidth]{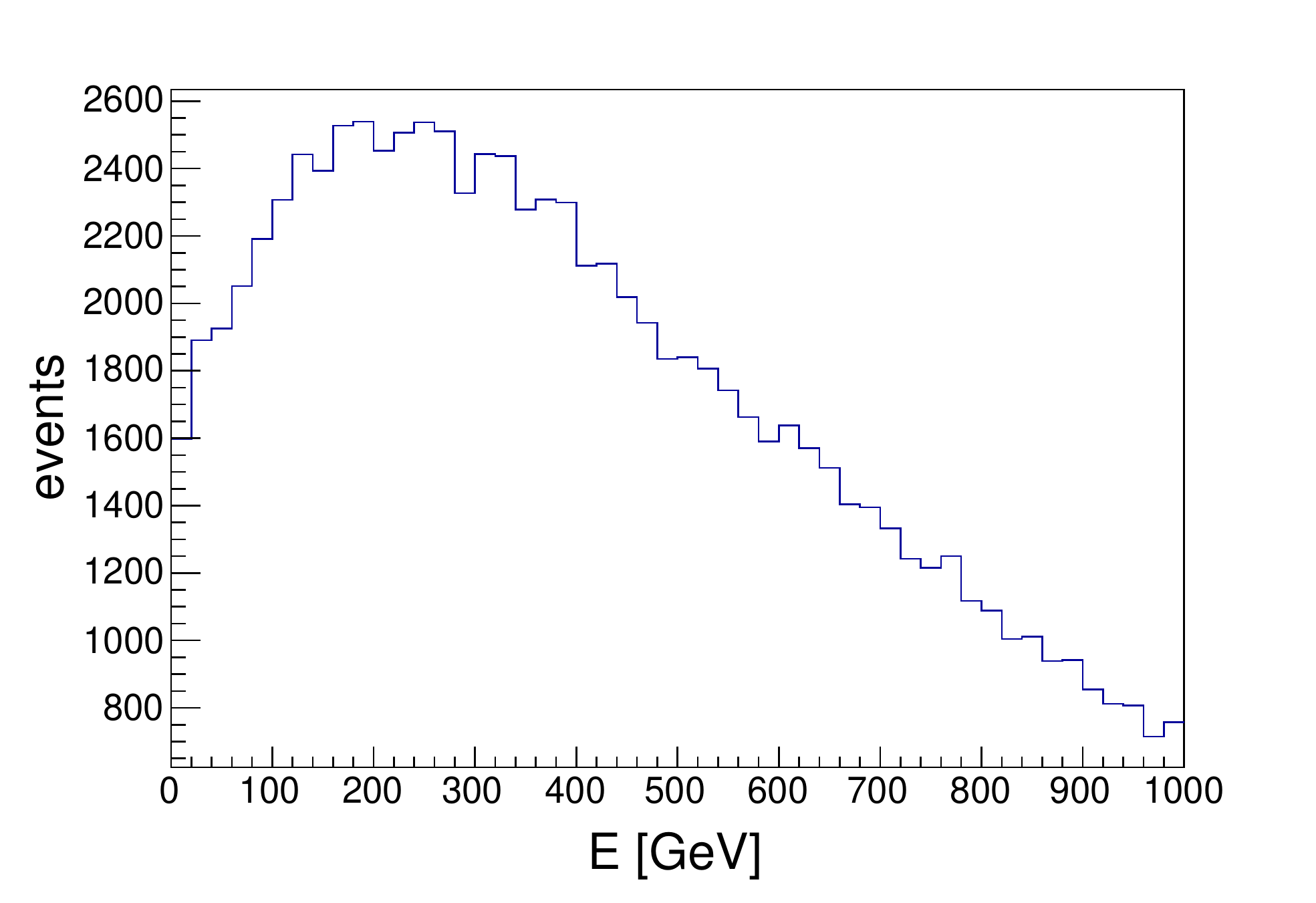}    \\
  \caption{\small{$p_T-\eta$ and energy distributions of the forward jets produced in a single production of the top partner with a mass $600\,\GeV$.}} 
  \label{fig:forwardjet}
\end{figure}

\item $\Xtt$

 $\Xtt$ is also light and therefore  easier to produce than the heavier partners. At the leading order, as eq.~(\ref{coup4}) shows, it couples with strength $c_1 g_{\psi}$ to the 
Higgs and  $Z$ bosons. The dominant decay channels are thus $\Xtt\rightarrow Zt$ and $\Xtt\rightarrow ht$ and 
$\BR(\Xtt \to  Z \, t) \approx \BR(\Xtt \to h\, t) \approx 0.5$. In model \fA\ the coupling to $Wb$ vanishes exactly, while in model \fB\ the coupling is non-zero 
but  suppressed by  $\epsilon\sim v/f$. The decay $\Xtt\to Wb$ is therefore typically sub-dominant and can 
become relevant only in a corner of  parameter space characterized by low mass,  $y \epsilon=O(1)$ and  $c_1<1$. Given that 
$\Xtt\to ht$ is probably difficult to detect
(see however Ref.~\cite{ht} for recent analyses), the search for $\Xtt$ must rely on the decay mode $\Xtt\to Zt$, with $Z$ further decaying to charged leptons. An extra suppression from the small branching ratio must then be payed.
This disfavors the $\Xtt$ signal compared to that of  $\Xft$, for which the branching ratio needed 
to reach the leptonic final state is close to one.

$\Xtt$ is produced  in pairs via QCD interactions and  singly via the $Z\Xtt t$ coupling,. In the latter case a top quark is produced in association. Both production modes 
lead to a resonant $\Xtt\to Zt$ plus one top of opposite charge.  In the 
case of single production there will be a forward jet, as previously discussed in the case of $\Xft$. In the case of pair production there will be  either a Higgs or a $Z$ from the other partner. 
Another possible single production mode, in association 
with a $b$ quark rather than a $t$, is strictly forbidden in model  \fA\ and  is suppressed by the small coupling to $Wb$ in model \fB. However  single 
production in association with a $b$ is kinematically favored over that with  $t$. Kinematics then compensates the suppressed coupling and makes  the two rates typically comparable in model \fB, as shown in  Fig.~\ref{fig:compareWBZT}.  By comparing with Fig.~\ref{fig:prod}, we see that, in the case of  $\Xtt$,  single production in association with a $t$ is suppressed compared to the case of  $\Xft$\footnote{Even though the two plots correspond to different models the couplings of the $\Xft$ and $\Xtt$ do not differ at  leading order in models \fA\ and \fB.\ }. This is mainly due to the $\sqrt{2}$ factor in charged current versus neutral current vertices, see
eq.~(\ref{coup4}). Moreover, the difference between the $W$ and $Z$ couplings, taking into 
account $u$- and the $d$-type valence quark content of the proton, further enhances  by a  $\sim 1.2$ factor the virtual $W$ emission rate with respect to the $Z$ rate. Combining this enhancement with the factor of $2$ in the squared coupling, one
 explains the relative sizes of  the $\Xft$ and $\Xtt$ production cross sections.

\begin{figure}[t]
\centering
  \includegraphics[width=0.5\textwidth]{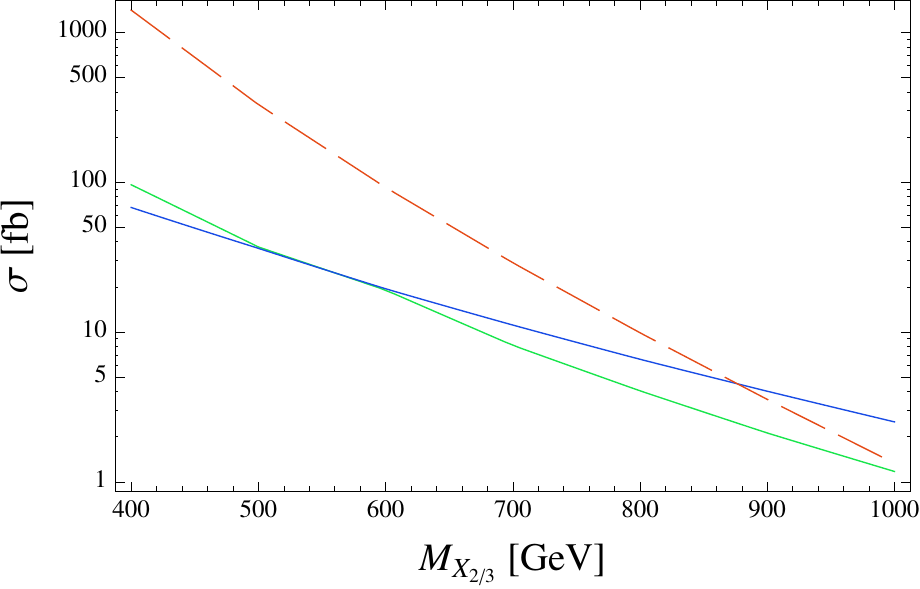}  \\
  \caption{\small{Cross sections of the $X_{2/3}$ pair (red dashed line) and single production in association with a $t$ (blue line) and with a $b$ (green line) for the 
  parameters choice:  $y=1$, $c_1=1$, $\xi=0.2$ in the model $\fB$. }}
\label{fig:compareWBZT}
\end{figure}

\item $T$ 

$T$  is systematically heavier than  $\Xtt$, but the  phenomenology is very similar. Therefore 
it will merely give a subdominant contribution to the $\Xtt$ channels described in the previous paragraph. Indeed,  by eq.~(\ref{coup4}),  also  $T$ 
couples at  leading order  with equal strength to the Higgs and to the $Z$, leading to $\BR(T \to  Z \, t) \approx \BR(T \to h\, t) \approx 0.5$. The coupling 
to $Wb$ arises at order $\epsilon$, and it can be relevant, as explained for  $\Xtt$ above, thanks to the favorable kinematics of associated production with a $b$. 

One may in principle consider  chain  decays seeded by  $T\to \Xtt Z$, $T\to \Xtt h$ or  $T\to \Xft W$, given these channels are normally kinematically open. However the corresponding couplings
are generically smaller than those controlling the direct decays to $t_R$. This is  a straightforward consequence of the equivalence theorem and  of $SU(2)$ selection rules. The decays to $t_R$, involve longitudinally polarized vectors and $h$, living in the linear Higgs doublet $H$: given the top partners are  $SU(2)$ doublets and $t_R$ is a singlet,   the coupling respects  $SU(2)$ and so it arises at zeroth order in $\epsilon$. On the other hand, the transitions among top partners living in different $SU(2)$ doublets obviously require an extra insertion of the Higgs vacuum expectation value. The resulting amplitudes are therefore suppressed by one power of $\epsilon$ and the corresponding branching ratios negligible.

\item $B$

$B$ is even heavier than  $T$, though the mass difference,  $m_B-m_T \sim {y^2 v^2 / 4 m_B}$ (see eq.~(\ref{mdtb})), is typically rather small. The most relevant decay mode is 
 $B\to W t$, mediated by the coupling $\sim c_1 g_\Psi$ in eq.~(\ref{coup4}).
Like in the  case of $T$, $SU(2)$ selection rules suppress the decay to  $W\Xtt $. Moreover, the decay   $B\to WT$, when kinematically allowed, proceeds either via a transverse $W$, with SM gauge coupling $g<g_\Psi$, or via a longitudinal $W$, with effective coupling suppressed by $\epsilon$. Therefore also this decay is significantly suppressed.
The decay  $B\to Zb$ is forbidden because, as we explained in sect.~\ref{trilinear}, flavor-changing neutral couplings are absent in the charge $-1/3$ sector.  The $B\to hb$ channel is forbidden in model \fA\ and suppressed by $\epsilon$ in model \fB. 
 In the latter model it can play a role, but only in a corner of the parameter space.

Single production, since the $ZBb$  vertex  is  absent, is  always accompanied by a top quark. The  signature of  single $B$ production  is therefore   a resonant  $B\to W t$ plus an  opposite charge top, the same final states of single $\Xft$ production. In the end, $B$ production, single and pair, has  the same signatures as $\Xft$
production: same sign leptons   or trileptons plus jets.

\end{itemize}

Let us now switch to models \oA\ and \oB, where the only new heavy fermion is the $\tilde T$.

\begin{itemize}
\item $\widetilde T$

$\widetilde T$ has a very rich phenomenology because it can be copiously produced through all the three mechanisms described above. We see in 
eq.~(\ref{coup1}) that  $\widetilde T$ couples  to both $Zt$ and  $Wb$, with a coupling of order $y\sim y_t/c_2$. It can therefore  be singly produced either in association 
with a top or with a bottom quark. Notice that in the range $c_2\sim 1$ suggested by power counting, the trilinear coupling is of order $y_t $, which  is expected to be generically smaller than the strong sector coupling $g_\psi$
that controls the single production of top partners in a $(2,2)$. The bands in the left panel of Fig.~\ref{fig:ttildesp2}, indicate the single prooduction cross section\footnote{By fixing $m_t$, $\xi$, $c_2$ and  $m_{\tilde T}$ the result for 
model \oB\  and  \oA\ coincide. Indeed, by
comparing the lagrangians~(\ref{scalarlagr1A}) and~(\ref{scalarlagr1B}), one notices that the gauge vertices  and the  mass spectrum of  model \oB\  equal those of  model \oA\  when  the equality $y^{\text{\oA}} \sin \epsilon = y^{\text{\oB}} \sin 2\epsilon /2$ holds.} for $0.5<c_2<2$: comparing the blue band  to the corresponding case of $\Xtt t$ and $\Xft t $ production  in models \fA\ and \fB\ , one notices, as expected, a typically smaller rate  for models  \oA\ and  \oB.\
While $y\sim y_t$ ($c_2\sim 1$) is favored by naive power counting, one can entertain the possibility of choosing $y> y_t$ ($c_2<1$), for which the single production rate can be sizeable. However, for a given value of $m_{\widetilde T}$ and $f$, there is a mathematical upper bound $y_{max}$ on $y$ determined by  eqs.~(\ref{minmassttilde}).
The right plot in  Fig.~\ref{fig:ttildesp2} shows that $y_{max}$ grows with $m_{\widetilde T}$ and that it is comparable in model \oA\ and model \oB.\ 
In the left panel  of Fig.~\ref{fig:ttildesp2}, the green line and the blue line  shows, respectively for $\tilde T b$ and $\tilde T t$,  the maximal allowed cross section, which basically coincides with the choice $y=y_{max}$  \footnote{Note that, for a given $m_{\widetilde T}$,  $y_{max}$ does indeed correspond to the maximal value of the $W \bar b \widetilde T$-coupling, while the coincidence is not exact in the case of the $Z \bar t \widetilde T$-coupling.}. For such maximal values the single production cross section can be quite sizeable.

Single production of a $\tilde T$-like partner was considered in the context of Little Higgs models
in Refs.~\cite{Perelstein, Han}, and more recently for 
composite Higgs models in Ref.~\cite{Vignaroli}, where it was  also considered the 
possibility of using a forward jet tag as a handle for this kind of searches. The total cross section in this channel is favored over single production with a $t$ by both kinematics and by 
 the $\sqrt{2}$ factor in charged current transitions. Indeed, as shown in  Fig.~(\ref{fig:ttildesp2}) associated $\tilde T b$ production dominates even over pair production in all the relevant mass-range while  single production with the $t$ is rather small.  The role of kinematics is especially important in this result, as the large $\tilde T b$ cross section is dominated by the emission of a soft $b$, with energy  in the tens of GeV, a regime obviously unattainable in the similar process wih a $t$. Indeed by performing a hard cut of order $m_t$ on the $p_T$ of the $b$, the $\tilde T b$ cross section would become comparable to that for $\tilde T t$. Unfortunately
the current LHC searches do not exploit the large inclusive rate of production with the $b$ quark  
 because they are designed to detect pair production. We will show in the following section that the acceptance of single production, with the  cuts presently adopted is extremely low. We believe there is space for  substantial improvement in the search strategy.

\begin{figure}[t]
\centering
  \includegraphics[width=0.48\textwidth]{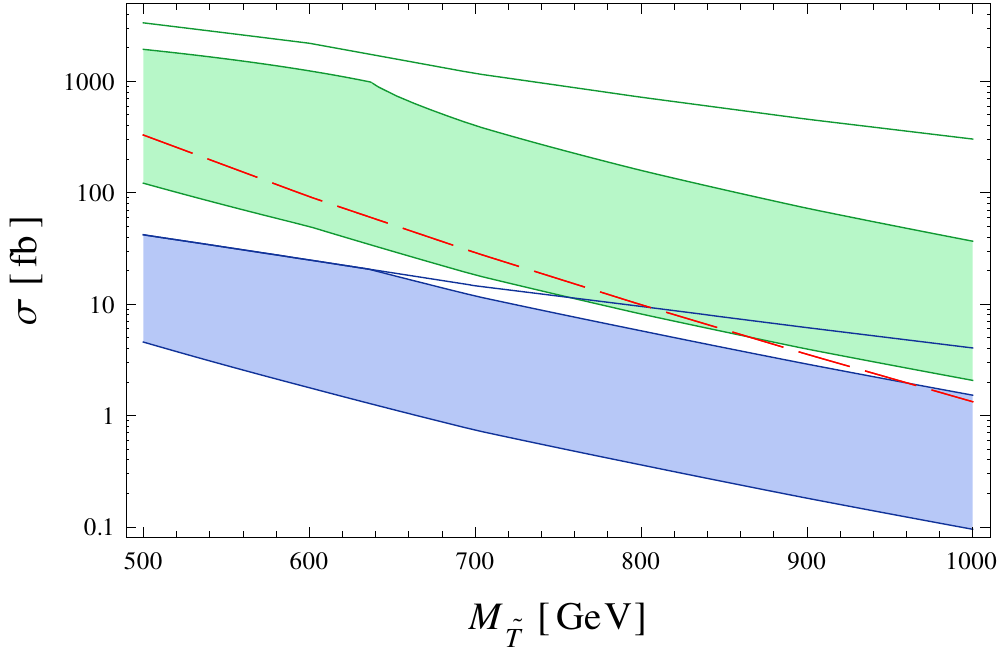}   \hfill
  \includegraphics[width=0.462\textwidth]{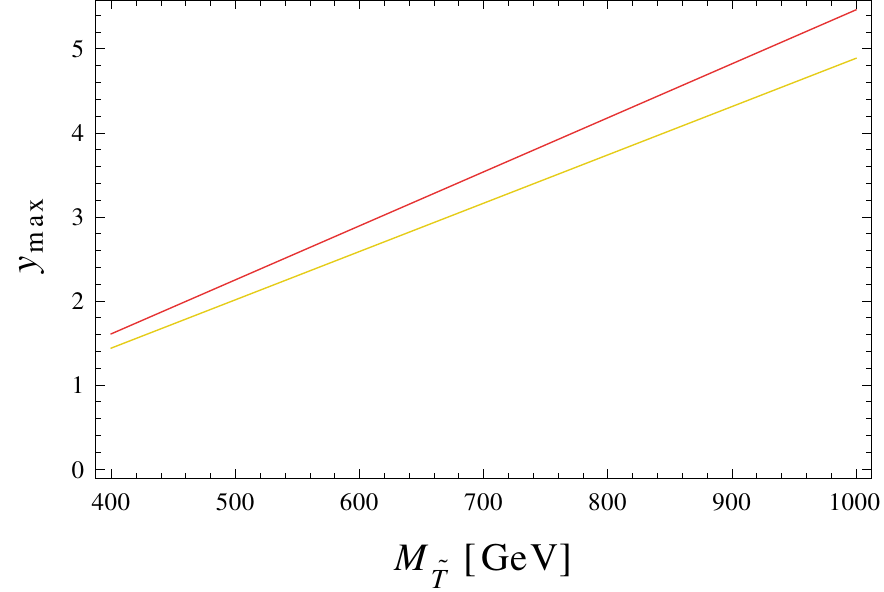}    \\
\caption{\small{Left panel: cross sections for the different production mechanisms of $\widetilde T$ for the models $\oA\ $and $\oB \ $for $\xi=0.2$. 
Red dashed: pair production;  green line: $\tilde T b$ production with the maximal allowed coupling, green band: $\tilde T b$ production
for $0.5<c_2<2$; blue line: $\tilde T t$ production for the maximal allowed coupling, blue band: $\tilde T t$ production
for $0.5<c_2<2$. Right panel: maximal allowed $y$ for the models $\oA\ $(in yellow) and $\oB \ $(in red).} }
\label{fig:ttildesp2} 
\end{figure}

Also concerning decays, all the possible channels are important in the case of  $\widetilde T$. It decays to $W b$, $Z t$ and  $h t$ at zeroth order in $\epsilon$, with 
a fixed ratio of couplings. By looking at eq.~(\ref{coup1}) we obtain 
 $\BR(\widetilde T \to Z\,t) \approx \BR(\widetilde T \to h\, t) \approx {1 \over 2}\BR(\widetilde T \to W\, b)\approx 0.25$. Actually the branching fraction to $Wb$ 
 is even further enhanced by the larger phase space, though this is only relevant for low values of $m_{\widetilde T}$. Given that the branching fraction is 
 larger, ideally the resonant $Wb$ production would be the best channel to detect the $\widetilde T$. However one should manage to design a search 
 strategy to reject the background while retaining the signal. In particular one should retain as much as possible the contribution from the large single production in association  with the $b$. A possibly cleaner decay channel could then  be 
  $\widetilde T \to Z\,t$ with leptonic $Z$.

\end{itemize}

\section{LHC Bounds} 
\label{sect:bounds}

In this section we derive bounds on our models using  the presently available LHC searches. Given that the top partners are heavy 
fermions coupled to top and bottom, we focus on the experimental searches for $4^{th}$ family quarks, which present a somewhat similar phenomenology 
\footnote{Significant bounds on the top partners could also emerge from unrelated studies like the searches of SUSY performed with the "razor" variable 
\cite{razor}. We thank M.~Pierini for suggesting this possibility, obviously this is an interesting direction to explore.}. 
We will make use of the following searches for $4^{th}$ family quarks performed by CMS: 1) $b'\to Wt$ with  same-sign dileptons or trileptons in the final state \cite{cmsBWt}; 2) $t'\to Zt$ with trileptons in the final state \cite{cmsTZt};
3) $t'\to Wb$ with two leptons in the final state \cite{cmsTWb}.

In what follows, we quantify the impact of these three searches on our models, by adopting the 
following strategy. We compute separately the production cross-sections of the top partners, the
branching fractions into the relevant channels and the efficiencies associated with the selection 
cuts performed in each experimental search. The cross-sections and the branching fractions 
at each point of the parameter space are encapsulated in semi-analytical formulae as 
described in section~\ref{sec:TPP}. 
The efficiencies must instead be obtained numerically through a Monte Carlo simulation. Not having at our disposal a reliable tool to estimate the response of the detector, a fully realistic
simulation of the hadronic final states would not be useful. Therefore we decided not to include
showering and hadronization effects in our analysis, and we stopped at the parton level.  We applied the reconstruction (e.g., of $b$-jets and leptons) and selection cuts on the partonic events in order to get an estimate of the kinematical acceptance. Moreover, we included the efficiencies for $b$-tagging, lepton reconstruction and trigger through universal reweighting factors extracted from the experimental papers. In oder to account for the possible merging of soft or collinear partons in single jet we applied the anti-$k_T$ clustering algorithm \cite{antikt} for jet reconstruction with distance parameter $\Delta R=0.5$.

\subsection{Search for $b^{\prime}\to W\,t$} 

This search applies only to models \fA, \fB.
The analysis of Ref.~\cite{cmsBWt} aims at studying a $4^{th}$ family $b'$ that is pair produced by QCD and is assumed to decay to $Wt$ with unit branching fraction. 
The search is performed in the final state with at least one tagged $b$-jet and either same-sign dileptons or trileptons ($e$ or $\mu$). Three or two additional 
jets are required, respectively, in the dilepton and trilepton channels. Apart from the usual isolation, hardness and centrality cuts for the jets and the leptons, a 
hard cut is required on the scalar sum of the transverse momenta of the reconstructed object and missing $p_T$. Ref.~\cite{cmsBWt} reports the observed number 
of events in the two categories and the expected SM background. From these elements, given the efficiency of the signal in the two channels, one  
puts a bound on the pair production cross-section and eventually on the mass of the $b'$. With $4.9\,fb^{-1}$ of data at $7$~TeV the bound is $611$~GeV 
at $95\%$ confidence level. Below we will quantify the impact of this search on the parameter space of our models. 

The top partners contributing to the signal are  $\Xft$ and  $B$ because, as shown in the previous section, they lead to two tops of opposite charge 
and to at least one extra $W$  in both  pair and  single production. 
To derive the bound we must compute, for each partner and production mode, 
the efficiency of the signal in the dilepton and the trilepton channels as a function of the partner's mass. The total production cross-sections are computed semi-analytically at each point of the parameter space. Combining the cross-sections with the efficiencies we obtain the 
signal yield in the two channels that must be compared with the observed number of the events and with the expected background. We perform this comparison 
by computing the confidence level of exclusion (CL) defined through the $CL_s$ hypothesis test \cite{CLs}, as explained in some detail in Appendix~C. At the 
practical level it is important that at the end of this procedure we obtain an \emph{analytical} expression for the $CL$ as a function of the fundamental parameters 
of our model. This makes very easy and fast to draw the exclusion bounds even if we work in a multi-dimensional parameter space.

\subsubsection*{Efficiencies} 

The first step is to simulate the signal processes. Rather than employing our complete model we have used a set of simplified {\sc{MadGraph}} 
models containing the SM fields and interactions plus the two relevant new particles -- $\Xft$ and $B$ -- with the appropriate couplings to $Wt$
responsible for the single production and the decay. 
We will employ the right-handed $\Xft Wt_R$ or $BWt_R$ vertices
because, as we have shown in section~\ref{gc},  the top partners couple mainly to the $t_R$.
However,  to make contact with Ref.~\cite{cmsBWt}, we simulated also  the case of left-handed vertices
because for a $4^{th}$ family $b'$  the coupling originates from an off-diagonal entry of the generalized 
$V_{CKM}$ matrix and it is purely left-handed. 
We will see that the chirality of the couplings  significantly affects the efficiencies.

We generated parton level events without showering, hadronization and detector simulation. The events were analyzed using the cuts and the 
identification/reconstruction efficiencies for $b$-tagging and leptons reported in \cite{cmsBWt}. 
We also included the trigger efficiency as an overall multiplicative factor. 
Not having enough information on 
how the $\tau$ leptons were treated in the analysis we have accounted only for the missing energy from the tau decays while the jets and leptons 
candidates coming from taus were simply rejected.  
We checked that the inclusion of $\tau$-jets does not introduce appreciable differences, but
$\tau$-leptons might affect our results.
We have found that the most severe cut is the one on the transverse momenta 
of the leptons candidates, $p_T>20\, \GeV$. This is because most of the events which could contribute to di-(tri-)leptons contain exactly 2(3) 
charged lepton candidates, thus loosing only one of them causes the loss of the event. The number of generated jets per event is instead larger than 
the minimally required one and therefore the impact of the jet cut is less prominent.

The signal efficiency is defined as the product of the cut efficiencies with the branching ratios of the $t$ and the $W$ to the required final states.
The results are given in Tables~\ref{tab:efficienciessimple2l} and \ref{tab:efficienciessimple3l} for different mass points. In these tables, 
the efficiencies of the pair-produced $b'$ obtained in Ref.~\cite{cmsBWt} are compared with
the ones obtained for a left-handed coupling with our method.
 The accuracy of our simplified 
treatment of QCD radiation and  detector effects is quantified by the level of agreement between these results. 
We see that the discrepancy is below $10\%$ in the dilepton channel and around $30\%$ in the case of trileptons. 
In view of these results we have decided to be conservative and to present our results by showing exclusion 
limits computed using our efficiency and also using an efficiency reduced respectively by   $10\%$ for dileptons  and $30\%$ for trileptons. 
From Tables~\ref{tab:efficienciessimple2l} and \ref{tab:efficienciessimple3l} we also see that the efficiency in our model
is significantly larger than the one for the  $4^{th}$ family $b'$. This is because the right-handed top (and the left-handed anti-top) produced in the decay in our models tends to 
produce more energetic charged leptons than a left-handed top. The lepton $p_T$ distribution is therefore harder and the cut $p_T>20\, \GeV$ is more easily 
satisfied. Finally, we notice, somewhat surprisingly, that the efficiencies for the $\Xft$ and for the $B$ partners are substantially identical. One would have 
expected some difference at least in the dilepton channel, since the two leptons come from the decay of a single heavy particle in the first case while they 
have a different origin in the second one. However this makes no difference in practice.

\begin{table}
\begin{center}
\begin{tabular}{ | c | c | c |c | c | }
\hline
  $M$   $[\textrm{GeV}]$    &  $\Xft$ partner $[\%]$ & $B$ partner $[\%]$&     4$^{\rm th}$ family $b'$ $[\%]$ & 
   $b'$ Ref.~\cite{cmsBWt}  $[\%]$\\
  \hline
 450   & $1.90\pm0.05$ & $1.93\pm0.05$& $1.65\pm0.04$& $1.52\pm0.13$\\ 
 550  & $1.97\pm0.05$  & $1.98\pm0.05$ & $1.72\pm0.05$& $1.71\pm 0.14$\\
 650   & $1.96\pm0.05$ & $1.96\pm0.05$ & $1.85\pm0.05$& $1.71\pm 0.15$\\ 
  \hline
\end{tabular}
\end{center}
\caption{ 
\small{Efficiencies for the pair produced $B$ and  $X_{5/3}$ going to same-sign dileptons. Efficiencies contain the cuts losses, b-tagging performance and BR's of W boson. }}
\label{tab:efficienciessimple2l}
\end{table}

\begin{table}
\begin{center}
\begin{tabular}{ | c | c | c |c | c | }
\hline
   $M$   $[\textrm{GeV}]$    &  $\Xft$ partner $[\%]$ & $B$ partner $[\%]$&     4$^{\rm th}$ family $b'$ $[\%]$ & 
   $b'$ Ref.~\cite{cmsBWt}  $[\%]$\\  \hline
 450  & $0.88\pm0.02$ & $0.84\pm0.02$ & $0.69\pm0.02$& $0.47\pm 0.05$\\ 
 550 & $0.98\pm0.02$  & $0.94\pm0.02$ & $0.81\pm0.02$& $0.56\pm 0.05$\\
 650  & $1.04\pm0.03$  & $1.07\pm0.02$& $0.82\pm0.02$& $0.63\pm 0.06$\\ 
  \hline
\end{tabular}\end{center}
\caption{\small{Efficiencies for the pair produced $B$ and $X_{5/3}$ going to trileptons containing two opposite-sign leptons. Efficiencies contain the cuts losses, b-tagging performance and BR's of W boson.}}
\label{tab:efficienciessimple3l}
\end{table}

\subsubsection * {Plots and results} 

Now using the event analysis algorithm shortly described above we can compute the signal efficiencies for same-sign dileptons and trileptons in the framework of the model with a totally composite top right and top partners in a four-plet. For this we again employ simplified models with only two top partners but this time we use exact couplings corresponding to typical points in the parameter space. Therefore apart from the right-handed coupling which is still dominant there is a small admixture of the left-handed one.  We present the results for $\Xft$ and $B$ masses in a range $400 - 1000\, GeV$ in tables \ref{tab:efficienciesPPtR} and \ref{tab:efficienciesSPtR} for pair and single production respectively.

\begin{table}
\begin{center}
\begin{tabular}{ | c | c | c |c | c | }
\cline{2-5}
 \multicolumn{1}{c}{} &  \multicolumn{2} {|c} { dilept eff.}   & \multicolumn{2}{|c|}  {trilept eff.}\\
 \hline
 $$M$ [\textrm{GeV}]$ &  for $B$  $[\%]$   & for $\Xft$ $[\%]$  &  for $B$ $[\%]$   & for $\Xft$  $[\%]$  \\
  \hline
 400   & $1.67\pm0.03$ & $1.61\pm0.04$ & $0.66\pm0.01$ & $0.67\pm 0.02$\\ 
 600   & $1.96\pm0.03$ & $2.02\pm0.04$ & $0.93\pm0.01$ & $0.93\pm 0.01$\\
 800   & $1.81\pm0.03$ & $1.86\pm0.04$ & $0.98\pm0.02$ & $0.97\pm 0.02$\\ 
 1000 & $1.63\pm0.03$ & $1.63\pm0.04$ & $0.99\pm0.02$ & $0.96\pm 0.02$\\ 
  \hline
\end{tabular}\end{center}
\caption{\small{Efficiencies for the pair produced $B$ and $X_{5/3}$ going to dileptons and trileptons containing two opposite-sign leptons. Efficiencies contain the cuts losses, b-tagging performance and BR's of W boson.}}
\label{tab:efficienciesPPtR}
\end{table}

\begin{table}
\begin{center}
\begin{tabular}{ | c | c | c |c | c | }
\cline{2-5}
 \multicolumn{1}{c}{} &  \multicolumn{2} {|c} { dilept eff.}   & \multicolumn{2}{|c|}  {trilept eff.}\\
 \hline
 $$M$ [\textrm{GeV}]$ &  for $B$  $[\%]$   & for $\Xft$ $[\%]$  &  for $B$ $[\%]$   & for $\Xft$  $[\%]$  \\
  \hline
 400   & $0.50\pm0.01$ & $0.49\pm0.01$ & $0.12\pm0.01$ & $0.12\pm 0.01$\\ 
 600   & $0.68\pm0.01$ & $0.69\pm0.01$ & $0.22\pm0.01$ & $0.22\pm 0.01$\\
 800   & $0.65\pm0.01$ & $0.74\pm0.01$ & $0.26\pm0.01$ & $0.25\pm 0.01$\\ 
 1000 & $0.63\pm0.01$ & $0.70\pm0.01$ & $0.28\pm0.01$ & $0.27\pm 0.01$\\ 
  \hline
\end{tabular}\end{center}
\caption{\small{Efficiencies for the single produced $B$ and $X_{5/3}$ going to dileptons and trileptons containing two opposite-sign leptons. Efficiencies contain the cuts losses, b-tagging performance and BR's of W boson.}}
\label{tab:efficienciesSPtR}
\end{table}

Now, by using the obtained efficiencies  together with the method elaborated above for computing the cross sections, one can compute the number of signal events in dileptons and trileptons  and check if it falls into the  region allowed by Fig.~\ref{fig:exclusionN2N3}.

\begin{figure}[t]
\centering
  \includegraphics[width=0.49\textwidth]{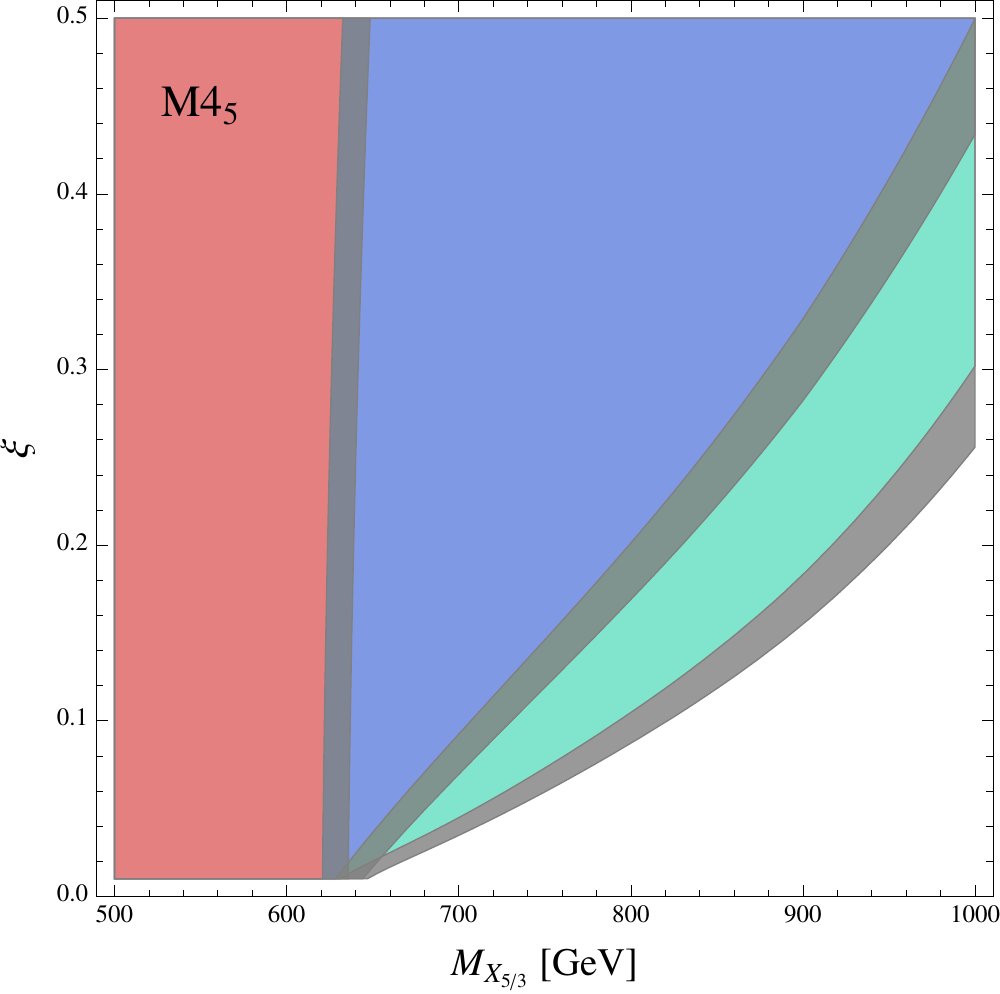}   \hfill
  \includegraphics[width=0.49\textwidth]{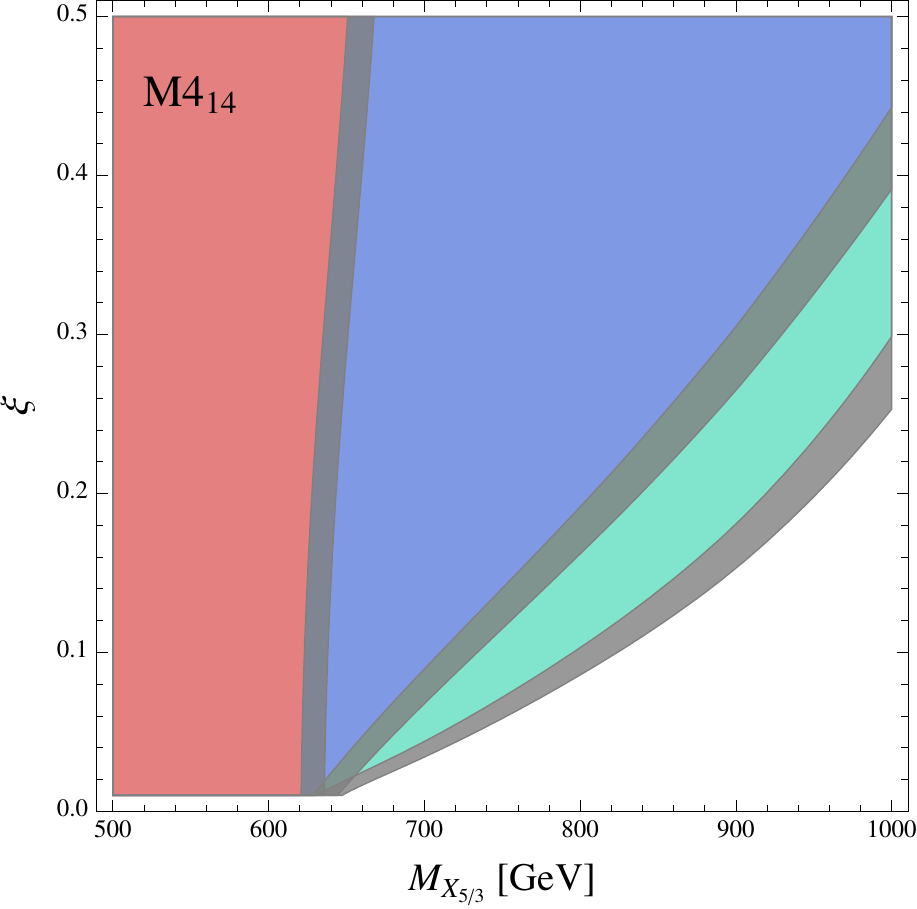}    \\
  \caption{\small{Excluded (95\%CL) regions in the $(M_{X_{5/3}},\xi)$ plane for the models $\fA\ $and $\fB$, 
  using the search for $b^{\prime}\to W\,t$.
  In red: $c_1=0.3$ and $y=3$ ($M_B \gg M_{X_{5/3}}$), in blue: $c_1=3$ and $y=3$ ($M_B \gg M_{X_{5/3}}$), in green: $c_1=3$ and $y=0.3$ ($M_B \gtrsim M_{X_{5/3}}$ for $\xi \gtrsim 0.1$, $M_B \gg M_{X_{5/3}}$ for $\xi \ll 0.1$). Gray regions correspond to a variation of the dileptons and trileptons signal of approximately  $10\%$ and $30\%$ respectively (see text for details).}}
  \label{fig:exclusionxiM}
  \end{figure}

\begin{figure}[t]
\centering
  \includegraphics[width=0.49\textwidth]{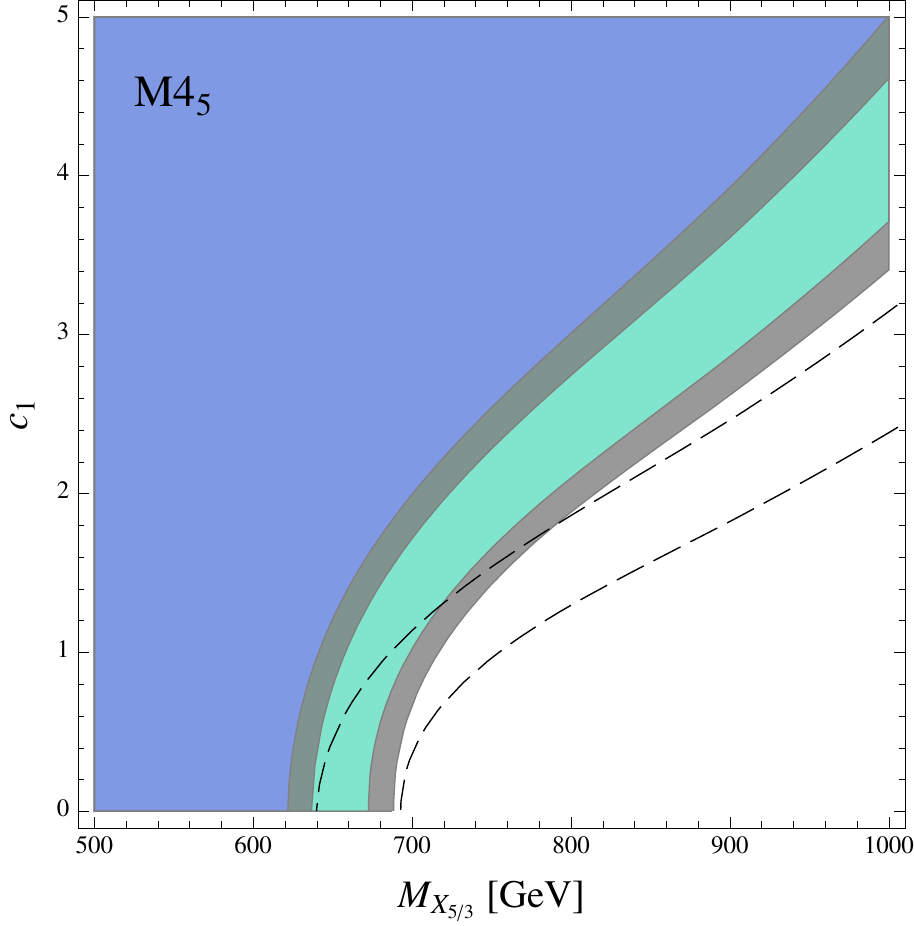}   \hfill
  \includegraphics[width=0.49\textwidth]{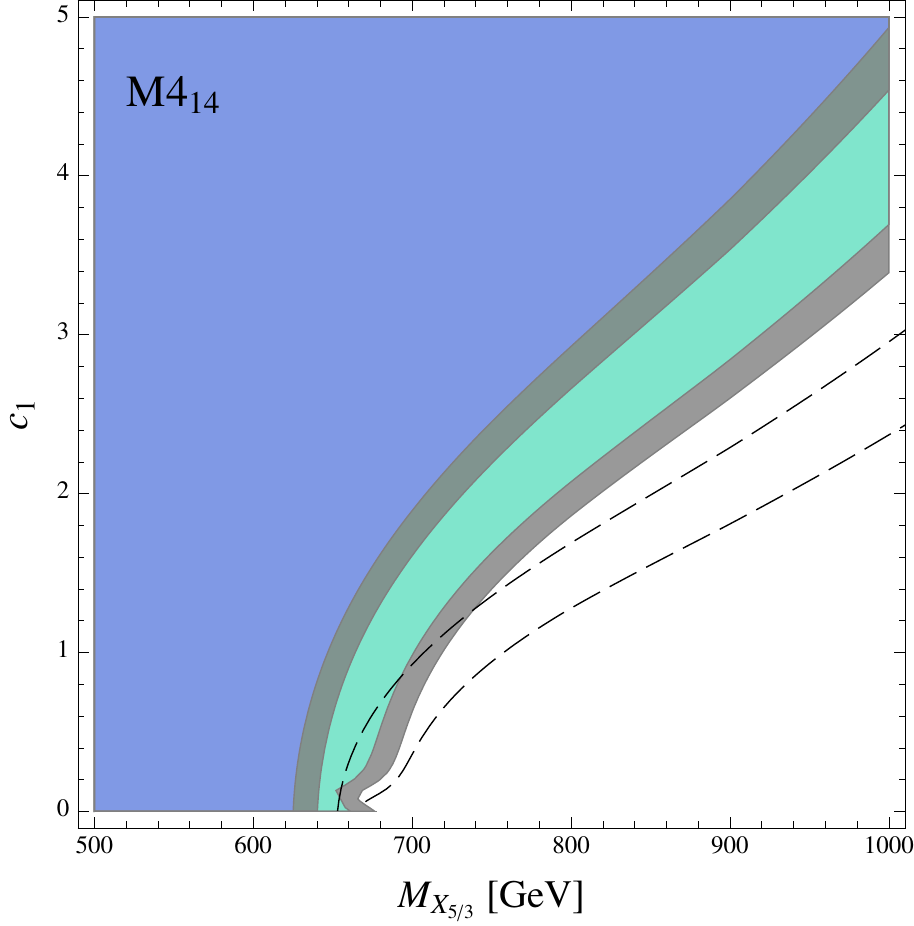}    \\
\caption{\small{Excluded (95\%CL) regions in  the $(M_{X_{5/3}},c_1)$ plane for $\xi=0.2$ for the models $\fA\ $and $\fB$,  using the search for $b^{\prime}\to W\,t$. In blue: $y=3$ ($M_B \gg M_{X_{5/3}}$), in green:  $y=0.3$ ($M_B \gtrsim M_{X_{5/3}}$). Black dashed lines correspond to the exclusions with $\xi=0.4$. Gray regions correspond to a variation of the dileptons and trileptons signal of approximately  $10\%$ and $30\%$ respectively (see text for details).} }
\label{fig:exclusioncM} 
\end{figure}

In Fig.~\ref{fig:exclusionxiM} we show the excluded region in the ($\xi$,$M_{X^{5/3}}$) plane, where $\xi = ( {v \over f})^2$, depending on whether the single production is suppressed ($c_1=0.3$) or enhanced ($c_1=3$) and whether also $B$  contributes to the signal ($M_B \gtrsim M_{\Xft}$, $y=0.3$) or not ($M_B \gg M_{\Xft}$, $y=3$).  Fig.~\ref{fig:exclusioncM} shows the exclusion in terms of $M_{\Xft}$ and $c_1$. Since, as was discussed in  sect.~\ref{gc}, the  leading contribution to single production couplings is the same for models \fA\ and \fB, the excluded regions are also similar for both models. A difference shows up when $c_1\ll 1$ and the $h \bar B b$ vertex of model $\fB\ $ becomes important thus decreasing BR($B \to W t$) and also when ${y \over g_{\psi}}\epsilon = {\cal O} (1)$ and higher order effects modify  the single production couplings. The excluded regions are almost symmetric with respect to $c_1 \to -c_1$, which can be understood as follows.  When only $\Xft$ production matters, the single production rate  is proportional to $|c_1|^2$ at  lowest order in $\epsilon$. Higher order terms only matter in the region of small $|c_1|$ where the single production 
rate is anyway negligible and the bound is driven  by pair production which is insensitive to  $c_1$. When $B$ production matters, that is because $m_B-m_\Xft \ll m_\Xft $, corresponding to $y\ll g_\psi$. From  eq.~(\ref {coup4}) it is then evident that in this regime  the couplings of both particles are approximately $\propto c_1$, so that  the signal yield is again symmetric under  $c_1 \to -c_1$.  The gray regions on the plots correspond to an estimate of the mentioned above error in the determination of the efficiencies. 

\subsection{Search for $t^{\prime}\to Z\, t$ } 

The search in Ref.~\cite{cmsTZt} is designed to detect an up-type $4^{th}$ generation quark $t^{\prime}$ pair-produced by QCD and decaying to $Zt$. 
The search is performed in the trilepton channel, with two same-flavor and opposite-charge leptons with an invariant mass around the $Z$ pole. Moreover, 
at least two jets are required. Apart from the usual hardness and isolation cuts for jets and leptons an important event selection is performed with the variable 
$R_T$, defined as the scalar sum of the reconstructed momenta \emph{without} including the two hardest leptons and the two hardest jets. $R_T$ is required 
to be above $80$~GeV. With $1.14\, fb^{-1}$ of $7$~TeV data the bound on the $t'$ is of  $475$~GeV.

All the top partners of charge $2/3$ can contribute to this final state \footnote {Actually because of the quite loose cuts on the invariant mass of leptonic $Z$ used in this search also the $X^{5/3}$ and the $B$ could contribute, however this effect is subdominant.}, these are  $\Xtt$ and  $T$ in models \fA\ and \fB\ and  $\widetilde{T}$ in models \oA\ and \oB. Remember however that  the masses and the couplings of  $\Xtt$ and of  $T$ are closely tight to those of, respectively,  $\Xft$ and  $B$. Namely, the masses are similar (or equal) and the couplings at the leading order (see eq.~(\ref{coup4})) differ by a factor of $\sqrt 2$.  Therefore the search of charge $2/3$ states will constrain the same combinations of the fundamental parameters of the model.
But the bound on the charge $2/3$ partners which can be obtained using Ref.~\cite{cmsTZt} is by far less stringent than the one from the $b'\to Wt$ search \cite{cmsBWt} described above. 
Given approximately 5 times less of analyzed data the limit on the production cross section of Ref.~\cite{cmsTZt} 
is significantly looser than the one of the Ref.~\cite{cmsBWt}.
Moreover in our model the production yield of the charge-$2/3$ states is typically lower than the one of the $\Xft$ and of the $B$. This is because of the branching ratio suppression to reach the $tZ$ final state and because the single production rate is smaller (see section~\ref{mrc}). 
Therefore we can safely ignore Ref.~\cite{cmsTZt} when constraining  models \fA\ and \fB. And moreover, for the reasons described above, we  expect that, even by updating the search of Ref.~\cite{cmsTZt} to the same integrated  luminosity of Ref.~\cite{cmsBWt},  it would not become more important.  

Hence in the following we will only use the search for $t^{\prime}\to Z\, t$ to constrain the parameters of  models \oA\ and \oB\ .
As in the previous section, we will obtain semi-analytical formulae for the signal yield by computing the efficiencies at each mass point and multiplying with the 
production cross section computed in Section~3.1. Differently from the previous case, the search is performed in a single channel. Therefore we will not 
need any statistical analysis, we will just compare the computed signal yield with the $95\%$ CL limit obtained in Ref.~\cite{cmsTZt} which corresponds to $9.6$ signal events.

\subsubsection*{Efficiencies}

The efficiencies are computed with a {\sc{MadGraph}} model which incorporates ${\widetilde{T}}$ and its couplings to $Zt$, $ht$ and $Wb$. 
These are responsible for the two single production modes and for the decay. The ${\widetilde{T}}$ couples only to left-handed quarks, therefore in this case 
we will employ left-handed couplings to compute the efficiencies. Our results can thus be directly compared with the efficiencies reported in Ref.~\cite{cmsTZt} 
because the coupling is left-handed also in the case of a $4^{th}$ family quark.

The ${\widetilde{T}}$ can contribute to the signal both in the pair and in the single production mode, provided that at least one ${\widetilde{T}}$ decays to $Zt$ paying a branching fraction of around $1/4$ (see Section~\ref{mrc}). In the case of pair production all the decay modes of the second produced ${\widetilde{T}}$ ($Zt$, $Wb$ or $ht$) are potentially relevant. We have computed separately the efficiencies in all three cases. The methodology of the analysis, and in particular the treatment of the experimental efficiencies, closely follows the one of the previous section. The results are shown in Table~\ref{tab:efficienciesPPSPTZt}, our efficiencies contain the cut losses and the $W$, $Z$, $h$ branching fractions to the required final state. The efficiencies listed in the first column of the table can be directly compared with the ones of Ref.~\cite{cmsTZt}, we have checked that the discrepancy is around $25\%$ which corresponds to approximately $1.5\sigma$ of the signal uncertainty obtained in the Ref.~\cite{cmsTZt}. 

We see in Table~\ref{tab:efficienciesPPSPTZt} that the efficiency for the single production with the $b$ is extremely low, below $1$~\permil. This is because 
the single production signal (see Figure~\ref{spd}) is characterized by three leptons plus one hard ($b$) jet from the top decay, plus one forward jet from the 
virtual $W$ emission and a ${\overline{b}}$ from the gluon splitting. But the gluon splitting is enhanced in the collinear region, therefore the $b$-jet emitted 
from the gluon is also preferentially forward and with low $p_T$. In order for the event to pass the selection cut, that requires at least two jets with 
$p_T>25$~GeV and $|\eta|<2.4$, at least one of the two preferentially forward jets must be central and hard enough, implying a significant reduction of the cross-section. However this 
is not yet the dominant effect, the main reduction of the signal is due to the cut $R_T>80$~GeV discussed before. Indeed $R_T$ is computed without including the two hardest leptons and  the two hardest jets, which in our case means, since we have only $3$ leptons and typically only $2$ jets, that the momentum of the softest lepton must be above $80$~GeV. Therefore in the end the signal is completely killed. The situation is better for the single production with the $t$ since one typically has more particles produced in this case and therefore the efficiencies are comparable with the ones of pair production. 
\begin{table}
\begin{center}
\begin{tabular}{ | c | c | c |c | c | c |  }
\cline{2-6}
 \multicolumn{1}{c}{} &  \multicolumn{3} {|c} {pair prod. eff.  $[\%]$}   & \multicolumn{2}{|c|}  {single prod. eff.  $[\%]$}\\
 \hline
 $M$ $[\textrm{GeV}]$ &  $T \bar T \to Zt\, Z\bar t$ &  $T \bar T \to Zt\, W\bar b$&  $T \bar T \to Zt\, h\bar t$    & $T\, \bar t\, j$  & $T\, \bar b\, j$   \\
  \hline
 300   & $1.78$& $1.22$ & $1.51$ & $\,\,\,\,\,\,1.13\,\,\,\,\,\,$ & $0.03$ \\ 
 350   & $1.93$& $1.47$ & $1.64$ & $1.17$ & $0.03$ \\
 450   & $2.21$& $1.81$ & $1.81$ & $1.25$ & $0.05$ \\ 
 550   & $2.34$& $1.93$ & $1.95$ & $1.30$ & $0.06$ \\ 
 650   & $2.40$& $2.12$ & $1.96$ & $1.35$ & $0.08$ \\ 
  \hline
\end{tabular}\end{center}
\caption{\small{Cuts efficiencies for  the charge 2/3 top partners going to trileptons for the case of pair production and different decay channels, and a single production for the cases of Z-t fusion($4^{th}$ column) and W-b fusion($5^{th}$ column). Efficiencies contain the cuts losses and BR's of W, Z and the SM Higgs boson.}}
\label{tab:efficienciesPPSPTZt}
\end{table}

The situation is better for the single production with the $t$, the efficiencies are comparable with the ones of pair production (see Table \ref{tab:efficienciesPPSPTZt}). However, we have seen in 
section~\ref{mrc} (see fig.~\ref{fig:ttildesp2}) that the rate of pair production is typically larger than the one of single production with the top, in the relevant mass range. Since the efficiencies are 
comparable  we do not expect a sizable contribution from this process.
The signal is totally dominated by the pair production
and  the $\BR(\widetilde T \to Z\,t)$ is fixed to be about 1/4, as discussed in section 
\ref{mrc}.
Therefore the bounds one can infer are mainly on $m_{\Tt}$, but
a mild dependence on the other parameters ($\xi$ and $y$) is still residual in the BR.
The resulting bound is about $m_{\Tt}\gtrsim 320$ GeV in both models $\oA, \oB$, 
and it is maximized at large $\xi$ and small $y$ $m_{\Tt}\gtrsim 350$ GeV in model $\oB$.
These bounds are not competitive with those coming from the $t^{\prime}\to W\, b$ search,
as we are going to discuss next.

\subsection{Search for $t^{\prime}\to W\, b$ } 

\begin{figure}[t]
\centering
  \includegraphics[width=0.48\textwidth]{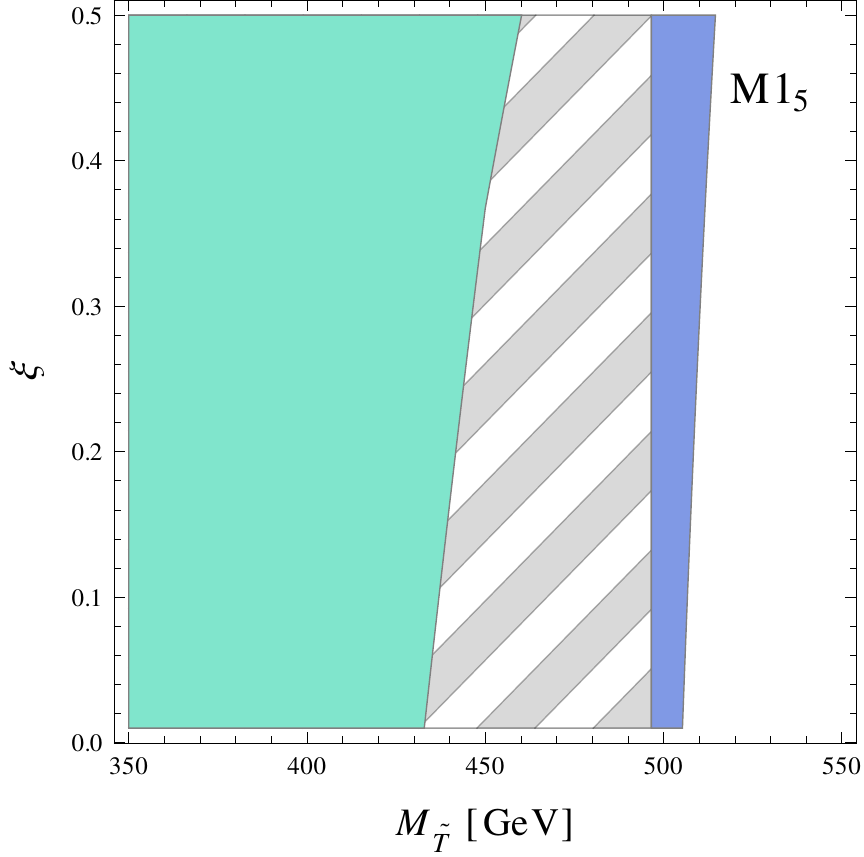}   \hfill
  \includegraphics[width=0.48\textwidth]{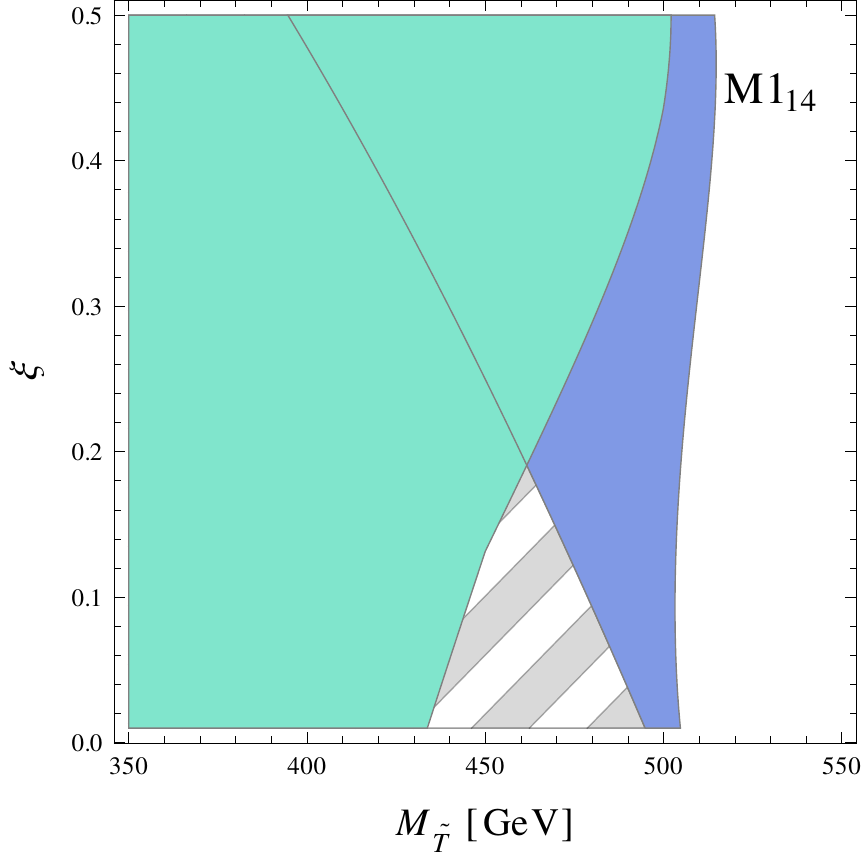}    \\
 \caption{\small{Excluded (95\%CL) regions  in the $(M_{\widetilde T},\xi)$ plane, using the search $t^{\prime} \to W b$,  for the models $\oA$ and $\oB$ for $y=0.5$ (green), $y=2$ (blue) (corresponding approximately to  $c_2\simeq 2$ (green), $c_2\simeq 0.5$ (blue)). In the gray dashed region there are no solutions for $M_{\widetilde T}(y,\xi)$ when $y=2$. }}
 \label{fig:exclusionxiMsing}
  \end{figure}
  
The last experimental study that we are going to consider is the search for a $4^{th}$ generation $t^{\prime}$ quark decaying to $W b$ \cite{cmsTWb}. 
The search is performed in the channel of two opposite sign leptons (away from the $Z$ pole) with two tagged bottom quarks. A very important selection 
cut, which is needed to suppress the background from the top quark production, is that the invariant mass of all the lepton and $b$-jet pairs, 
$M_{l b}$, is above $170$~GeV. This forbids that the lepton and the $b$ originate from the decay of a top quark. Using   data from $5 fb^{-1}$ of integrated luminosity, a lower bound of $557\GeV$ was set on the $t^{\prime}$ mass \cite{cmsTWb}.

In our models, only $\widetilde{T}$ can decay to $Wb$ with a sizable branching fraction. We will therefore use Ref.~\cite{cmsTWb} to put constraints 
on models \oA\ and \oB \footnote{Also pair produced $B$ and $\Xft$ decaying to $Wt$ contributes to the  final states considered in Ref.~\cite{cmsTWb}. However the resulting bound on these states is lower than the one obtained using   Ref.~\cite{cmsBWt}. In addition the signature used in Ref.~\cite{cmsTWb}  is insensitive to  single production. Thus we do not expect any improvement of the bounds on the models \fA\ and \fB\ from this search.}. The single production mode with the $b$ is definitely not relevant in this case because it only leads to one lepton. The one with the $t$ is also irrelevant because the second lepton would come from the 
decay of the top quark and it would not satisfy the cut on $M_{lb}$. We are therefore left with pair production. Moreover, because of the $M_{bj}>170$~GeV 
cut, and  as was explicitly checked in  Ref.~\cite{Berger:2012ec}, pair production  contributes to the signal only if both $\widetilde{T}$'s decay to $Wb$. We are then left with the same channel, 
$\Tt\overline{\Tt}\to WbWb$, considered in Ref.~\cite{cmsTWb}. The chirality of the coupling responsible for the decay is also the same as in the 
$4^{th}$ family case. Therefore the efficiencies can be extracted directly from Ref.~\cite{cmsTWb} without any need for additional simulations. Given the efficiency and 
taking into account that the branching fraction of the $\widetilde{T}$ to $Wb$, we can easily compute the signal yield and compare it with the bound
obtained in \cite{cmsTWb}.

\subsubsection*{Plots and results}

We show the excluded regions of the parameter space in terms of $\xi$ and $M_{\widetilde T}$ on the Fig.~\ref{fig:exclusionxiMsing}. The exclusion is stronger for larger $y$ (and smaller $c_2$) due to a larger $BR(\widetilde T \to W b)$ in this case.
As was already discussed in the Section~\ref{gc} the gauge interactions of the model $\oB$ are similar to the ones of the model $\oA$ and therefore the excluded regions are also similar. The difference is sizable in the region close to $\xi=0.5$ where in the model $\oB$ interactions with a Higgs boson vanish according to Eq.~(\ref{scalarlagr1B}) and therefore the $BR$ of the competitive decay to $W b$ increases. The regions without solutions for $\widetilde T(y,\xi)$ when $y$ is large correspond to those defined by the Eq.~(\ref{minmassttilde}).

Due to a larger amount of data analyzed and a higher $BR$ of the $\widetilde T\to W\,b$ decay mode the search of Ref.~\cite{cmsTWb} gives a better constraint on the parameters of our models than the previously considered search $\widetilde T\to Z\,t$~\cite{cmsTZt}. However one may expect that with increased amount of analyzed data the search for $\widetilde T\to Z\,t$ can become competitive due to its sensitivity to single production. 

    \begin{figure}[t]
\centering
  \includegraphics[width=0.49\textwidth]{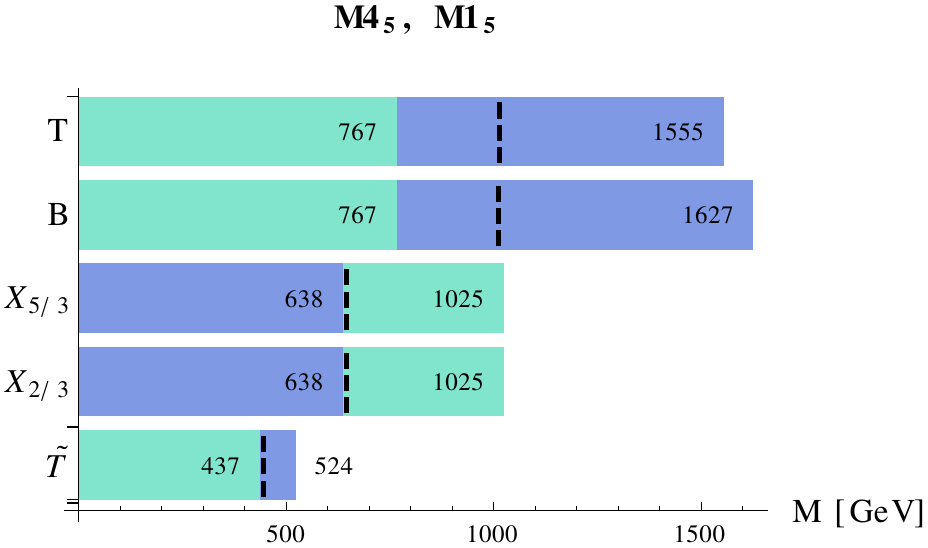}   \hfill
  \includegraphics[width=0.49\textwidth]{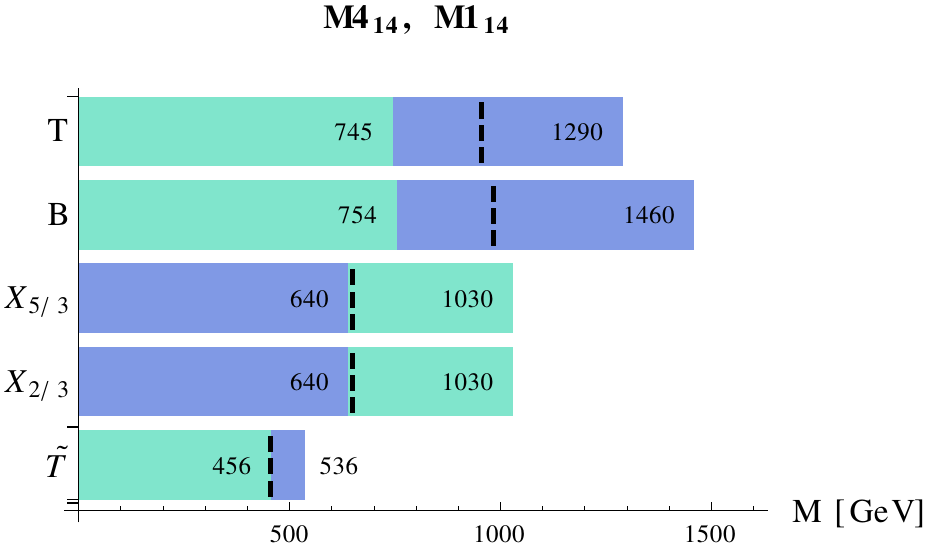}    \\
 \caption{\small{Maxmal and minimal bounds on the masses of top partners for $y\in[0.3,3]$, $c_1\in[0.3,3]$ and $\xi\in[0.1,0.3]$ for the models $\fA,\ $$\oA\ $(left pannel) and $\fB,\ $$\oB\ $(right pannel). Blue and green bars correspond respectively to high and low values of $y$. Black dashed lines correspond to the exclusions for the reference values $\xi=0.1$, $c_1=1$, $y=1$.}}
 \label{fig:exclusionMchart}
  \end{figure}

\subsection{Summary of exclusions}

The results of the searches described above can be conveniently summarized by scanning
over the values of the model parameters and selecting the most and the least stringent  bounds on the top-partners' masses.
The highest excluded masses of $X_{5/3}$ and $X_{2/3}$ correspond to the lowest value of $y$ and highest $c_1$ and $\xi$, and the opposite for the lowest exclusion. For $T$ and $B$ the highest exclusion corresponds to the highest $y$, $c_1$ and $\xi$ and the opposite for the lowest exclusion. Maximal $\widetilde T$ mass exclusion is reached when $y$ and $\xi$ are maximal and the minimal exclusion is obtained for minimal $y$ and $\xi$. 
In Fig.~\ref{fig:exclusionMchart}, we show our results for the maximal and minimal exclusions obtained by varying the parameters in the  ranges: $y\in[0.3,3]$, $c_1\in[0.3,3]$ and $\xi\in[0.1,0.3]$.

\section{Conclusions}
\label{conclusions}

In this paper we described an approach to  systematically construct  the low-energy effective lagrangian for the lighest colored fermion multiplet related to the UV completion of the top quark sector: the top partner. Our construction is based on 
robust assumptions, as concerns symmetries, and on plausible assumptions, as concerns the dynamics. Our basic dynamical assumption, following Ref.~\cite{silh}, is that the electroweak symmetry breaking sector, or at least the fermionic sector, is broadly decribed by a coupling $g_*$ and a mass scale $m_*$. This assumption implies a well definite power counting rule. In particular the derivative expansion is controlled by inverse powers of $m_*$. In the technical limit where the top partner multiplet $\Psi$, is parametrically much lighter than the rest of the spectrum ($M_\Psi\ll m_*$), our power counting provides a weakly coupled effective lagrangian description of the phenomenology of  $\Psi$. The basic idea is that, in this case, the effects of the bulk of the unknown spectrum at the scale $m_*$ can be systematically described by an expansion in powers of $M_\Psi/m_*$. The lagrangian obtained in this limit  defines our
simplified description of the top parters.
One should however keep in mind that  the  most likely physical situation is one where $m_*-M_\Psi \sim M_\Psi$, where an effective lagrangian is formally inappropriate. In practice, however, we expect it to be more than adequate for a first semi-quantitative description of the phenomenology and certainly to assess experimental constraints. The comparison with explicit constructions supports this expectation.

%

As concerns the symmetries of the strong sector,  we considered the minimal composite Higgs based 
on the  $SO(5)/SO(4)$ coset. Furthermore we focussed on the simplest possibility where the right-handed top quark $t_R$ is itself a composite fermion. The leading source of breaking of  $SO(5)$ is thus identified with top quark Yukawa coupling $y_t$. In our construction, we have fully exploited the selection rules obtained by treating $y_t$ as a small spurion with definite transformation properties. For instance
the structure of the mass spectrum and the couplings are greatly constrained by symmetry and  selection rules. In particular the pNGB nature of the Higgs doublet implies the couplings originating from the strong sector are purely derivative: at high energy, or for heavy on-shell fermions, these couplings are effectively quite sizeable and yet they do not affect the spectrum even accounting for  $\langle H\rangle \not =0$. If the Higgs were not treated as a pNGB 
 a large trilinear would be associated with a large Yukawa coupling and the spectrum would  necessarily be affected when $\langle H\rangle \not = 0$. 
 
Depending on the quantum numbers of the top partner multiplet $\Psi$ and of the composite operator ${\cal O}$ that seeds the top Yukawa in the microscopic theory, one can then consider a variety of models. We focussed on the four possibilities shown in Table \ref{models}, which could be considered the simplest ones. Our method can 
however be directly applied to perhaps more exotic possibilities. For instance one exotic, but not implausible, case
would be ${\cal O}={\mathbf{14}}_{\mathbf{2/3}}$ 
 with $\Psi$ in the symmetric traceless tensor of $SO(4)$, that is  $\Psi={\mathbf{9}}_{\mathbf{2/3}}$. This case involves a top partner with electric charge $8/3$, performing a spectacular chain decay to $3W^+ +b$.
 
Our effective lagrangian depends on a manageable number  of parameters. Once the top mass is fixed, beside the Goldstone decay constant $f$ and the partner mass $M_\Psi$, there remain, depending on the model, only one or two additional parameters. 
These parameters, $c_{1,2}$, control the size of the trilinear couplings between $\Psi$, third family fermions, and vector bosons or Higgs.  They thus control the decay and the single production  of top partners.
Moreover, naive power counting suggests a preferred $O(1)$ range for these parameters. This fact, coupled with the constraints due to symmetry, robustly implies a definite structure for the interactions vertices in each model. For instance, for the case where $\Psi$ spans an $SO(4)$  quadruplet, the trilinear coupling to the Higgs doublet involves mostly a $t_R$ and is expected to be of the order of a strong sector coupling $g_\Psi=M_\Psi/f$. Moreover it grows with $M_\Psi$ making simple production even more important in the range of heavy $\Psi$. In the case of a singlet $\Psi$, the trilinear is of order $y_t$ and involves the left handed doublet $(t,b)_L$. These details, including the chirality of the top, affect the collider phenomenology of the models and, consequently,  the constraints from searches.

Not  only have we a few lagrangian parameters, but also they  mainly affect  phenomenology 
via their contribution to  the trilinear couplings. Using this property we devised a semi-analytical way to efficiently
simulate the contribution of single production to the signal.  For any given mass $M_\Psi$, we numerically simulated single production once for all, assigning trilinear coupling equal to unity. The physical cross section  was then obtained by folding this numerical result with the analytical dependence of the physical trilinear coupling on the model parameters. 
Thus, once the efficiencies associated with a given experimental search are known, the constraint 
in parameter space can be obtained analytically. We implemented the calculation of the cross-sections in a  
{\sl{Mathematica}} notebook which is available on request.

We applied our results to the presently available LHC searches. We focussed on the search for $4^{th}$ family fermions, that have signatures similar to those of top partners, and recast them to constrain our models. The main results can be read from Fig.~\ref{fig:exclusioncM} and Fig.~\ref{fig:exclusionxiMsing}. The former figure shows that, in the relevant region $c_1=O(1)$, single production  has a mild but non-negligible impact on the bounds. 

In the course of our analysis, it became evident that there exists significant space for improvement in the search strategy if one wants to best constrain this class of models. The searches we used were tailored to pair production of heavy quarks, while single production of top partners has different features. First of all, in single production there is only one hard decaying object, with the $t$ or $b$ produced in association beeing often less hard (much more so in  the case of a $b$): the cuts performed in present searches tends to penalize single production. Secondly, single production is always associated with a very forward jet, originating from the collinear splitting  of a typically valence quark into a longitudinally polarized vector boson. The resulting forward jet has exactly the same features of the tag jets
of $WW$ scattering. There is a good chance the use of the same tag in the searches for singly produced top partners
would significantly extend the sensitivity.  We also realized that the single production with the $b$ is typically very large 
(see fig.~\ref{fig:ttildesp2}) in the case of a singlet top partner $\widetilde{T}$,
tagging this production mechanism would increase 
significantly the LHC sensitivity to this kind of particles. Ideally the best channel of detection would be the 
resonant $Wb$ production from the $\widetilde{T}$ decay, accompanied by one forward jet from the longitudinal vector 
boson emission. The second $b$ quark which is present in the reaction, which comes from the gluon splitting as 
in fig.~\ref{spd}, is typically quite soft both in $p_\bot$ and in energy. Thus it is probably strongly 
affected by QCD initial state radiation and  difficult to detect.
Building upon these considerations, it would be worth to undertake a thorough experimental analysis,
including the effect of radiation and detector simulation, suitably designed for the search of singly produced
top-partners.

In the test of weak scale naturalness, the search for all possible fermionic top partners represents the other half of the sky. In this paper we have introduced a first systematic description of top partner phenomenology.  The simplicity of the result  should hopefully serve as a basis for future systematic experimental studies. As seen from our theorist's analysis, the present searches have already advanced well into the region suggested by naturalness. But there is no doubt somebody out there can do  better.

\section*{Acknowledgments}
We thank Sascha Joerg for collaboration on part of this work
and Francesco Tramontano for help with MCFM.
The work of O.M. is supported by the European Programme Unication in the LHC Era, 
contract PITN-GA-2009-237920 
(UNILHC). 
The work of R.R.  is partially sponsored by the Swiss National Science Foundation under grant 200020-138131.
The work of A.W. is supported in part by the ERC Advanced Grant no.267985 Electroweak Symmetry Breaking, Flavour and Dark Matter: One Solution for Three Mysteries (DaMeSyFla).

\appendix

\section{Explicit CCWZ construction for $\mathbf{{\textrm{SO}}(5)/{\textrm{SO}}(4)}$}
\label{append}

\subsubsection*{Generators and Goldstone Matrix}

The generators of $\textrm{SO}(5)$ in the fundamental representation are conveniently chosen to be
\begin{equation}
(T^\alpha_{L,R})_{IJ} = -\frac{i}{2}\left[\frac{1}{2}\varepsilon^{\alpha\beta\gamma}
\left(\delta_I^\beta \delta_J^\gamma - \delta_J^\beta \delta_I^\gamma\right) \pm
\left(\delta_I^\alpha \delta_J^4 - \delta_J^\alpha \delta_I^4\right)\right]\,,
\label{eq:SO4_gen}
\end{equation}
\begin{equation}
T^{i}_{IJ} = -\frac{i}{\sqrt{2}}\left(\delta_I^{i} \delta_J^5 - \delta_J^{i} \delta_I^5\right)\,,
\label{eq:SO5/SO4_gen}
\end{equation}
where $T^{\alpha}_{L,R}$ ($\alpha = 1,2,3$) are the $\textrm{SO}(4) \simeq \textrm{SU}(2)_L \times \textrm{SU}(2)_R$ unbroken generators, while $T^{i}$ ($i = 1, \ldots, 4$) are the broken ones and parametrize the coset $\textrm{SO}(5)/\textrm{SO}(4)$. An equivalent notation for unbroken generators which we will use is $T^{a}$ with $a = 1,\ldots,6$. The indices $IJ$ take the values $1, \ldots, 5$. The normalization of the $T^{A}$'s is chosen as ${\rm Tr}[T^A, T^B] = \delta^{AB}$. 

The $T^{\alpha}_{L}$ and $T^{\alpha}_{R}$ generators span respectively the $\textrm{SU}(2)_L$ and $\textrm{SU}(2)_R$ subgroups, and obey the standard commutation relations
\begin{equation}
\left[T^{\alpha}_{L,R}, T^{\beta}_{L,R}\right] = i \varepsilon^{\alpha\beta\gamma}\, T^{\gamma}_{L,R}\,.
\end{equation}
The $T_L$'s are therefore identified as the generators of the SM $\textrm{SU}(2)_L$. Notice that in our parametrization the unbroken $T^a$'s are block-diagonal
\begin{equation}
T^a=\left(\begin{array}{cc}t^a &0 \\ 0 &0 \end{array}\right)\,,
\end{equation}
and the generators obey the following commutation relation
\begin{equation}
\left[T^{a},T^{i}\right]=\left(t^{a}\right)_{{j}{i}}T^{j}\,.
\label{crb}
\end{equation}

With these generators, the parametrization of the Goldstone boson matrix is explicitly given by
\be
U= U(\Pi) = \exp\left[i \frac{\sqrt{2}}{f} \Pi_{i} T^{i}\right]=\left(\begin{matrix}
\textbf{1}_{4\times 4}-{\vec\Pi\vec\Pi^T\over \Pi^2} \left(1-\cos{\Pi\over f}\right)&
{\vec\Pi\over \Pi}\sin{\Pi\over f}\\
-{\vec\Pi^T\over \Pi}\sin{\Pi\over f}&\cos{\Pi\over f}\\
\end{matrix}\right) 
\label{gmatr}
\ee
where $\Pi^2 \equiv \vec{\Pi}^t \vec{\Pi}$. Under $g\in \textrm{SO}(5)$, the Goldstone matrix transforms as 
\begin{equation}
U(\Pi)\;\rightarrow\; U(\Pi^{(g)})=g\cdot  U(\Pi)\cdot h^t(\Pi ; g)\,,
\label{gtrans}
\end{equation}
where $h(\Pi ; g)$ is block-diagonal in our basis
\begin{equation}
h=\left(\begin{matrix}h_4 &0 \\ 0 &1 \end{matrix}\right)\,,
\label{hd}
\end{equation}
with $h_4\in \textrm{SO}(4)$. Under the unbroken $\textrm{SO}(4)$ the $\Pi$'s transform \emph{linearly}, using eq.~(\ref{crb}) we get $\Pi^{i}\rightarrow {(h_4)^{i}}_{\ j}\Pi^{j}$. Given our embedding of the SM group, the $\Pi$ four-plet can be rewritten as
\begin{equation}
\vec{\Pi}=\left(\begin{matrix}\Pi_1\\ \Pi_2\\ \Pi_3\\ \Pi_4\end{matrix}\right)=
\frac1{\sqrt{2}}\left(\begin{matrix}-i\,(h_u-h_u^\dagger)\\ 
h_u+h_u^\dagger
 \\ i\,(h_d-h_d^\dagger)\\ 
h_d+h_d^\dagger
 \end{matrix}\right)\,,
\end{equation}
\label{Higgsfield}
where
\be
H=\left(\begin{array}{c} h_u\\ h_d\end{array}\right)\,,
\ee
is the standard Higgs doublet of $+1/2$ Hypercharge. 

In the unitary gauge, in which 
\be
h_u=0,\,\qquad h_d\equiv \frac{h}{\sqrt{2}}=\frac{\langle h\rangle+\rho}{\sqrt{2}}\,,
\label{ugauge}
\ee
where $\rho$ is the canonically normalized physical Higgs field, the Goldstone boson matrix of eq.~(\ref{hd}) simplifies and becomes 
\be
 U=\left(\begin{matrix}
1&0&0&0&0\\
0&1&0&0&0\\
0&0&1&0&0\\
0&0&0&\cos\frac{h}{f}&\sin\frac{h}{f}\\
0&0&0&-\sin\frac{h}{f}&\cos\frac{h}{f}\\
\end{matrix}\right) \,.
\label{uvev}
\ee

Given that we will have to gauge the SM subgroup of $\textrm{SO}(5)$, we must consider also \emph{local} transformations, $g=g(x)$, in the above equation. We also have to define \emph{gauge sources} $A_\mu^A$
\begin{equation}
A_\mu = A_\mu^A T^A\;\rightarrow\;A_\mu^{(g)}=g\left[A_\mu + i\partial_\mu \right]g^t\,,
\end{equation}
some of which we will eventually make dynamical while setting the others to zero. Explicitly, the dynamical part of $A_\mu$ will be
\begin{equation}
A_\mu = \frac{g}{\sqrt{2}}W^+_\mu\left(T_L^1+i T_L^2\right)+\frac{g}{\sqrt{2}}W^-_\mu\left(T_L^1-i T_L^2\right)+g \left(\cw Z_\mu+\sw A_\mu \right)T_L^3+g' \left(\cw A_\mu-\sw Z_\mu \right)T_R^3\,,
\label{gfd}
\end{equation}
where $\cw$ and $\sw$ denote respectively the cosine and the sine of the weak mixing angle and $g$, $g'$ are the SM 
couplings of $\textrm{SU}(2)_L$ and $\textrm{U}(1)_Y$. 
Notice that $A_\mu$ belongs to the unbroken $\textrm{SO}(4)$ subalgebra, this will simplify the expression for the $d$ and $e$ symbols that we will give below.

\subsubsection*{The ${\mathbf{d}}$ and ${\mathbf{e}}$ symbols}

Still treating $A_\mu$ as a general element of the $\textrm{SO}(5)$ algebra, we can define the $d$ and $e$ symbols as follows. Start from defining 
\begin{equation}
\bar{A}_\mu\equiv A_\mu^{(U^t)}=U^t\left[A_\mu+i\partial_\mu \right] U\,,
\end{equation}
this transforms under $\textrm{SO}(5)$ in a peculiar way
\begin{equation}
\displaystyle
\bar{A}_\mu\;\rightarrow\;A_\mu^{(h\cdot U^t\cdot{g^t}\cdot{ g})}=\bar{A}_\mu^{(h)}=h\left[\bar{A}_\mu+i\partial_\mu\right] h^t
\end{equation}
Since $h=h(\Pi ; g)$ is an element of $\textrm{SO}(4)$ as in eq.~(\ref{hd}), the shift term in the above equation, $ih\partial_\mu h^t$, lives in the $\textrm{SO}(4)$ subalgebra. Therefore, if we decompose $\bar{A}_\mu$ in broken and unbroken generators
\begin{equation}
\bar{A}_\mu\equiv -\,d_\mu^{i} T^{i}-\,e_\mu^a T^a\,,
\label{defde}
\end{equation}
we have that $d_\mu^{i}$ transforms linearly (and in the fourplet of $\textrm{SO}(4)$) while the shift is entirely taken into account by $e_\mu^a$. We have
\begin{equation}
d_\mu^{i}\;\rightarrow \left(h_4\right)^{i}_{\ {j}}d_\mu^{j}\; \;\;\;\;\;\textrm{and}\;\;\;\;\;e_\mu\equiv e_\mu^a t^a\;\rightarrow\;h_4\left[e_\mu-i\partial_\mu\right]h_4^t\,.
\label{trrule}
\end{equation}

Let us now restrict, for simplicity, to the case in which $A_\mu$ belongs to the $\textrm{SO}(4)$ subalgebra, as for our dynamical fields in eq.~(\ref{gfd}). It is not difficult to write down an explicit formula for $d$ and $e$, these are given by
\begin{eqnarray}
d_\mu^{i}&&=\sqrt{2}\left(\frac1{f}-\frac{\sin{\Pi/f}}\Pi\right)\frac{\vec{\Pi}\cdot \nabla_\mu\vec{\Pi}}{\Pi^2}\Pi^{i}+\sqrt{2}\frac{\sin{\Pi/f}}\Pi\nabla_\mu\Pi^{i}\nonumber\\
e_{\mu}^a&&=-A_\mu^a+4\,i\,\frac{\sin^2{(\Pi/2f)}}{\Pi^2}\vec{\Pi}^t t^a\nabla_\mu\vec{\Pi}
\label{dande}
\end{eqnarray}
where $\nabla_\mu\Pi$ is the "covariant derivative" of the $\Pi$ field:
\begin{equation}
\nabla_\mu\Pi^{i}=\partial_\mu\Pi^{i}-i A_\mu^a\left(t^a\right)^{i}_{\  j}\Pi^{ j}\,.
\end{equation}

The first use we can make of the $d_\mu$ symbol is to define the $\textrm{SO}(5)$-invariant kinetic Lagrangian for the Goldstone bosons, this is given by
\be
{\mathcal{L}}_\pi=\frac{f^2}4 d_\mu^{i} d^\mu_{i}\,.
\label{hkt}
\ee
In the unitary gauge of eq.~(\ref{ugauge}) and using eq.~(\ref{gfd}) for $A_\mu$ the Goldstone Lagrangian becomes
\be
{\mathcal{L}}_\pi=\frac{1}2 (\partial h)^2 + \frac{g^2}4 f^2 \sin^2{\frac{h}{f}} \left(|W|^2+\frac1{2c_w^2}Z^2\right)\,,
\label{hktug}
\ee
from which we can check that the field $\rho$ is indeed canonically normalized and read the $W$ and $Z$ masses 
$m_W=g/2f\sin{\frac{\langle h\rangle}f}$, $m_Z=m_W/c_w$. This fixes relation among $\langle v\rangle$ 
and the EW scale $v=246$~GeV
\be
v\,=\,f\,\sin{\frac{\langle h\rangle}f}\,.
\label{VEV}
\ee

The $e_\mu$ symbol can instead be used to construct the CCWZ covariant derivatives, because the shift term in its transformation rule of eq.~(\ref{trrule}) compensates for the shift of the ordinary derivative. Consider for instance the field ${\Psi}$ defined in eq.~(\ref{4plet}) of the main text, which transforms in the ${\mathbf4}$ of $\textrm{SO}(4)$, {\it{i.e.}} like $\Psi\rightarrow h_4\cdot\Psi$. The covariant derivative is
\begin{equation}
\nabla_\mu\Psi \,=\,\partial_\mu\Psi+i\,e_{\mu}^at^a\Psi\,.
\label{covder}
\end{equation}

\subsubsection*{The CP symmetry}

By looking at eq.~(\ref{Higgsfield}) and remembering that CP acts as $H(x)\rightarrow H^*(x^{(P)})$ on the Higgs doublet we 
immediately obtain the action of the CP transformation on the Goldstone fields $\Pi$ and on the Goldstone matrix $U$. It is
\be
\vec{\Pi}(x)\;\rightarrow\;C_4\cdot\vec{\Pi}(x^{(P)})\,,\qquad U(x)\;\rightarrow\; C_5\cdot U(x^{(P)})\cdot C_5\,,
\ee
where $C_4$ and $C_5$ are respectively a $4\times4$ and a $5\times5$ diagonal matrices defined as
\be
C_4={\textrm{diag}}(-1,+1,-1,+1)\,,\qquad C_5={\textrm{diag}}(-1,+1,-1,+1,+1)\,.
\ee
In the above equations the superscript ``$(P)$'' denotes the action of ordinary spatial parity. 
Similarly, the ordinary action of CP on the SM gauge fields in eq.~(\ref{gfd}) is recovered if we take
\be
A_\mu\;\rightarrow\;C_5\cdot A_\mu^{(P)}\cdot C_5\,.
\ee
From the above equations it is straightforward to derive the CP transformations of the $d$ and $e$ symbols defined in eq.~(\ref{defde}),
\be
d_\mu^i\;\rightarrow\;{C_4}^{i}_{\;j}(d_\mu^{(P)})^j\,,\qquad e_\mu\;\rightarrow\;C_4\cdot (e_\mu^{(P)})\cdot C_4\,.
\ee

In the fermionic sector, adopting for definiteness the Weyl basis, the CP transformation of the $q_L$ and of the $t_R$ are the usual ones
\be
\chi(x)\;\rightarrow\;\chi^{(CP)}=i\gamma^0\gamma^2\psi^*(x^{(P)})\,,
\label{orcp}
\ee
for $\chi=\{t_L,b_L,t_R\}$. For the top partners, in the case in which they transform in the fourplet of ${\textrm{SO(4)}}$ as in eq.~(\ref{4plet}), it is natural to define CP as
\be
\Psi_i\;\rightarrow\;{C_4}_{i}^{\;j}\Psi_j^{(CP)}\,,
\ee
while for the case of the singlet we simply have $\Psi\;\rightarrow\;\Psi^{(CP)}$. Notice that with this definition the charge eigenstate fields $\{T,B,\Xtt,\Xft\}$ defined in eq.~(\ref{4plet}) have ``ordinary'' CP transformation as in eq.~(\ref{orcp});

\section{Fermion Couplings}
\label{ferc}

In this appendix we report the explicit form of the fermion couplings to gauge bosons and to the Higgs, in the unitary gauge defined 
by eq.~(\ref{ugauge}), that arise in our four models \fA, \fB, \oA \ and \oB, defined respectively in eq.s~(\ref{eq:lagrangian2}), 
(\ref{eq:lagrangian214}) and (\ref{eq:lagrangian211}). All the couplings are given \emph{before} the rotations that diagonalize the mass-matrices.

The first two terms, which are relevant for the models \fA \ and \fB, are
\bea
i\,{\overline{\Psi}}_R^i\slashed{d}_it_R=&&\frac{g}{\sqrt{2}}\sin{\frac{h}{f}}({\overline{X}_{5/3}})_R\slashed{W}^+t_R
-\frac{g}{\sqrt{2}}\sin{\frac{h}{f}}{\overline{B}}_R\slashed{W}^-t_R
-\frac{g}{2c_w}\sin{\frac{h}{f}}{\overline{T}}_R\slashed{Z}t_R\nonumber\\
&&
-\frac{g}{2c_w}\sin{\frac{h}{f}}({\overline{X}_{2/3}})_R\slashed{Z}t_R
+i\,\left[({\overline{X}_{2/3}})_R-{\overline{T}}_R\right]\frac{\slashed{\partial}\rho}{f}t_R\,,
\label{dcoup}
\eea
and the term with the $e_\mu$ symbol which we combine, for convenience, with the one from the covariant derivative in 
eq.~(\ref{cder})
\bea
{\overline{\Psi}}\left(\frac23 g'\slashed{B}-\slashed{e}\right)\Psi=&&
\frac{g}{c_w}(-\frac12+\frac13 s_w^2){\overline{B}}\slashed{Z}B+\frac{g}{c_w}(\frac12-\frac53 s_w^2)
{\overline{X}}_{5/3}\slashed{Z}\Xft\nonumber\\
&&+\frac{g}{c_w}(\frac12\cos{\frac{h}{f}}-\frac23 s_w^2){\overline{T}}\slashed{Z}T+\frac{g}{c_w}(-\frac12\cos{\frac{h}{f}}-\frac23 s_w^2){\overline{X}_{2/3}}\slashed{Z}\Xtt\nonumber\\
&&+\left\{\frac{g}{\sqrt{2}}{\overline{B}}\slashed{W}^-\left[\cos^2{\frac{h}{2f}}T+\sin^2{\frac{h}{2f}}\Xtt\right]\right.\nonumber\\
&&\left.
+\frac{g}{\sqrt{2}}{\overline{X}}_{5/3}\slashed{W}^+\left[\sin^2{\frac{h}{2f}}T+\cos^2{\frac{h}{2f}}\Xtt\right]\,+\,h.c.\right\}\nonumber\\
&&+\; {\textrm{``photon couplings''}}
\,.
\label{ecoup}
\eea
The couplings to the photon are not reported explicitly in the above equation because they are simply the standard ones, being 
completely 
fixed by the ${\textrm{U}}(1)_{\textrm{em}}$ residual gauge symmetry.

In addition to the ones in eq.~(\ref{ecoup}), non-derivative couplings with the Higgs field emerge in model \fA\ from the terms
\bea
y f\,(\overline Q^{\textbf{5}}_L)^{I} U_{I\, i}\Psi_{ R}^i&=&y f\,
\bar b_L B_R+y f\,{\overline{t}}_L\left[\cos^2{\frac{h}{2f}} T_R+\sin^2{\frac{h}{2f}}(\Xtt)_R\right]\,,
\nonumber\\
y c_2 f\, (\overline Q^{\textbf{5}}_L)^{I} U_{I\, 5}t_R&=&
-\frac {y c_2 f}{\sqrt{2}}\sin{\frac{h}{f}} {\overline{t}}_L t_R\,,
\label{mc5}
\eea
while in model \fB\ we have
\bea
y f\,({{\overline{Q}}_L^{{\mathbf{14}}}})^{I\,J} U_{I\, i} U_{J\, 5} \,\Psi_R^i&=&
y f\,\cos{\frac{h}{f}} \bar b_L B_R+
\frac {y f} 2{\overline{t}}_L\left[(\cos{\frac{h}{f}}+\cos{\frac{2h}{f}})T_R+
(\cos{\frac{h}{f}}-\cos{\frac{2h}{f}})(\Xtt)_R\,,
\right]\nonumber\\
{y c_2 f \over 2}({{\overline{Q}}_L^{{\mathbf{14}}}})^{I\,J} U_{I\, 5}U_{J\, 5} \,t_R&=&
-\frac {y c_2 f} {2 \sqrt{2}}\sin{\frac{2h}{f}}{\overline{t}}_L t_R\,.
\label{mc14}
\eea

For the models with the singlet, \oA\ and \oB, the only gauge-fermion interactions come from the covariant derivative
\bea
\frac23g'{\overline{\Psi}}\slashed{B}\Psi&=&-\frac23\frac{g}{c_w} s_w^2{\overline{\widetilde{T}}}\slashed{Z}{\widetilde{T}}
+\frac23e\, {\overline{\widetilde{T}}}\slashed{A}{\widetilde{T}}
\,.
\eea
The couplings with the Higgs come instead from
\bea
y f\,({ {\overline{Q}}_L^{{\mathbf{5}}}})^{I} U_{I\, 5}  \Psi_R &=& -\frac {y f}{\sqrt{2}}\sin{\frac{h}{f}}\overline{t}_L{\widetilde{T}}_R\,,\nonumber\\
y c_2 f\,({{\overline{Q}}_L^{{\mathbf{5}}}})^{I} U_{I\, 5} \,t_R &=&  -\frac {y c_2 f}{\sqrt{2}}\sin{\frac{h}{f}}\overline{t}_L t_R\,,
\label{scalarlagr1A}
\eea
for model \oA\ and from
\bea
{y\over 2} f\,({ {\overline{Q}}_L^{{\mathbf{14}}}})^{I\;j} U_{I\, 5} U_{J\, 5}  \Psi_R &=& -\frac {y f}{2\sqrt{2}}\sin{\frac{2h}{f}}\overline{t}_L{\widetilde{T}}_R\,,\nonumber\\
{y c_2  \over 2} f\,({{\overline{Q}}_L^{{\mathbf{14}}}})^{I\;j} U_{I\, 5} U_{J\, 5}  \,t_R &=&  -\frac {y c_2 f}{2\sqrt{2}}\sin{\frac{2h}{f}}\overline{t}_L t_R\,.
\label{scalarlagr1B}
\eea
for model \fB.

\section{Statistical tools} 

In the analysis performed in the Ref.~\cite{cmsBWt} the $CL_s$ method is used to obtain the exclusion confidence intervals for the mass of the $b^{\prime}$ quark. However this exclusion is made in terms of the pair production cross section assuming some fixed ratio between the yield in dileptons and trileptons channels. In our case this ratio depends on the relative strength of the single and pair production and can significantly deviate from the one used in the Ref.~\cite{cmsBWt}. Thus we want to re-do part of the experimental analysis in order to extract a more model-independent exclusion in terms of the number of di- and trileptons separately.  Though we are not restricted to using the $CL_s$ only, we think that this method is well suited for constraining the parameter space of our model.

To use the $CL_s$ we first construct a test statistics $q$ as a log-ratio of probability density for the signal+background hypothesis to the background hypothesis:
\be
q = -2 \log \prod_{i=2l,3l}{P \left( n_i | s_i+ b_i \right) \over  P \left( n_i | b_i \right)}
\ee
where $n_i$ - number of observed in the pseudo-experiment di- and trilepton events, $s_i$ and $b_i$ - number of the signal and background events respectively. The distribution $P$ for a small number of events can be taken as a Poissonian modified due to the presence of the uncertainties. The largest uncertainty in the experimental analysis comes from the background estimation and can be accounted for by taking a marginal probability density defined as
\be
P \left( n_i | s_i+ b_i \right) = \int {\cal P} \left( n_i | s_i + \nu_i b_i \right) ln{\cal N}(\nu_i,1,\delta_{\nu_i}) d\nu_i
\ee
where ${\cal P}$ stands for a Poissonian distribution of the observed number of events and $ln{\cal N}$ for a log-normal distribution of the nuisance parameters $\nu_i$ centered at the value 1 with a variance corresponding to a relative error in the background estimation $\delta_{\nu_i}^2=\delta_{b_i}^2 / b^2_i$. The analogous definition is taken for the background-only probability distribution $P \left( n_i |  b_i \right)$.

The confidence level of the signal+background (backround only) hypothesis is defined as
\be
CL_{sb(b)} = \int_{q^{obs}}^{\infty} P_{sb(b)}(q) dq
\ee
where $P_{sb(b)}(q)$ is a probability density of $q$ which corresponds to $n_i$ distributed according to the signal+background (background only) hypothesis and $q^{obs}$ corresponds to the observed number of events $n_i^{obs}$. Finally the exclusion confidence level for the signal $s_i$ is:
\be
CL_{excl} = 1-{ CL_{sb} \over CL_{b}}
\ee

Obtained in this way confidence intervals coincide with those given in the Ref.~\cite{cmsBWt} with a relative deviation of excluded pair-production cross section less than 5\%. The difference can be caused by our simplified treatment of the nuisance parameters, i.e. neglecting the signal uncertainties and assuming that the backgrounds of di- and trileptons are completely uncorrelated. 

\begin{figure}[t]
\centering
  \includegraphics[width=0.5\textwidth]{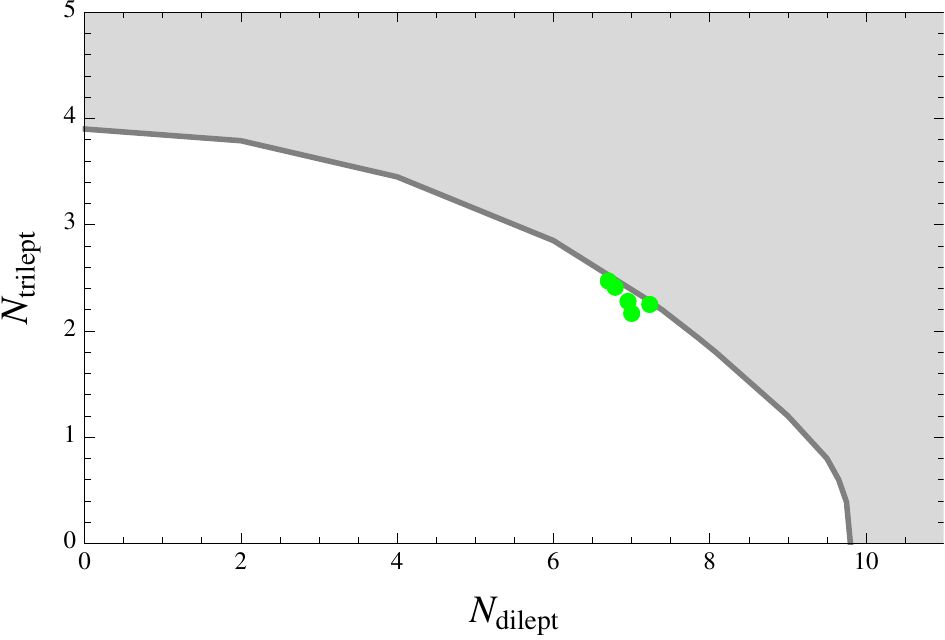}    \\
  \caption{\small{Excluded with 95\%CL values of the signal in same-sign dileptons and trileptons channels (gray area) and the maximal allowed values of the di- and trileptons yields according to Ref.~\cite{cmsBWt} (green points).}}
  \label{fig:exclusionN2N3}
\end{figure}

Using the given above definition we find a region in a plane $(s_2,s_3)$ excluded with 95\%CL (Fig.~\ref{fig:exclusionN2N3}).

\end{document}